\documentclass[twocolumn, trackchanges]{aastex7}

\begin{document}

\title{Photometry of Kuiper belt objects in inner and outer mean motion resonances with Neptune}

\author[orcid=0000-0002-1506-4248,gname='Audrey', sname='Thirouin']{Audrey Thirouin}
\affiliation{Lowell Observatory, 1400 W Mars Hill Road, Flagstaff, Arizona, 86001, United States of America.}
\email[show]{thirouin@lowell.edu}  

\author[orcid=0000-0003-3145-8682,gname=Scott S., sname='Sheppard']{Scott S. Sheppard} 
\affiliation{Earth \& Planets Laboratory, Carnegie Institution for Science, \\ 5241 Broad Branch Road NW, Washington, District of Columbia, 20015, United States of America.}
\email{ssheppard@carnegiescience.edu}

\begin{abstract}

We present a time-resolved photometric survey of Kuiper Belt Objects (KBOs) trapped in the inner and outer Neptune resonances. Lightcurves and rotational properties provide constraints on the shapes, internal structures, collisional histories, and formation environments of these primitive bodies. Our survey was conducted using the \textit{Magellan-Baade} telescope and the \textit{Lowell Discovery Telescope}, allowing us to go fainter than past works and yielding lightcurve studies for 41 KBOs in the 1:1, 5:4, 4:3, 11:5, 7:3, 12:5, and 5:2 resonances. We obtained complete lightcurves for objects displaying large amplitudes, with two KBOs in the 5:2 resonance (2001 XQ$_{254}$, 2013 RZ$_{108}$), one object in the 7:3 (2013 TJ$_{159}$), and one object in the 12:5 (2015 AR$_{293}$). Their lightcurve amplitudes range from 0.26 to 0.46mag, with rotational periods between 5.35 and 11.63 h. 2013 TJ$_{159}$ has the fastest rotation, largest amplitude, and most asymmetric lightcurve in our sample. Combining our results with published studies, we find that inner and outer resonant KBOs generally display lower lightcurve amplitudes than the dynamically Cold Classical population and the 5:3 and 7:4 resonances. The absence of the amplitude--period trends observed in those populations, and their surface color distributions, further suggests that inner and outer resonances do not share the same origin or evolutionary history as the Cold Classicals and overlapping resonances. Based on the literature and our survey, only one nearly equal-sized contact binary, 2004 TT$_{357}$, is currently known among the surveyed populations, implying a lower limit to the contact binary fraction of $\sim$12-15\% in the 5:2 resonance.
 
\end{abstract}

\keywords{\uat{Photometry}{1234} --- \uat{Resonant Kuiper belt objects }{1396} --- \uat{Trans-Neptunian objects}{1705}}


\section{Introduction} 
\label{sec:intro}

Kuiper Belt Objects (KBOs) are among the most primitive remnants of Solar System formation, preserving key information about the physical conditions and processes that governed planetesimal formation in the outer Solar System \citep{Morbidelli2020}. The Kuiper Belt is dynamically structured and comprises several distinct subpopulations: (1) the \textit{Classical} belt, including dynamically Hot and Cold objects between $\sim$40–48~au; (2) \textit{Scattered Disk Objects} (SDOs), which occupy highly eccentric and inclined orbits with perihelia near Neptune; (3) \textit{Detached} objects, characterized by large perihelion distances that place them largely beyond Neptune’s direct gravitational influence; and (4) \textit{Resonant} objects, which are trapped in mean motion resonances with Neptune \citep{Volk2024, Gladman2021, Gladman2008}.

\begin{figure}
 \includegraphics[width=9.5cm,angle=0]{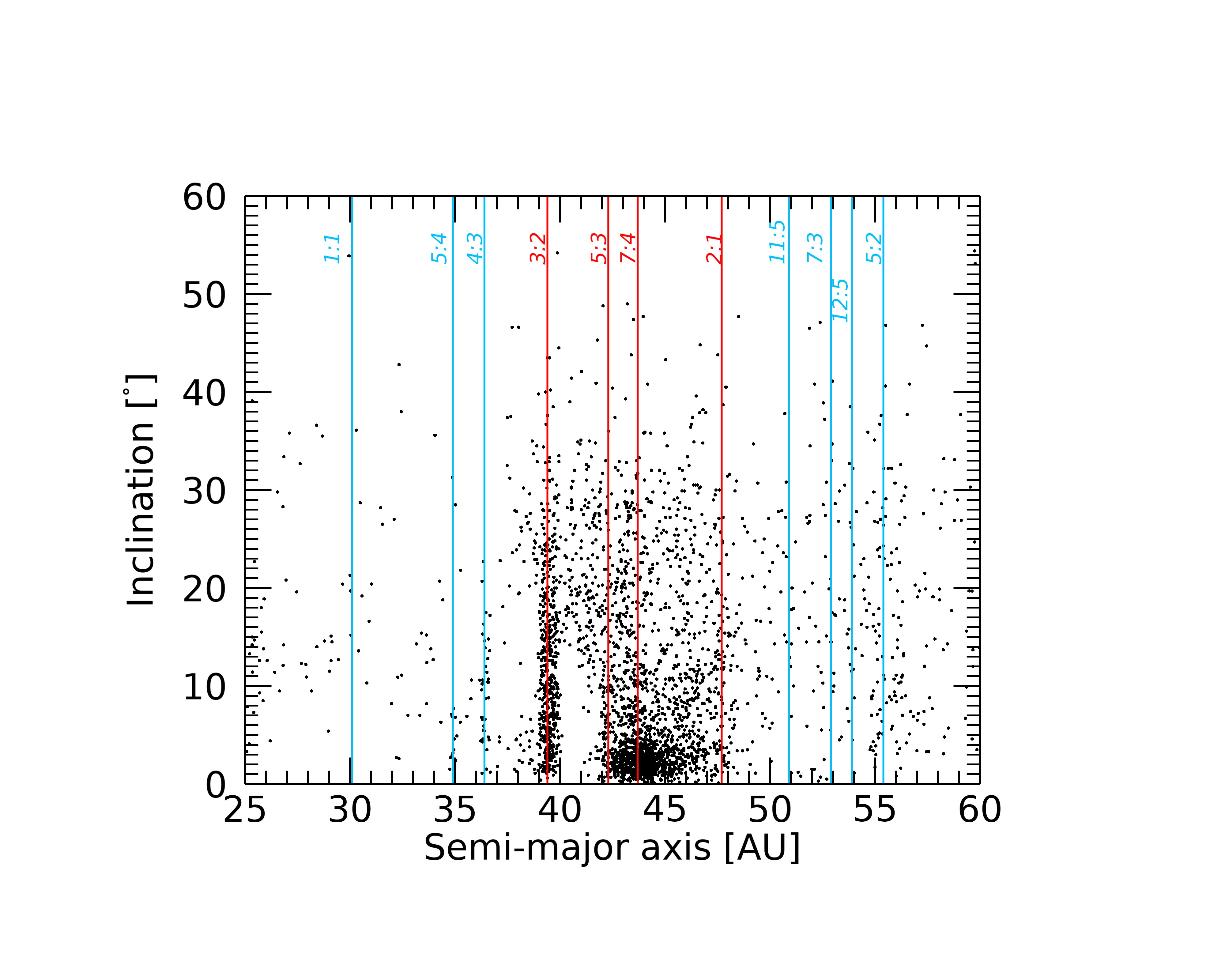} 
\caption{Outer Solar System small bodies have been plotted with black dots while some mean motion resonances with Neptune are indicated with colored vertical lines. The 3:2, 5:3, 7:4, and 2:1 mean motion resonances (red vertical lines) have already been surveyed and discussed in \citet{ThirouinSheppard2018, ThirouinSheppard2022, ThirouinSheppard2024}. Inner and outer mean motion resonances, highlighted by cyan vertical lines, are the focus of this work. We note that additional resonances are present at and beyond these semi-major axes, but they are not plotted for clarity. See \citet{Gladman2021} for more details about other resonances.   }
\label{fig:OrbElements}
\end{figure}

Resonant objects—the focus of this work—were likely captured into mean motion resonances during the outward migration of Neptune \citep{Levison2008}. An object is in resonance when its orbital period forms a ratio of integers with that of Neptune. The corresponding semi-major axis of an $x:y$ resonance can be approximated as:
\begin{equation}
a_{x:y}= a_{\mathrm{Neptune}}\left(\frac{x}{y}\right)^{2/3}
\end{equation}
where $a_{\mathrm{Neptune}}$ is Neptune’s semi-major axis and $x$ and $y$ are positive integers. Strong and weak mean motion resonances with Neptune span the Kuiper Belt (Figure~\ref{fig:OrbElements}; \citet{Gladman2008}). Resonances such as the 5:3 and 7:4 overlap with the Classical subpopulation, while inner (outer) resonances lie before (beyond) the Classical belt, which is delimited by the 3:2 and 2:1 mean motion resonances. The different resonances exhibit diverse properties, including distinct surface color distributions, varying fractions of resolved and contact binaries, as well as population size \citep{ThirouinSheppard2024, Pike2023, Noll2020, Sheppard2012, MurrayClay2011}. While some resonances share similar characteristics, others are markedly different, suggesting that resonant populations sample bodies formed over a range of heliocentric distances and reflect differing formation and evolutionary pathways, likely shaped as Neptune migrated outward and implanted objects into the outer Solar System \citep{MurrayClay2011}. 

While surface colors and resolved binary fractions have been studied for several resonances, their rotational properties and their fractions of contact binaries remain comparatively poorly constrained \citep{Sheppard2012, Noll2020, Pike2023}. Recent surveys have begun to address this gap through lightcurve studies of objects in the 3:2, 5:3, 7:4, and 2:1 resonances \citep{ThirouinSheppard2018, ThirouinSheppard2022, ThirouinSheppard2024}. These surveys have estimated the lower fractions of contact binaries in the mentioned resonances, but have also demonstrated, for example, that the rotational properties of the 5:3 and 7:4 resonant KBOs are similar to those of the dynamically Cold Classical KBOs. Building on these efforts, we present a photometric survey of objects in the inner and outer resonances with Neptune, aimed at deriving partial and complete lightcurves and identifying candidate contact binaries while comparing the studied resonances.

  In this paper, we will present our survey and target selection as well as the data reduction and analysis (Section 2). In Section 3, we will present our main results, while Section 4 will focus on the comparison of the inner and outer resonances as well as the resonances overlapping with the Classical populations, and will estimate the fraction(s) of contact binaries in the studied resonances. Finally, Section 5 will summarize our main findings.  


\section{Photometric survey and Data} 
\label{sec:survey}
\subsection{Target Sample and Selection}

As of March 2026, the ongoing dynamical KBO analysis webpage of the \textit{Deep Ecliptic Survey (DES)}$\footnote{\url{https://www2.boulder.swri.edu/~buie/kbo/desclass.html}}$ reports 31 resonant KBOs trapped in the 1:1 mean motion resonance with Neptune (i.e, Neptune Trojans), 18 KBOs in the 5:4, 53 in the 4:3, 1 in the 11:5, 15 in the 7:3, 6 in the 12:5, and 79 in the 5:2 \citep{Elliot2005, Adams2014}. In total, this corresponds to 102 KBOs in the inner resonances and 101 in the outer resonances considered in this work. Targets were selected based on the following criteria: (1) apparent magnitude brighter than V$\sim$23–23.5~mag; (2) a broad size range, corresponding to absolute magnitudes H$\sim$6–9~mag; (3) no previously published time-resolved photometric studies; and (4) coverage of a range of eccentricity and inclination.

\subsection{Telescopes and Instruments}

We obtained photometric data between March 2017 and March 2026 using the 6.5~m \textit{Magellan-Baade} telescope at Las Campanas Observatory in Chile and the 4.3~m \textit{Lowell Discovery Telescope (LDT)} near Happy Jack (Arizona) in the USA, enabling coverage of both hemispheres. We used the Inamori-Magellan Areal Camera and Spectrograph (IMACS; 27.4$\arcmin$ field of view, 0.20$\arcsec$/pixel) at \textit{Magellan-Baade} and the Large Monolithic Imager (LMI; 12.5$\arcmin \times 12.5\arcmin$ field of view, 0.12$\arcsec$/pixel unbinned) at the \textit{LDT}.

To maximize signal-to-noise while minimizing fringing, we primarily used broadband filters that covered mostly the Sloan $g'$ and $r'$-band or Johnson-Kron-Cousins V and R-band wavelengths: WB4800–7800 at \textit{Magellan-Baade} and VR at the \textit{LDT}. One observing run used the Sloan $r'$ filter to mitigate moonlight. Exposure times were adjusted based on telescope, target brightness, and observing conditions. The observing setup and strategy was the same as our previous works (see \citep{ThirouinSheppard2024, ThirouinSheppard2022}).

The observing log (Table~\ref{tab:Summary_photo}) summarizes the observing geometry for each run, along with the targets’ absolute magnitudes (H) and orbital elements (semi-major axis (a), eccentricity (e), and inclination (i)). Orbital elements and absolute magnitudes were retrieved from the Minor Planet Center (MPC) in March 2026.

\subsection{Data reduction and analysis}  

We typically scheduled 3 to 5 targets per night and cycled between them to maximize telescope efficiency. This strategy enables the construction of partial lightcurves over several hours, with sufficient sampling to place lower limits on rotation periods and lightcurve amplitudes. However, weather conditions and technical issues occasionally limited the temporal coverage. To confirm initially sparse lightcurves, we attempt to reobserve each target at least once, subject to scheduling and weather constraints. Objects exhibiting partial lightcurve amplitudes of $\gtrsim$0.30~mag are then prioritized for follow-up, with additional observations over multiple nights to obtain their complete lightcurves.  

Bias frames and dome flats were acquired each night to construct median bias and median dome flat calibration frames for image reduction. After calibration, target and comparison star fluxes were extracted using standard aperture photometry routines \citep{Thirouin2010, ThirouinSheppard2024}. Periodicity was analyzed with the Lomb periodogram \citep{Lomb1976}. The highest peak identifies the strongest periodic signal; however, the true rotational period may be a multiple of this value \citep{Thirouin2010}. For example, double-peaked lightcurves have periods twice that indicated by the periodogram \citep{Sheppard2008, ThirouinSheppard2024}. We also note that alias solutions may remain viable and require additional data to be confirmed or rejected. All complete lightcurves have been fitted with a second-order Fourier series \citep{ThirouinSheppard2024}.

Once a lightcurve is obtained, it can be used to infer basic physical properties of the object \citep{Sheppard2008, SheppardJewitt2004, Pravec2002, Leone1984}. Flat or low-amplitude lightcurves ($\Delta m \lesssim 0.2$~mag) are typically associated with nearly spherical or spheroidal bodies (Maclaurin spheroids), sometimes with surface albedo variations, as in the case of Pluto \citep{Buie1997}. Moderate-amplitude sinusoidal lightcurves (up to $\sim$0.4~mag) are generally attributed to elongated triaxial ellipsoids (Jacobi ellipsoids) with axes $a>b>c$ rotating about the $c$-axis, as is the case for Haumea \citep{Lacerda2008}. Such lightcurves are well reproduced by second-order Fourier fits, as used in this work. In contrast, large-amplitude, non-sinusoidal lightcurves displaying broad U-shaped maxima and sharp V-shaped minima are characteristic of near equal-sized contact binaries, as is the case of 2001~QG$_{298}$ \citep{Lacerda2011, SheppardJewitt2004}. Further details on lightcurve interpretation can be found in \citet{ThirouinSheppard2024, Sheppard2008, Lacerda2007, Pravec2002, Leone1984}.

As shown by \citet{Binzel1989}, the elongation ($a/b$) of a triaxial body can be estimated from its lightcurve amplitude assuming a viewing geometry. For the simplest case of an equatorial view ($90^{\circ}$), the relation between amplitude ($\Delta m$) and elongation becomes $\Delta m = 2.5 \log(a/b)$. Because the viewing angle is unknown, the derived $a/b$ ratio represents a lower limit. Combining the rotational period with this lower limit on $a/b$, one can also estimate a lower limit to the bulk density ($\rho$). Assuming the object is a triaxial ellipsoid in hydrostatic equilibrium observed equatorially, the density can be derived following \citet{Chandrasekhar1987}, where regions in the density-rotation rate space where different and allowed figures of equilibrium are studied. We emphasize that from the lightcurve amplitude at a single epoch, we can constrain an object's lower limit on its elongation ($a/b$), but the c-axis is derived assuming a figure of equilibrium \citep{Chandrasekhar1987}. Multiple lightcurves at significantly different epochs and viewing geometries are required to better determine the full set of axis ratios.

\section{Observations and Analysis} 
\label{sec:obs}

In this section, we present the short-term variability results of our survey. We first describe objects with complete lightcurves, including estimates of their rotational periods and amplitudes. We then examine partial lightcurves that show some variability over the observed time span. Finally, we briefly discuss flat lightcurves that exhibit no detectable variability over the observation's duration. Partial and flat lightcurves are available in the Appendix~\ref{sec:appA}, while the complete (or full) lightcurves are in the main paper. Photometry is available in Appendix~\ref{sec:appB}.

\subsection{Complete lightcurves}  

\begin{figure}
    \includegraphics[width=9.5cm,angle=0]{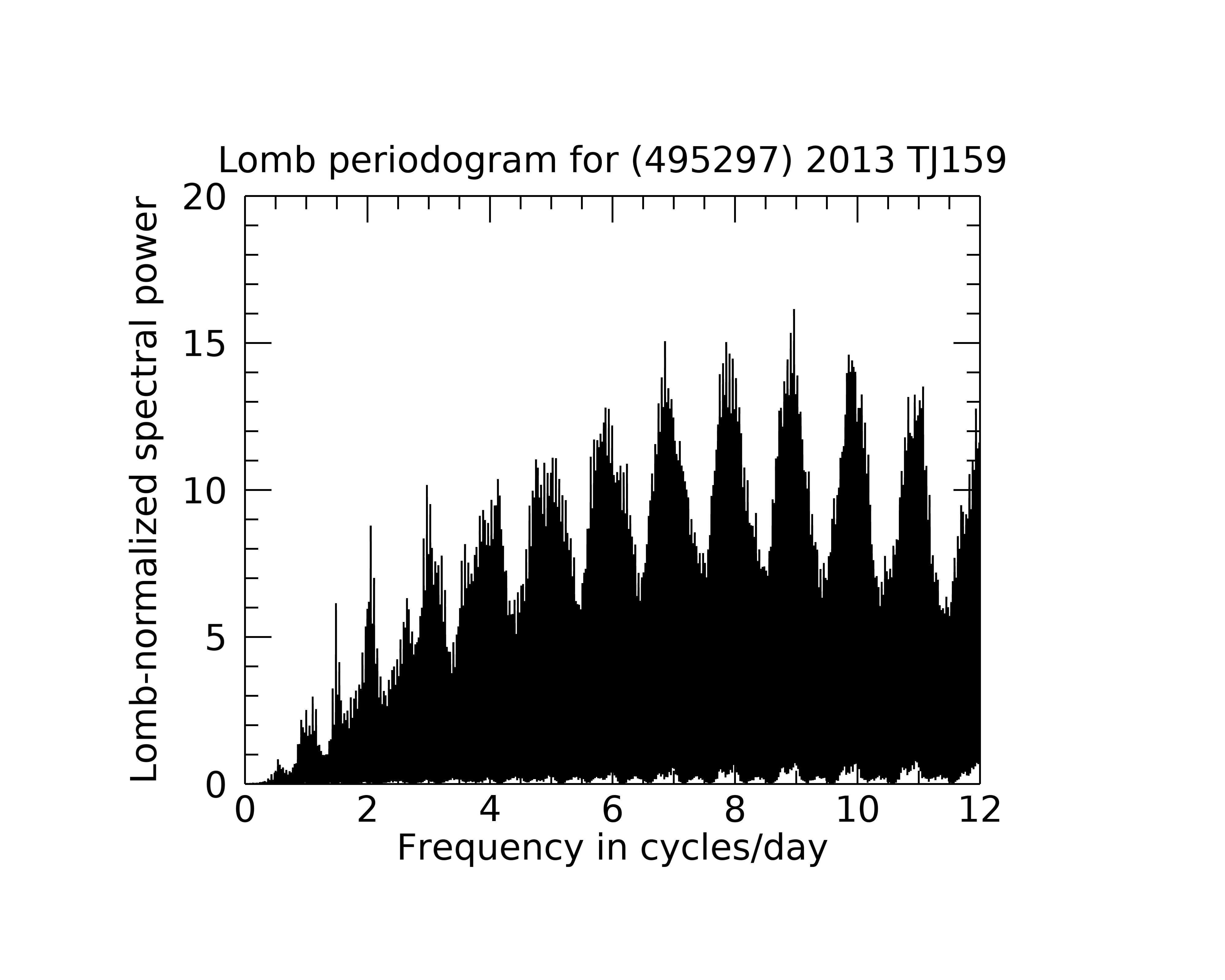} 
    \includegraphics[width=9.5cm,angle=0]{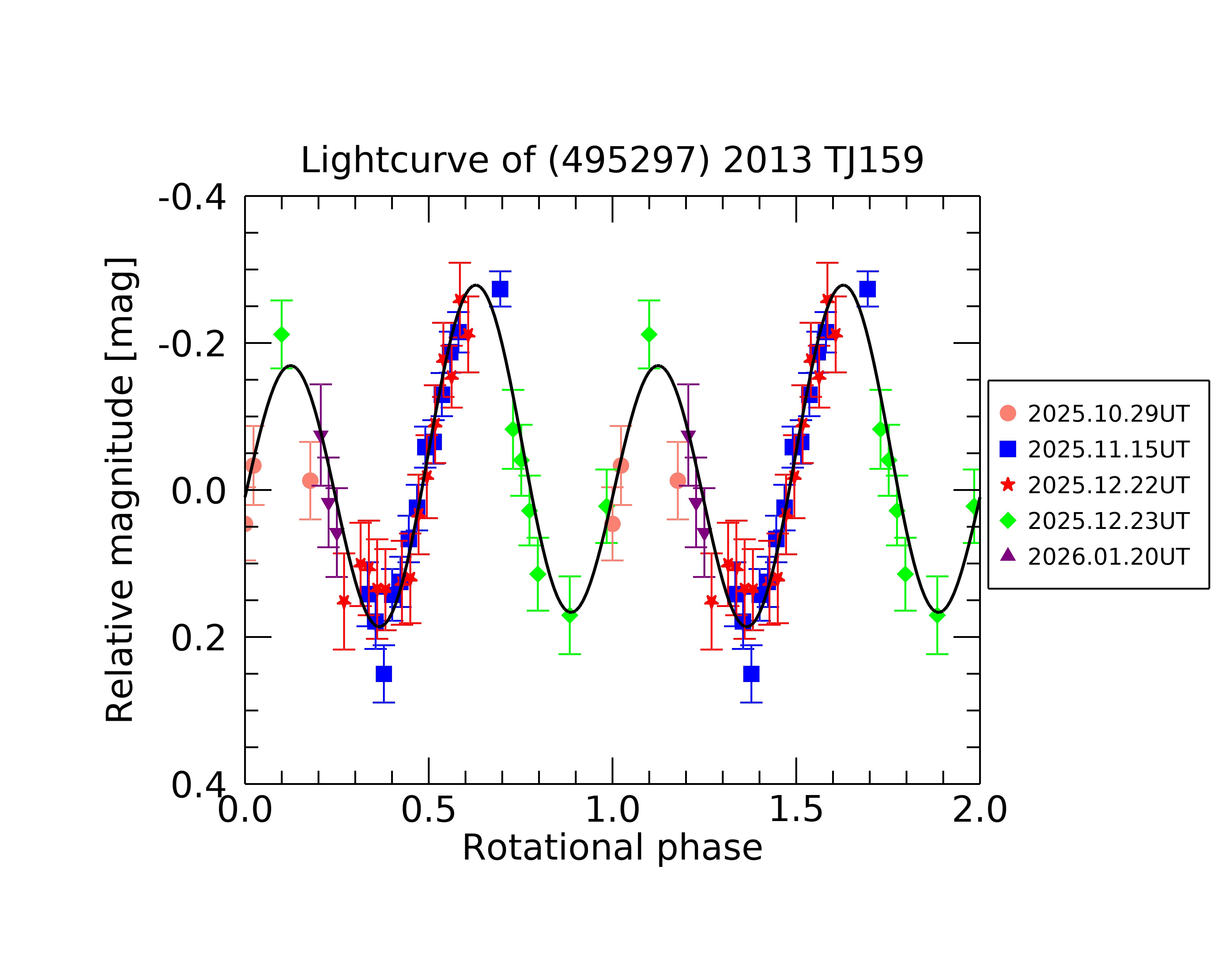} 
\caption{The main peak of the Lomb periodogram (upper plot) is located at 8.96~cycles/day. The double-peaked lightcurve of 2013~TJ$_{159}$ with a rotational period of 5.35$\pm$0.04~h has
an amplitude of 0.46$\pm$0.03~mag (lower plot). }
\label{fig:TJ159}
\end{figure}

\paragraph{(495297) 2013~TJ$_{159}$} This KBO, in the outer 7:3 mean motion resonance with Neptune, was observed over five nights between October 2025 and January 2026 with the \textit{Magellan-Baade} telescope, yielding its first reported lightcurve. It is the fastest rotator and exhibits the largest amplitude in our sample. The Lomb periodogram (Figure~\ref{fig:TJ159}) indicates a very fast double-peaked rotational period of 5.35$\pm$0.04~h and a peak-to-peak amplitude of 0.46$\pm$0.03~mag. 

The lightcurve is strongly asymmetric, with a $\sim$0.12~mag difference between maxima. Assuming an equatorial viewing geometry and hydrostatic equilibrium, we infer a highly elongated triaxial shape ($a>b>c$), with axis ratios $b/a = 0.65$ and $c/a = 0.46$, and a lower limit to the density of $\rho \gtrsim 1.46$~g~cm$^{-3}$ \citep{Chandrasekhar1987}. 

 The fastest rotator known to date in the Kuiper belt is Haumea, with a double-peaked rotational period of 3.92~h \citep{Rabinowitz2006, Lacerda2008, Thirouin2016}. With a rotational period of 5.35~h, 2013~TJ$_{159}$ is among the fastest rotators in our resonant sample and lies at the short-period end of the known distribution for KBOs. The mean rotational period of KBOs tends to be about 10-11~h, with most above about 8 hours, and thus the fast rotation of 2013~TJ$_{159}$ stands out \citep{ThirouinSheppard2024, Thirouin2016}. While similarly fast rotators have been identified in other KBO subpopulations, rotations below 6 hours are very rare, suggesting that 2013~TJ$_{159}$ represents the high-spin, highly elongated tail of the distribution. Future observations are needed to determine if the amplitude increases or decreases and to determine its pole orientation as the viewing geometry of the object changes due to its orbital movement.

\begin{figure}
    \includegraphics[width=9.5cm,angle=0]{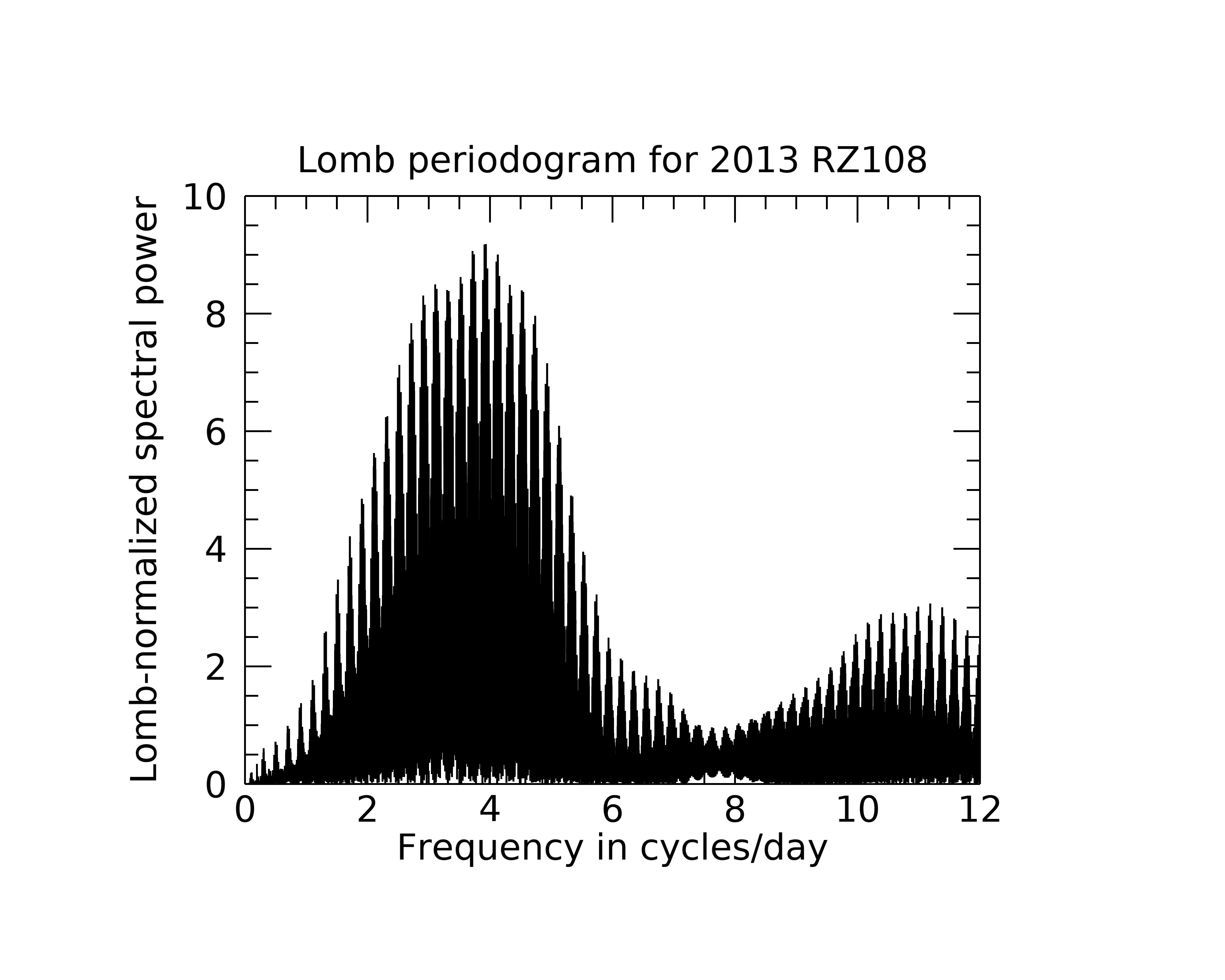} 
    \includegraphics[width=9.5cm,angle=0]{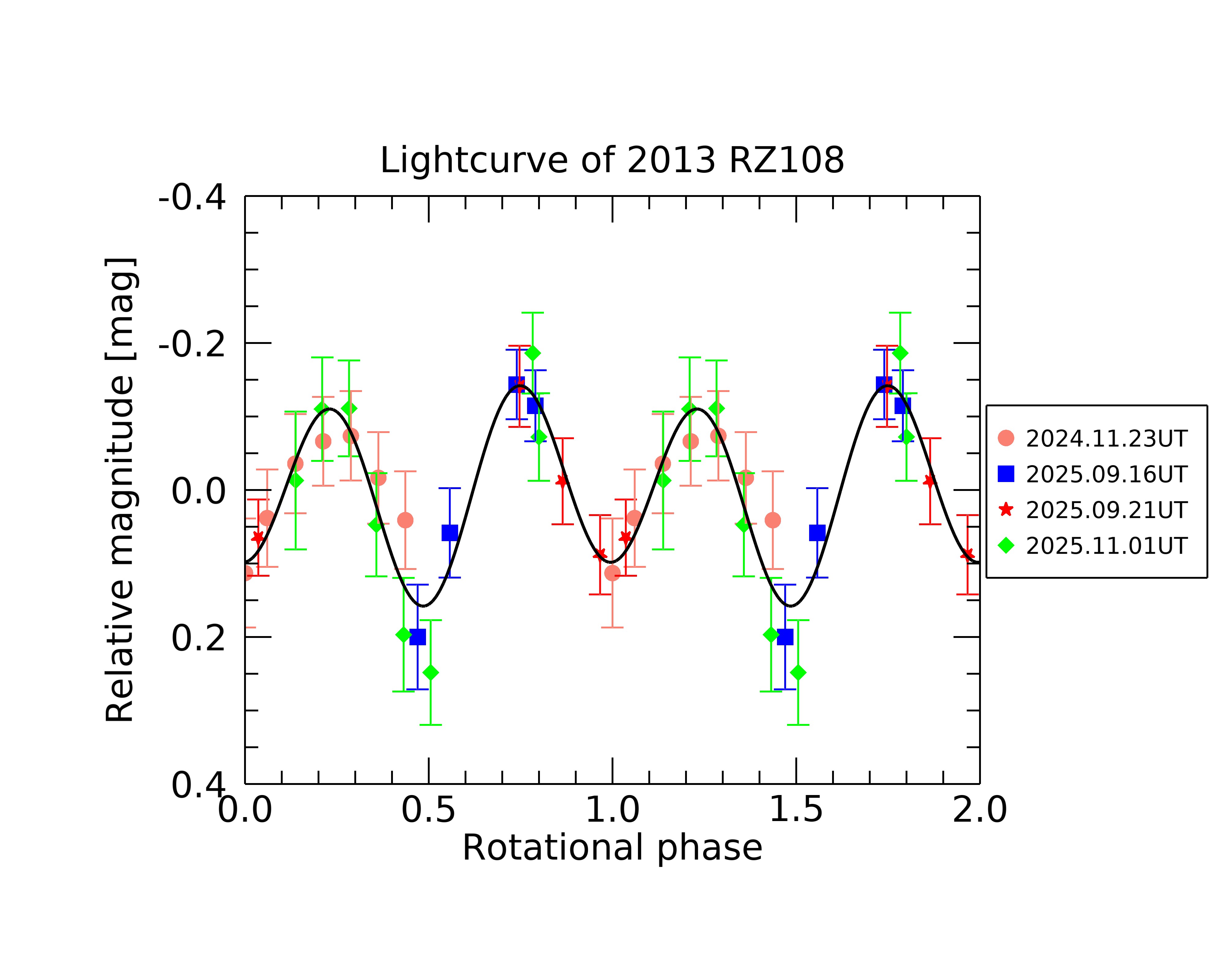} 
 \caption{The KBO 2013~RZ$_{108}$ has an asymmetric double-peaked lightcurve with a period of 11.63~h (main peak located at 4.13~cycles/day, upper plot) and a peak-to-peak lightcurve amplitude of 0.30~mag (lower plot).}
\label{fig:RZ108}
\end{figure}

\paragraph{2013~RZ$_{108}$} This KBO, in the 5:2 mean motion resonance with Neptune, was observed over four nights between November 2024 and October 2025 with the \textit{LDT}. The combined dataset yields a double-peaked rotational period of 11.63$\pm$0.04~h and a lightcurve amplitude of 0.30$\pm$0.03~mag (Figure~\ref{fig:RZ108}). The lightcurve is slightly asymmetric, with the first peak $\sim$0.05~mag lower than the second one.

The amplitude and near-sinusoidal shape suggest an elongated triaxial body ($a>b>c$). Assuming an equatorial viewing geometry and hydrostatic equilibrium, we derive axis ratios of $b/a = 0.76$ and $c/a = 0.50$, and a lower limit to the density of $\rho \gtrsim 0.30$~g~cm$^{-3}$ \citep{Chandrasekhar1987}.

\begin{figure}
    \includegraphics[width=9.5cm,angle=0]{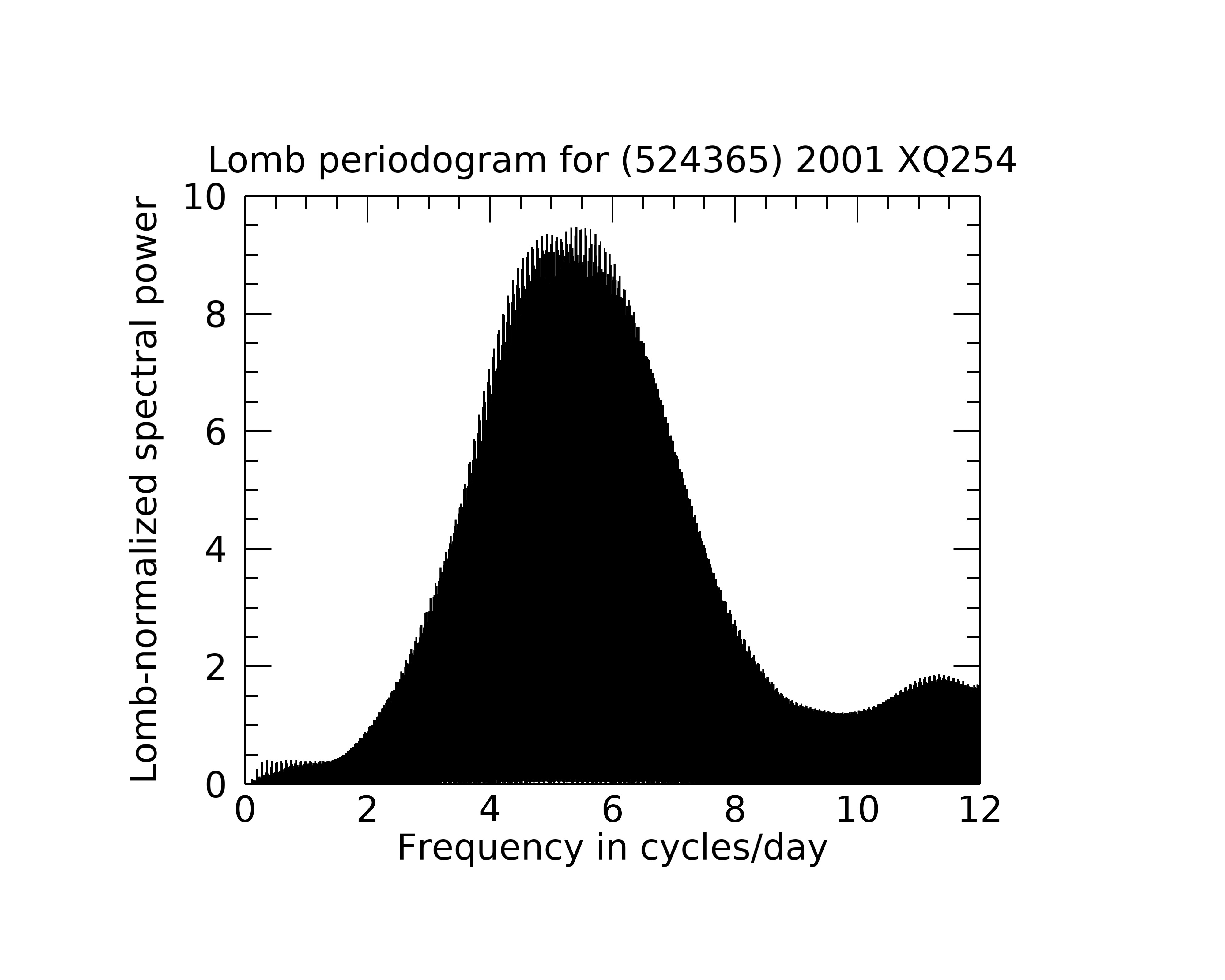} 
    \includegraphics[width=9.5cm,angle=0]{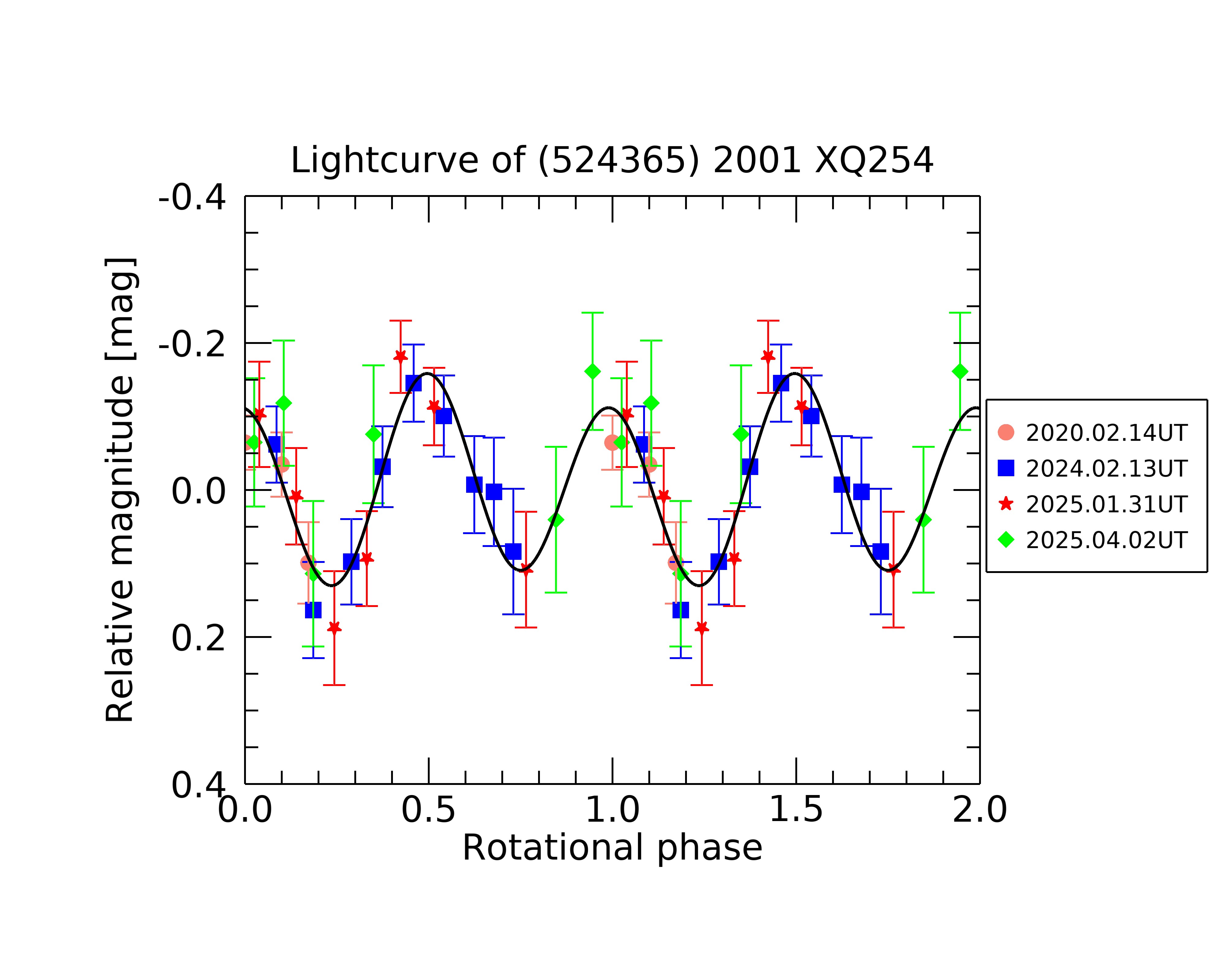} 
 \caption{The main peak of the Lomb periodogram (upper plot) is located at 5.25~cycles/day. The double-peaked lightcurve of 2001~XQ$_{254}$ with a rotational period of 9.15$\pm$0.05~h has
an amplitude of 0.29$\pm$0.04~mag (lower plot). A Fourier series fit (black continuous line) is overplotted, and we note an asymmetry of about 0.05~mag between the first and second maxima. }
\label{fig:XQ254}
\end{figure}

\paragraph{(524365) 2001~XQ${254}$} Located in the outer 5:2 mean motion resonance with Neptune, 2001~XQ${254}$ was observed over four nights in 2020, 2024, and 2025 with the \textit{LDT}. Combining all datasets, we derive a complete rotational lightcurve. The object exhibits a double-peaked lightcurve with a rotational period of 9.15$\pm$0.05~h and an amplitude of 0.29$\pm$0.04~mag from a second-order Fourier fit (Figure~\ref{fig:XQ254}).

The moderate amplitude suggests an elongated triaxial shape ($a>b>c$) rotating about the $c$-axis. Assuming an equatorial viewing geometry and hydrostatic equilibrium, we estimate axis ratios of $b/a = 0.77$ and $c/a = 0.50$, and a lower limit to the density of 0.48~g~cm$^{-3}$ \citep{Chandrasekhar1987}. The lightcurve is slightly asymmetric, with the second maximum exceeding the first by $\sim$0.05~mag, likely indicating surface albedo variations.

\begin{figure}
    \includegraphics[width=9.5cm,angle=0]{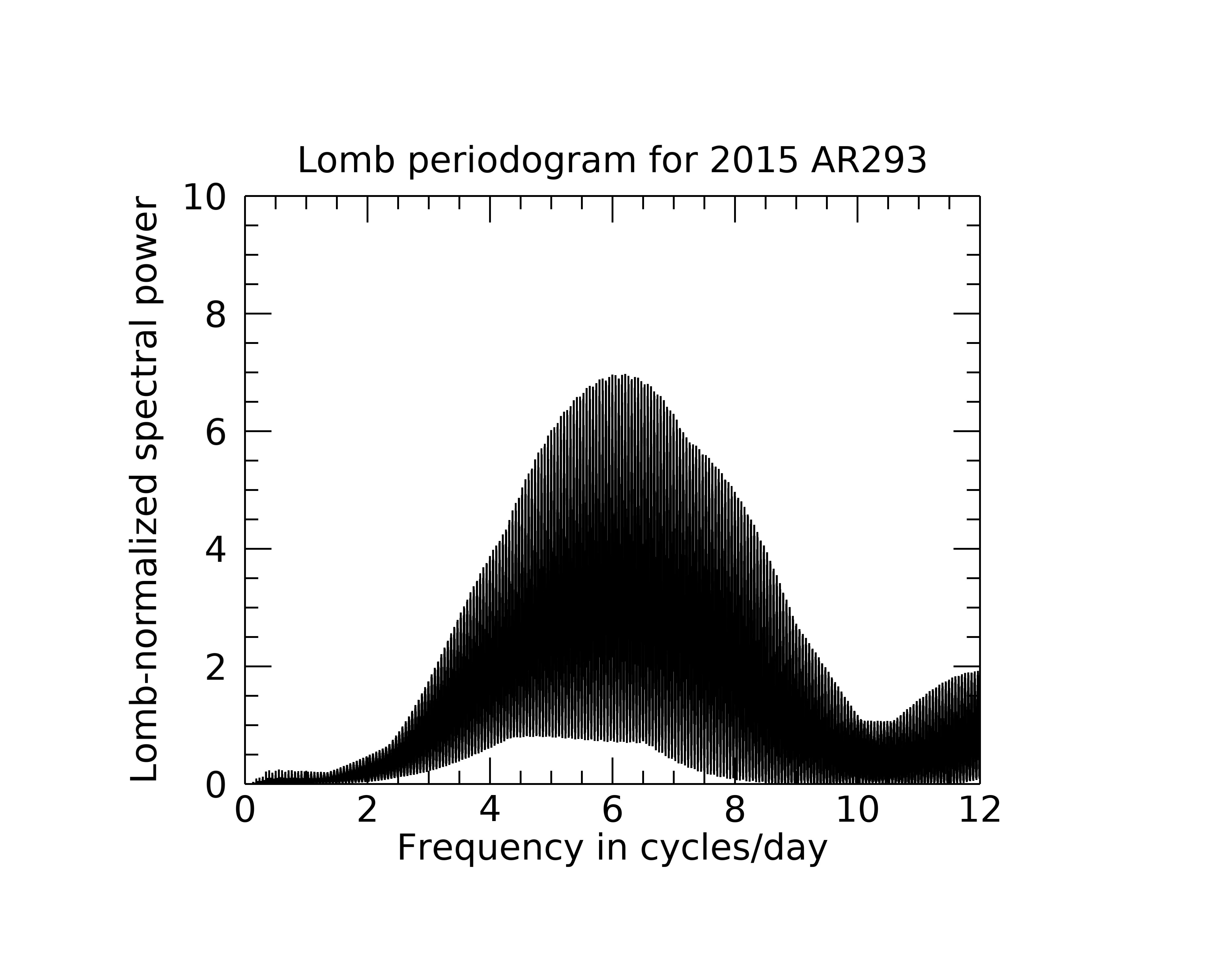} 
    \includegraphics[width=9.5cm,angle=0]{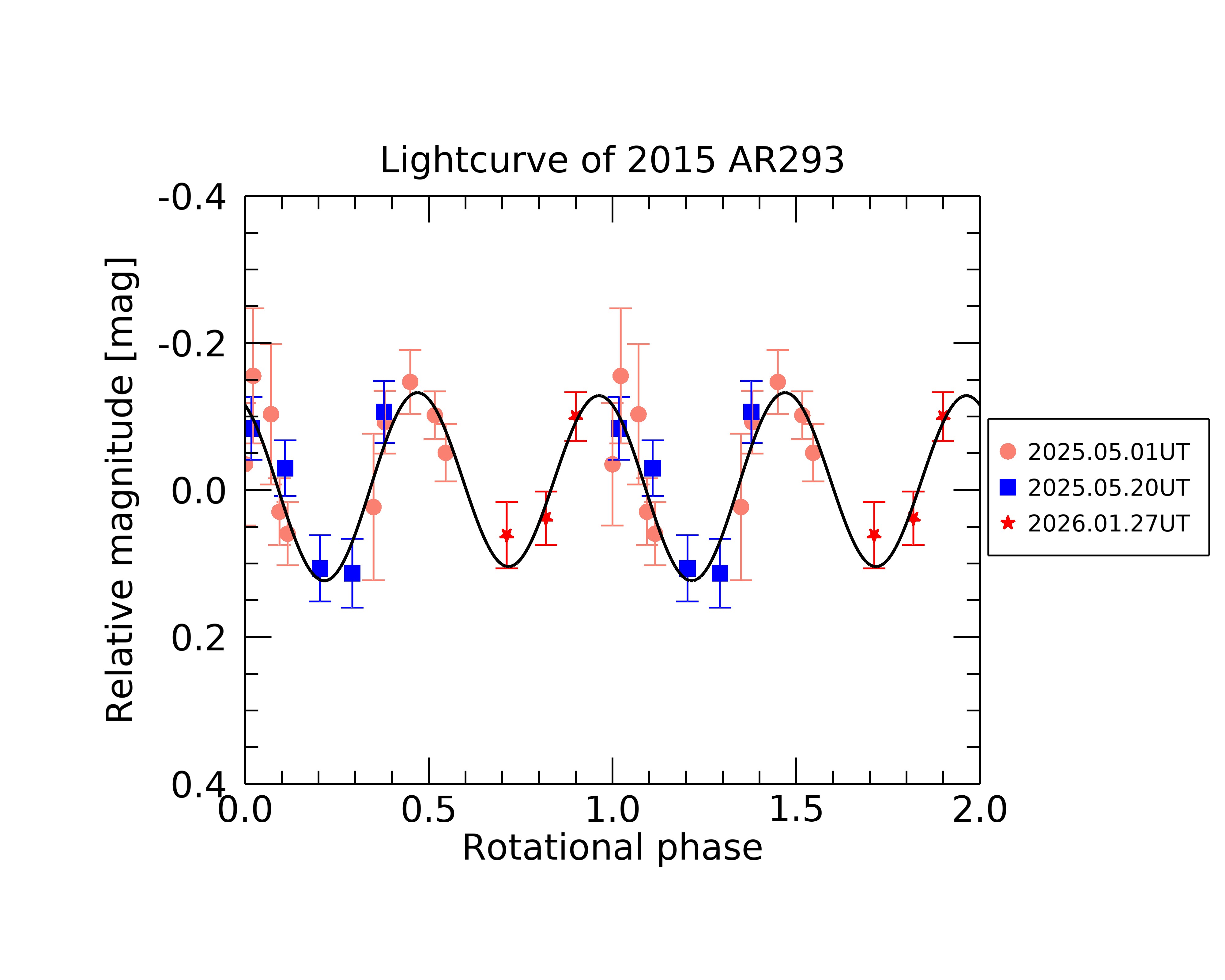} 
\caption{The main peak of the Lomb periodogram (upper plot) is located at 6.21~cycles/day. The symmetric double-peaked lightcurve of 2015~AR$_{293}$ with a rotational period of 7.73$\pm$0.07~h has an amplitude of 0.26$\pm$0.05~mag (lower plot).}
\label{fig:AR293}
\end{figure}

\paragraph{2015~AR$_{293}$} This object, in the 12:5 mean motion resonance with Neptune, was observed on three occasions between May 2025 and January 2026 with the \textit{LDT}. The combined dataset yields a symmetric double-peaked lightcurve with a rotational period of 7.73$\pm$0.07~h and a moderate amplitude of 0.26$\pm$0.05~mag (Figure~\ref{fig:AR293}).

The lightcurve morphology and amplitude indicate a moderately elongated triaxial shape ($a>b>c$). Assuming an equatorial viewing geometry and hydrostatic equilibrium, we derive axis ratios of $b/a = 0.79$ and $c/a = 0.51$, and a lower limit to the density of $\rho \gtrsim 0.66$~g~cm$^{-3}$ \citep{Chandrasekhar1987}.

\subsection{Partial lightcurves}

 \paragraph{1:1 mean motion resonance (Neptune trojans)} We observed 2006~RJ$_{103}$ for $\sim$7~h in 2024 with the \textit{LDT}. Over this interval, the object displayed a variability of 0.14~mag, indicating a rotational period longer than 7~h and a lightcurve amplitude exceeding 0.14~mag. Similarly, 2010~TS$_{191}$ was imaged over about 5.5~h and presents a lightcurve amplitude larger than 0.15~mag. 
 
  \paragraph{5:4 mean motion resonance} Based on a single observing block, 2003~QB$_{92}$ likely has a rotational period longer than 5~h and a lightcurve amplitude exceeding 0.22~mag. The KBO 2005~SC$_{278}$ was observed on four occasions with the \textit{LDT} in 2023 and 2024. During the third night, the data appear to cover both a minimum and a maximum, suggesting a lightcurve amplitude of $\sim$0.2~mag and a rotational period longer than 8~h. Although all datasets were combined to search for a complete rotational lightcurve, no significant periodic signal was detected.
  
 \paragraph{4:3 mean motion resonance} 1998~UU$_{43}$ and 2013~RQ$_{157}$ were each observed once with the \textit{LDT} over intervals of 4~h and 4.5~h, displaying variabilities of 0.13~mag and 0.17~mag, respectively. In contrast, 2013~RW$_{124}$ and 2014~WE$_{509}$ were observed during two separate runs, both yielding consistent results. We infer rotational periods longer than 4~h for both objects, with 2013~RW$_{124}$ exhibiting a larger variability ($>$0.2~mag) than 2014~WE$_{509}$ ($>$0.1~mag).
 
  \paragraph{7:3 mean motion resonance} The partial lightcurve of 2001~XT$_{254}$ suggests that this KBO rotates in more than 5.5~h and has a variability larger than 0.2~mag based on one observing night.
  
  \paragraph{12:5 mean motion resonance} We observed 1999~CC$_{158}$ for approximately 5~h with the \textit{LDT} in February 2026. Its partial lightcurve suggests a rotational period longer than 5~h and a lightcurve amplitude exceeding 0.11~mag.
  
 \paragraph{5:2 mean motion resonance} We obtained short observing blocks of $\sim$2~h for 2009~YG$_{19}$, 2014~JX$_{80}$, and 2015~BC$_{519}$ over one or more nights. Their lightcurve amplitudes exceed 0.15~mag, 0.2~mag, and 0.2~mag, respectively. 2013~GY$_{136}$ and 1999~HB${12}$ were observed over intervals of $\sim$3.5~h on one and two occasions, respectively. The variability of 1999~HB$_{12}$ reaches $\sim$0.2~mag, while 2013~GY$_{136}$ displays a smaller variability ($>$0.1~mag). Single-night observations spanning $\sim$4~h indicate amplitudes larger than 0.19~mag for 2013~LZ$_{36}$, 0.12~mag for 2014~SW$_{373}$, and 0.06~mag for 2014~YL$_{50}$. From two observing runs in 2021 and 2024, we infer that 2015~KH$_{174}$ has a rotational period longer than 4.5~h and a variability exceeding 0.2~mag. Similarly, observations of 2005~XN$_{113}$ obtained during three runs in 2019 and 2025 suggest a rotational period longer than 5~h and an amplitude greater than 0.2~mag. Two observing runs in 2020 and 2023 indicate that 2015~BD$_{519}$ has a rotational period exceeding 5.5~h and a lightcurve amplitude above 0.2~mag. Finally, observations of 2014~UW$_{224}$ obtained in 2024 and 2026 with the \textit{LDT} suggest a rotational period longer than 8~h and a variability greater than 0.2~mag.
  
\subsection{Flat lightcurves}    

Several objects observed in our survey exhibit extremely low to nonexistent variability over the durations of the observing blocks. Reasons to explain such flat lightcurves are multiple: (1) a spheroidal object with a homogeneous surface (i.e., no albedo spot(s) on the object's surface), (2) an object with a rotational period significantly longer than the observing block, and/or (3) a pole-on orientation. 
Flat lightcurves have been found in nearly all the surveyed resonances: 2007~VL$_{305}$, 2010~TT$_{191}$, 2011~SO$_{277}$, and 2012~UD$_{185}$ in the 1:1 resonance; 2014~UC$_{225}$ in the 5:4 resonance, 2003~SS$_{317}$, 2013~RH$_{109}$, 2013~RQ$_{109}$, 2014~DU$_{143}$, 2015~BM$_{518}$, and 2016~SJ$_{57}$ in the 4:3 resonance, 2013~RM$_{109}$ in the 11:5 resonance, 2004~DJ$_{71}$, 2008~CT$_{190}$ in the 7:3 resonance, and 2012~BA$_{155}$ in the 5:2 resonance.

 
\startlongtable
\begin{deluxetable*}{lcccccc|cc|cccc}
\tabletypesize{\scriptsize}
 \tablecaption{\label{tab:Summary_photo} Observing log for all targets in inner and outer resonances with Neptune observed with the \textit{LDT} and \textit{Magellan-Baade} telescope from 2017 to 2026. Heliocentric and geocentric distances (r$_h$ and $\Delta$) as well as phase angle ($\alpha$) at the time of the observations are indicated. \underline{Note}: 2014~SW$_{373}$ has an ambiguous classification of 5:2 resonant or Scattered Disk Object in the \textit{DES} classification. For our purpose, we will consider 2014~SW$_{373}$ as a 5:2 resonant KBO. }
\tablewidth{0pt}
\tablehead{Object  & UT-date  &$\Delta$  &  r$_h$ &   $\alpha$   & Filter & Telescope & Period & Amplitude & H & a &e & i \\
                               &           &    [au]  &  [au]  &  [$^{\circ}$]   & &  & [h] & [mag] & [mag]& [au] & & [$^\circ$] }
\startdata    
\multicolumn{13}{c}{1:1 mean motion resonance}                                                                                                                                                                                                                                                                                                                                                                                               \\
(613490) 2006~RJ$_{103}$ &	10/26/2024	&  29.128  &30.017 & 0.9&  VR 	&	LDT & $>$7 & $>$0.14 & 7.31&30.217 &0.032	& 8.2 \\
(527604) 2007~VL$_{305}$ &	11/30/2019 	&    27.303 & 28.202 &0.8 &  WB 	&	Magellan & - & $\sim$0.15& 8.46 &	30.200	&0.068	&	28.1	 	  \\
&	 12/01/2019 & 27.305 & 28.202 & 0.8 &  WB 	&	Magellan & ...&...&...&...		&...	&	...	 	  \\
&	 12/02/2019 & 27.308 & 28.202& 0.9 &  WB 	&	Magellan & ...&...&...&...		&...	&	...	 	  \\
(666739) 2010~TS$_{191}$ & 01/07/2021 &    27.989 & 28.647  & 1.5  & VR& LDT & $>$5.5  & $>$0.15  &  8.06 & 30.218	& 0.052	& 	6.6     \\  
2010~TT$_{191}$ & 01/07/2021 &  31.378 & 32.191   & 1.0  & VR& LDT &   - & $\sim$0.1  &7.93 &30.265	& 0.064	& 	4.3      \\  
& 02/13/2024 &   31.793&  32.203  & 1.6  & VR& LDT &   - & ...  &... &...	& ... 	& ...     \\  
(530664) 2011~SO$_{277}$ & 10/03/2024	& 29.707 & 30.522 & 1.1 & WB& Magellan 	& -	&$\sim$0.1 & 7.71 &30.367	&0.005	&9.6 \\ 
2012~UD$_{185}$ & 10/26/2024	& 30.682 & 31.537 & 0.9& VR& LDT 	& -	&$\sim$0.1 & 7.53 &30.372		&0.038	&28.2 \\  
& 10/22/2025	&   30.534 & 31.516 & 0.2& VR& LDT 	&...	&...&...&	...	&...	&... \\ 
\multicolumn{13}{c}{5:4 mean motion resonance}                                                                                                                                                                                                                                                                                                                                                                                               \\
(427581) 2003~QB$_{92}$ &  09/03/2024 &  37.136 & 38.105 & 0.4 & VR  & LDT & $>$5& $>$0.22 & 7.41& 35.121 & 0.086 & 3.5 \\  
(308460) 2005~SC$_{278}$ &  10/19/2023 &32.119  & 33.102   & 0.3  & VR  & LDT & $>$8& $\sim$0.2 & 7.31 & 35.118 & 0.071 & 1.5   \\  
&  10/05/2024 &32.145 & 33.056  & 0.7  & VR  & LDT &  -  & - & ... & ... & ... & ...   \\  
&  10/26/2024 &32.061 & 33.054 & 0.1  & VR  & LDT &  -  & - & ... & ... & ... & ...  \\  
&  11/23/2024 &32.155 & 33.050 & 07  & VR  & LDT &  - & - & ... & ... & ... & ...  \\  
2014~UC$_{225}$ &  10/22/2025 &  30.108 & 31.055 & 0.6 & VR  & LDT &  - & $\sim$0.1 & 7.52& 35.342 & 0.121 & 4.9  \\  
\multicolumn{13}{c}{4:3 mean motion resonance}                                                                                                                                                                                                                                                                                                                                                                                               \\
(523955) 1998~UU$_{43}$ &  02/09/2026 & 33.644 & 34.160 &  1.4  & VR  & LDT &  $>$4   & $>$0.13 & 7.23 & 36.634  & 0.129 & 9.6   \\  
(143685) 2003~SS$_{317}$ &  12/13/2023 & 27.966&  28.796  & 1.1  & VR  & LDT &  - & $\sim$0.2 & 8.16 & 36.803 & 0.242 & 5.9   \\  
&  11/23/2024 &27.963 & 28.932 & 0.4  & VR  & LDT &  ... & ... & ... & ... & ... & ...   \\  
2013~RH$_{109}$&  09/16/2025 &  37.571 & 38.554   &  0.3 & VR  & LDT & -  & $\sim$0.1 & 7.6 & 36.439 & 0.073 & 14.8  \\  
 &  09/21/2025 &37.586 & 38.553&  0.4  & VR  & LDT &  ... &... & ... & ... & ... & ...  \\  
2013~RQ$_{109}$ & 10/05/2024 &30.075 & 30.913 & 1.0  & VR  & LDT & -  & $\sim$0.1 & 7.92 & 36.689 & 0.158 & 14.5  \\  
2013~RQ$_{157}$ &  09/03/2024 &34.754 & 35.726 &  0.4  & VR  & LDT &  $>$4.5  & $>$0.17 & 7.89 & 36.634 & 0.075 & 15.5  \\  
2013~RW$_{124}$ &  09/03/2024 &34.278 & 34.850&  1.4  & VR  & LDT & $>$4 & $>$0.2 & 7.55 & 36.754 & 0.070 & 4.8  \\  
&  10/05/2024 &33.943 & 34.855&  0.7  & VR  & LDT & ... & ... & ... & ... & ... & ...  \\  
(533211) 2014~DU$_{143}$&  04/20/2025 &    38.280 & 39.255  &  0.4  & VR  & LDT &  -   & $\sim$0.1 & 6.68 & 36.425 & 0.135 & 13.6 \\ 
(535019) 2014~WE$_{509}$ &05/08/2024 &  33.387 & 33.997 &  1.4  & VR  & LDT &  $>$4   & $>$0.1 & 7.25 & 36.439  & 0.090 & 13.9 \\
&04/20/2025 &   33.233 & 34.066  &  1.0  & VR  & LDT &...  &... & ...& ... &...&... \\ 
(559178) 2015~BM$_{518}$ & 03/25/2026 &  33.455 & 34.428 &0.4  & VR  & LDT &  -  & $\sim$0.1 & 6.42 & 36.343 & 0.119 & 8.8 \\
2016~SJ$_{57}$ & 12/13/2023 &  33.098 & 33.913& 0.9  & VR  & LDT &  -  & $\sim$0.05 & 7.41 & 36.619  & 0.074 & 15.3 \\
\multicolumn{13}{c}{11:5 mean motion resonance}                       \\
 2013~RM$_{109}$ &  09/03/2024 &41.072 & 42.049  & 0.3 & VR  & LDT & ... & $\sim$0.15 & 6.72& 51.306 & 0.229 & 14.3 \\ %
\multicolumn{13}{c}{7:3 mean motion resonance}                      \\
(131696) 2001~XT$_{254}$ & 02/09/2026   &   35.433 &  36.398   &   0.3  & VR  & LDT & $>$5.5 & $>$0.2 & 7.54 & 52.645 & 0.319 &0.5  \\%
(183964) 2004~DJ$_{71}$ & 03/25/2026 & 32.477&  33.453 &0.4 & VR  & LDT &  -  & $\sim$0.1 &7.63 & 52.533 & 0.377 & 11.3 \\%
(523624) 2008~CT$_{190}$ & 03/25/2026 &  35.022&  35.800  &1.0 & VR  & LDT &  -  & $\sim$0.1 & 5.63 & 52.870 & 0.344 & 38.8 \\%
(495297) 2013~TJ$_{159}$ & 10/29/2025  &  37.884 & 38.852  & 0.3 & WB  & Magellan&   5.35  &  0.46$\pm$0.03 & 6.79 & 53.584 & 0.322 & 4.8  \\  
 & 11/15/2025  &  38.004 & 38.861 & 0.7   &  WB  & Magellan&   ... &  ...& ... & ... & ... & ...  \\  
  & 12/22/2025  & 38.504 & 38.881 & 1.3   &  WB  & Magellan&   ... &  ...& ... & ... & ... & ...  \\  
   & 12/23/2025  &  38.522  & 38.882 & 1.4   &  WB  & Magellan&   ... &  ...& ... & ... & ... & ...  \\  
    & 01/20/2026  & 39.014&  38.897 & 1.4   & WB  & Magellan &   ... &  ...& ... & ... & ... & ...  \\  
\multicolumn{13}{c}{12:5 mean motion resonance}                        \\
(79978) 1999~CC$_{158}$ &    02/09/2026   &47.141 & 48.064 & 0.4  & VR &  LDT  & $>$5 &$>$0.11 &5.74 &54.076&0.279&18.7   \\ 
2015~AR$_{293}$ & 05/01/2025  &  34.961 & 35.597   & 1.3    & VR  & LDT &  7.73   & 0.26$\pm$0.05 & 6.85 & 54.047& 0.344 & 24.4   \\
& 05/25/2025  &   35.266  & 35.593   & 1.6  & VR &  LDT  &... &... &... &...&... &...  \\  
& 01/27/2026  &  34.990 & 35.559    & 1.3  & VR &  LDT  &... &... &... &...&... &...  \\  
\multicolumn{13}{c}{5:2 mean motion resonance}                     
  \\	        
(38084) 1999~HB$_{12}$      &    03/10/2017      &32.092&  32.593 &  1.5 & r & LDT& $>$3.5& $>$0.2&7.06 &55.194 &0.410&13.2   \\ 
   	         &    06/08/2019      & 31.659 &32.565  &  0.8 & VR & LDT&  ... & ... & ...   &...&...&...\\
(524365) 2001~XQ$_{254}$ &02/14/2020  & 31.143 & 32.104 & 0.4  & VR &  LDT  & 9.15 & 0.29$\pm$0.04 & 7.89 &55.063 &0.437 & 7.1  \\ 
& 02/13/2024  & 32.003 & 32.972 & 0.3  & VR &  LDT  &... &... &... &...&... &...  \\  
& 01/31/2025  & 32.322 & 33.215 & 0.7  & VR &  LDT  & ... &  ...  &... &...&... &...  \\  
& 04/02/2025  &32.464  &33.258  & 1.1  & VR &  LDT  & ...& ... &... &...&... &...  \\  
  2005~XN$_{113}$   &  10/03/2019    &    37.004 & 37.480 & 1.4   & VR & LDT  & $>$5 & $>$0.2 & 6.5&	54.916 & 0.401  & 3.4  \\
 &  10/06/2019  &  36.958& 37.478&1.3   & VR & LDT  & ... & ...& ... &... & ...  & ... \\
  &  01/31/2025  &   35.360 &  35.970 &1.2   & VR & LDT  & ...& ... & ... &... & ...  & ... \\ %
2009~YG$_{19}$   &   10/16/2023        &   36.064 & 36.455  & 1.4 & VR & LDT  & $>$2& $>$0.15  & 6.2  & 55.222 & 0.404 & 5.2\\  
 (531017) 2012~BA$_{155}$   &  02/28/2019    & 38.023  & 38.983 &  0.4  & VR & Magellan&  - & $\sim$0.1  &6.04& 	55.429 & 	0.376  &14.4    \\
 	                            & 03/01/2019      & 38.023  & 38.982 &  0.4 & WB & Magellan&  ... & ...& ...  &...&...&...  \\
 	                             &    03/02/2019     &  38.024 & 38.981 &  0.4  & WB & Magellan&  ... & ...& ...   &...&...&... \\                                  
  2013~LZ$_{36}$ & 05/19/2020    &   36.060 & 36.979& 0.7  & VR  & LDT  & $>$4  & $>$0.19  &7.1 &	56.074 &0.351 & 13.9   \\ 
 2013~GY$_{136}$     &   05/14/2021     &  32.840 & 33.849  & 0.1  & VR & LDT & $>$3.5  & $>$0.1 & 7.8  &55.974 &0.419 &10.9  \\	
  2013~RZ$_{108}$        &   11/23/2024     & 29.682 & 30.662  & 0.2  & VR & LDT & 11.63  &  0.30$\pm$0.03 & 7.8  & 56.148 &0.460 &13.0  \\	
   &   09/16/2025     &  30.320 & 30.759 & 1.7  & VR & LDT & ... & ...  & ...  & ... &...&...  \\	 
      &   09/21/2025     &  30.247  &30.761  & 1.6  & VR & LDT & ... &... & ...  & ... &...&...  \\	 
    &   10/01/2025     & 29.834 & 30.775 & 0.6 & VR & LDT & ...&... & ...  & ... &...&...  \\	 
  (523713) 2014~JX$_{80}$    &    06/08/2019     & 35.145 &36.058 &  0.7 & VR & LDT& $>$2  & $>$0.2 &6.15   & 55.239 &0.352 &28.9\\ 
  &    08/16/2020     & 35.776  & 36.154  & 1.5 & VR & LDT& ...  & ... & ...   &...&...&... \\ 
    &    06/28/2023     &35.556 & 36.440  & 0.8& VR & LDT& ...  & ... & ...   &...&...&... \\ 
2014~SW$_{373}$ & 09/21/2025  &35.518&  36.222& 1.1 &VR  & LDT   & $>$4 &$>$0.12 & 7.08  & 	56.286 &0.359 & 20.8   \\ 
(544430) 2014~UW$_{224}$ & 02/13/2024    &35.935 & 36.912 & 0.2  &    VR & LDT &  $>$8 &  $>$0.2 &6.77 & 56.098& 0.410 & 2.9  \\  
&    02/09/2026    & 36.519 & 37.504 & 0.1  &    VR & LDT & ... &  ... &... & ...&... &...  \\  
2014~YL$_{50}$ &  12/02/2019    &  38.378  & 39.090& 1.0   &WB   & Magellan   & $>$4  &$>$0.06  & 7.02 & 55.977 & 0.330 & 29.1  \\
2015~BC$_{519}$  &    01/07/2021        & 34.285&  35.173&  0.7 & VR & LDT  &$>$2 & $>$0.2  & 6.80 & 55.701 &0.420 & 1.7   \\
  &    12/13/2023        & 33.909 & 34.433& 1.4 & VR & LDT  & ...  & ...  & ... & ... & ...& ...   \\
  &    03/21/2025       & 33.386  & 34.142 &  1.1 & VR & LDT  & ... & ...  & ... & ... & ...& ...   \\ 
(535991) 2015~BD$_{519}$ &   02/14/2020     &   44.284 & 45.262  & 0.2  & VR & LDT  &$>$5.5 & $>$0.2 &6.0& 54.776 &0.341 & 10.4  \\           
&   12/13/2023     &   43.656 & 44.117& 1.1  & VR & LDT  & ...&...&...& ... &... & ... \\           
 2015~KH$_{174}$    &  05/14/2021        &  33.108 & 34.073   & 0.5 &VR&LDT & $>$4.5 & $>$0.2 &   7.8  & 55.961  & 0.435 & 10.3  \\   
   & 06/01/2024    &32.327 & 33.328& 0.3  &    VR & LDT &  ... &  ... &... & ...& ... & ...\\                                              
\enddata
\end{deluxetable*}

 \section{Discussion} 
\label{sec:dis}

  \subsection{Literature}

Only a few of the inner and outer Neptune resonant KBOs studied here have had any KBO characterized through lightcurve studies (Table~\ref{Summary_resonant}). To date, only two KBOs in the 5:4 resonance, four in the 5:2, and one in the 11:5 have published complete or partial rotational lightcurves. No photometric studies have been published (refereed) for objects in the 1:1, 4:3, 7:3, or 12:5 resonances. 

Below, we discuss several results from the literature. The rotational lightcurve of 2002~TC$_{302}$ was revisited by \citet{Ortiz2020} using photometric, stellar occultation, and thermal data. They concluded that 2002~TC$_{302}$ in the 5:2 resonance is a slow rotator potentially hosting an unresolved (and undetected so far) satellite up to $\sim$300~km in diameter, and we adopt their results throughout this work.  Independent studies by \citet{SheppardJewitt2002} and \citet{Thirouin2013} found that the 5:2 resonant 1999~DE$_{9}$ exhibits a very low-amplitude lightcurve. \citet{Kern2006} reported a single-peaked rotational period of 4.0~h for 2002~GP$_{32}$ located in the 5:2 resonance. However, their lightcurve appears better described by a double-peaked solution, as some data points are not well reproduced by the single-peaked fit. Also, due to the lightcurve amplitude which is larger than the 0.15~mag threshold used in \citet{Sheppard2008, Thirouin2010} to distinguish from a single-peaked lightcurve of a spheroidal object and a double-peaked lightcurve due to an elongated object, we therefore adopt a double-peaked rotational period of 8~h for 2002~GP$_{32}$. The first and unique lightcurve so far for 2004~TT$_{357}$ published by \citet{Thirouin2017} infers that this object is a likely contact binary in the 5:2 mean motion resonance. The partial lightcurves of 2015~RN$_{278}$, 2015~RR$_{278}$ (both in the 5:4 resonance), and 2015~RU$_{278}$ in the 11:5 resonance ( three objects from the Outer Solar System Origins Survey) have been obtained by \citet{Alexandersen2019} over two consecutive nights with the \textit{Subaru} telescope.

\startlongtable
\begin{deluxetable*}{lccccc}
\tablecaption{\label{Summary_resonant} We summarize all published (refereed) complete and partial lightcurves of Kuiper Belt Objects trapped in inner and outer resonances with Neptune. Only resonances studied in this work are reported in this table. \underline{Note}: 2015~RU$_{278}$ is listed as a 11:5 resonant object in \citet{Alexandersen2019}, but this object is a Scattered Disk Object in the \textit{DES} classification. Absolute magnitudes (H) are from the MPC and were retrieved in March 2026. }
\tablewidth{0pt}
\tablehead{
 Object & Period   & Period & $\Delta m$  & H  & Reference \\ 
 & Single-peaked [h] & Double-peaked [h]& [mag] & [mag] &   \\ }
\startdata
\multicolumn{6}{c}{5:4 mean motion resonance}                 \\  
(593617) 2015~RN$_{278}$  & $>$5 & - & $>$0.22 & 8.51 & \citet{Alexandersen2019} \\ 
\\ 2015~RR$_{278}$  & $>$5 & - & $>$0.40 & 9.7 & \citet{Alexandersen2019} \\ 
\multicolumn{6}{c}{5:2 mean motion resonance}   \\
(26375) 1999~DE$_{9}$  & $>$12 & - & $<$0.10 & 4.89 & \citet{SheppardJewitt2002} \\
         & 12.33 &  -& 0.09$\pm$0.03  & ... & \citet{Thirouin2013} \\
(612350) 2002~GP$_{32}$  & 4.0$\pm$0.1 & - & 0.18$\pm$0.04 & 7.07 & \citet{Kern2006} \\
  (84522) 2002~TC$_{302}$  & 5.41 & - & 0.04$\pm$0.01 &3.92 & \citet{Thirouin2012} \\
  & - & $\sim$56.1 & 0.06$\pm$0.01 & ... & \citet{Ortiz2020} \\ 
(612891) 2004~TT$_{357}$  & 7.79 & - & 0.76$\pm$0.03 & 7.91  & \citet{Thirouin2017} \\
   \multicolumn{6}{c}{11:5 mean motion resonance}  \\ 
 2015~RU$_{278}$  & $>$5 & - & $>$0.31 & 7.01 & \citet{Alexandersen2019} \\   
\enddata
\end{deluxetable*}
  
   \subsection{Our survey and the literature}  
   
To provide the most comprehensive view of KBOs trapped in the inner and outer mean motion resonances with Neptune, we combine results from previous lightcurve studies (Table~\ref{Summary_resonant}) with those from our survey (Table~\ref{tab:Summary_photo}) in the following discussion.
     
   \subsubsection{Lightcurve amplitude distributions}  
   
\begin{figure}
   \includegraphics[width=9.5cm,angle=0]{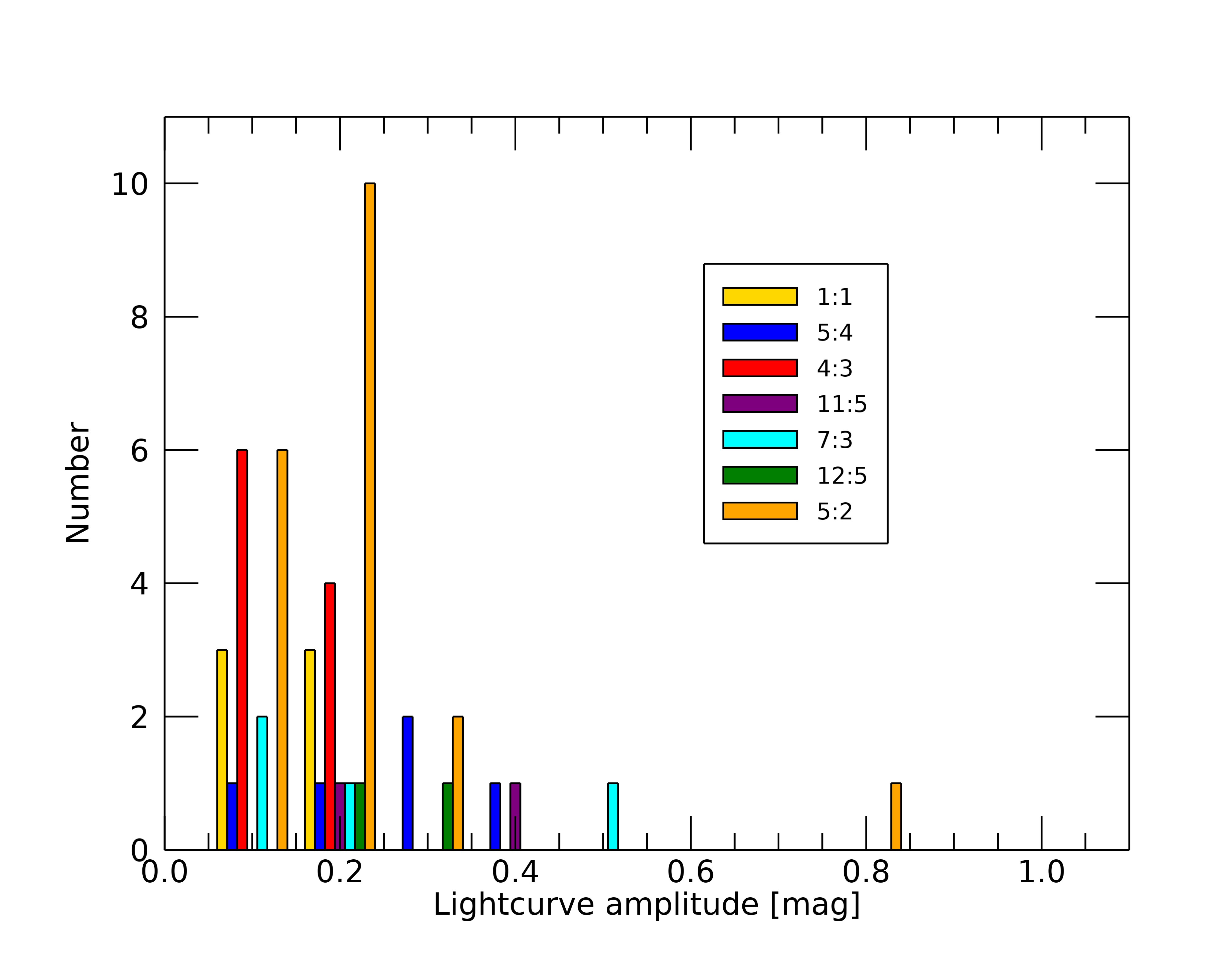} 
 \caption{This figure reports the amplitude distributions for the inner and outer resonances studied in this work. For clarity, we do not distinguish partial, flat, and complete lightcurves for each resonance. Therefore, all three categories are merged for each resonance. Most studied KBOs exhibit variability below 0.4~mag, except for two objects. }
\label{fig:HistoAmplitudes}
\end{figure}

Considering all resonances together, lightcurve amplitudes range from flat to 0.76~mag. Figure~\ref{fig:HistoAmplitudes} shows the amplitude distributions for each resonance, revealing that most KBOs exhibit variabilities below 0.4~mag. Only two objects, 2004~TT$_{357}$ and 2013~TJ$_{159}$, display amplitudes exceeding 0.4~mag. Across the full sample, the mean amplitudes are 0.30~mag, 0.18~mag, and 0.11~mag for complete, partial, and flat lightcurves, respectively. Complete lightcurves generally exhibit larger amplitudes than partial or flat ones, likely reflecting an observational bias toward targets showing detectable variability. Despite this bias, the average amplitudes in the resonances studied here remain relatively low compared to those reported for the 3:2, 2:1, 5:3, and 7:4 resonances, as well as the dynamically cold Classical population \citep{ThirouinSheppard2024, ThirouinSheppard2022, ThirouinSheppard2019, ThirouinSheppard2018}. The 7:3 resonance appears to host larger-amplitude objects, although this result is based on only one complete lightcurve (2013~TJ$_{159}$) and one partial lightcurve (2001~XT$_{254}$).

\begin{figure*}
   \includegraphics[width=10cm,angle=0]{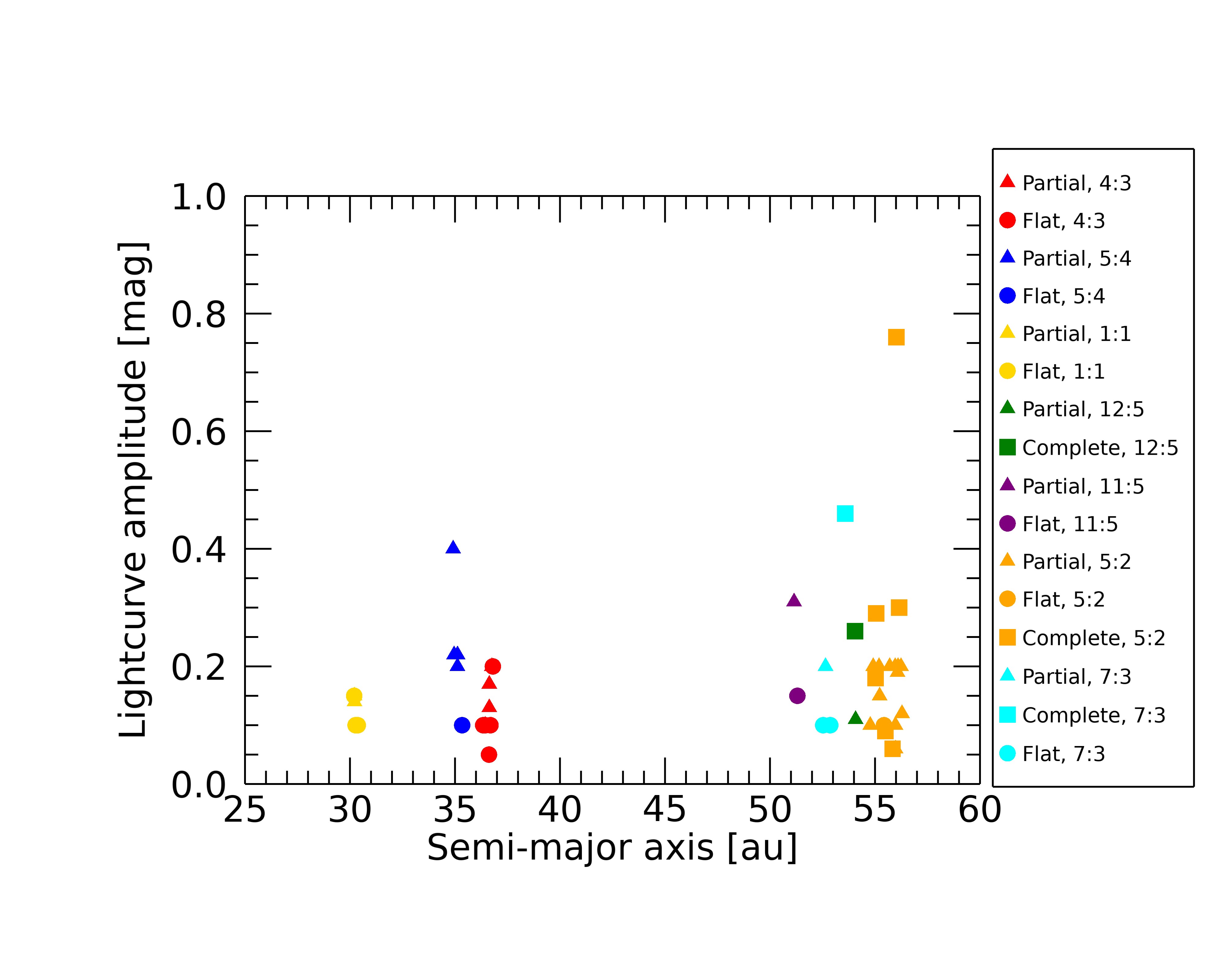} 
         \includegraphics[width=10cm,angle=0]{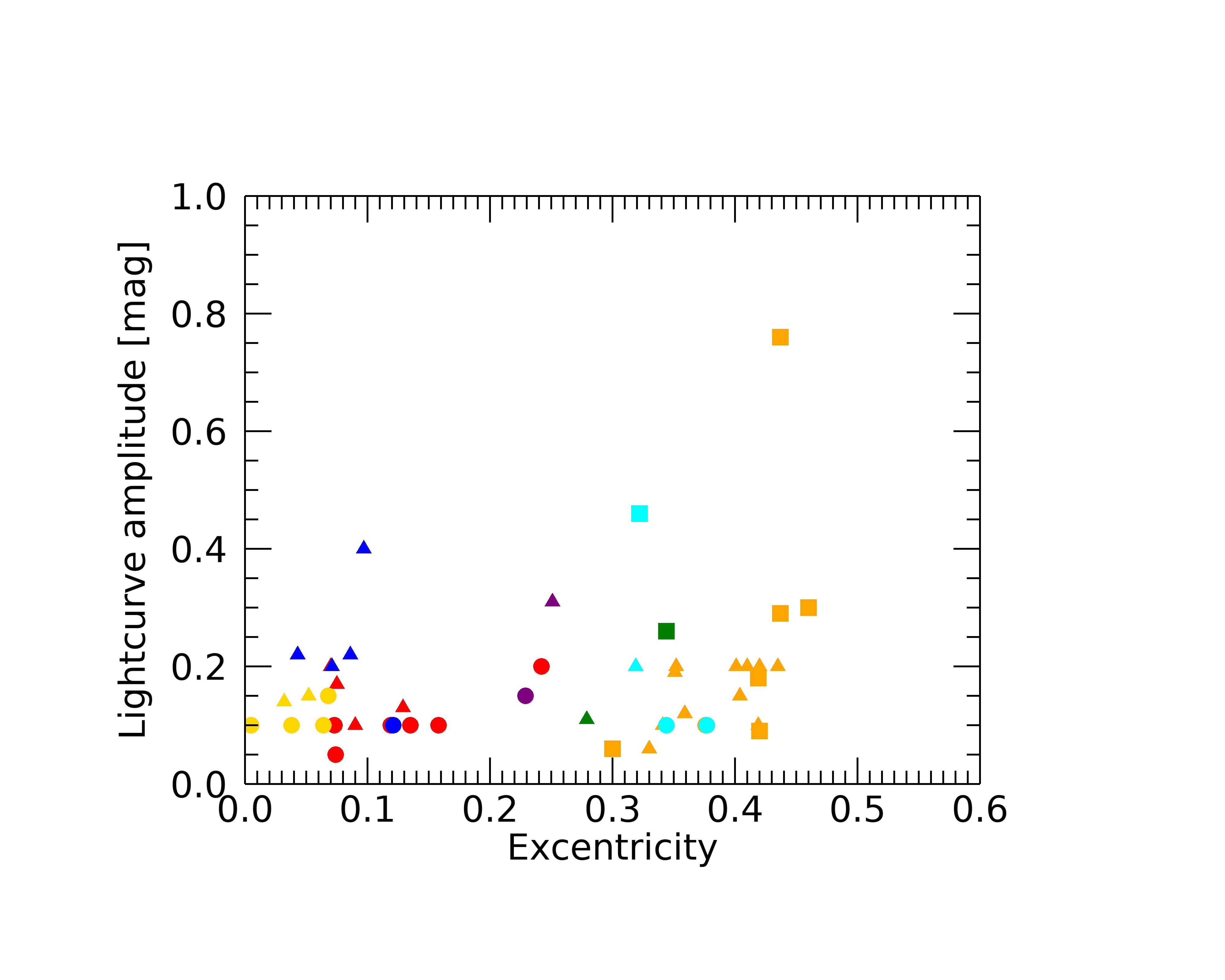}
     \includegraphics[width=10cm,angle=0]{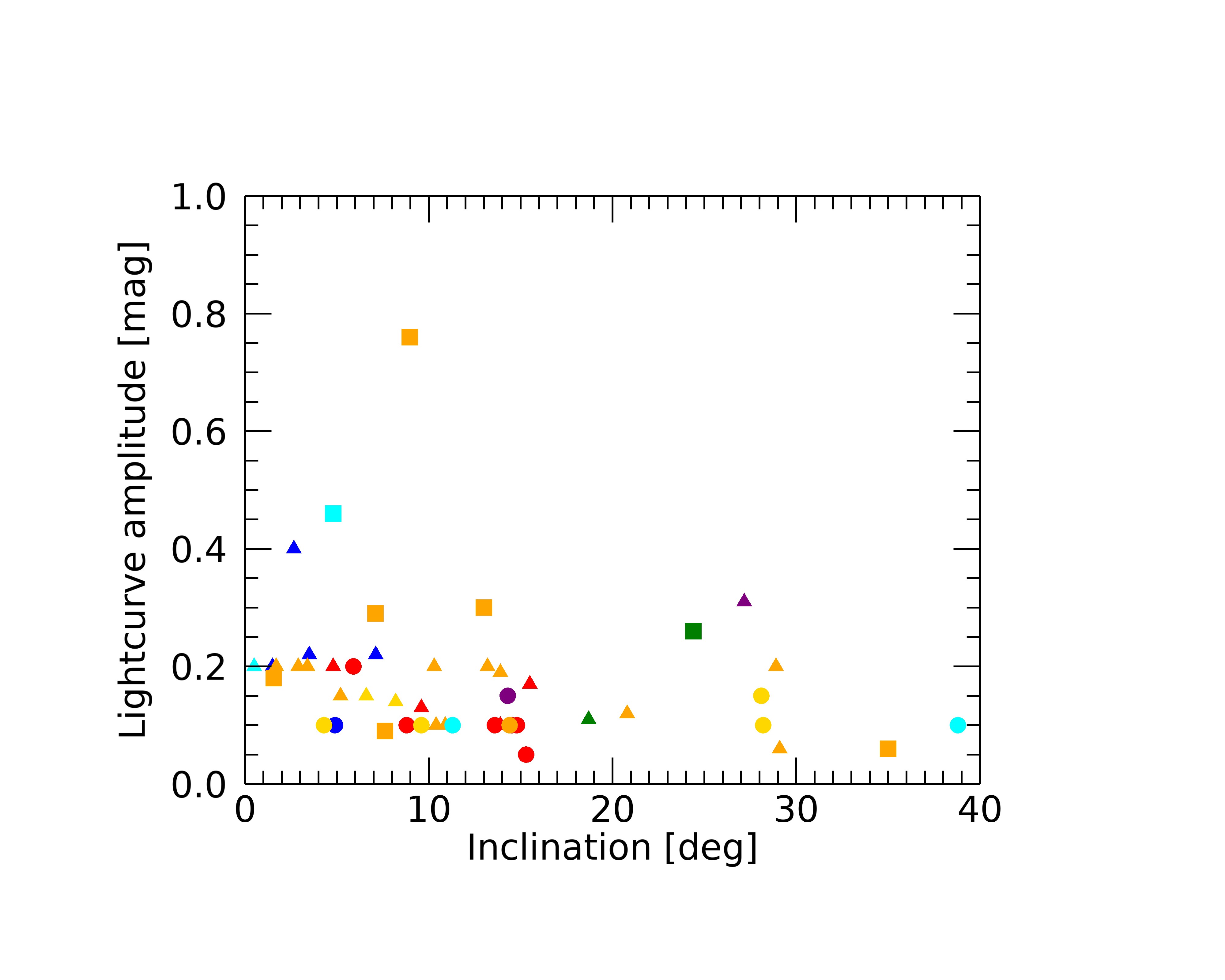}  
      \includegraphics[width=10cm,angle=0]{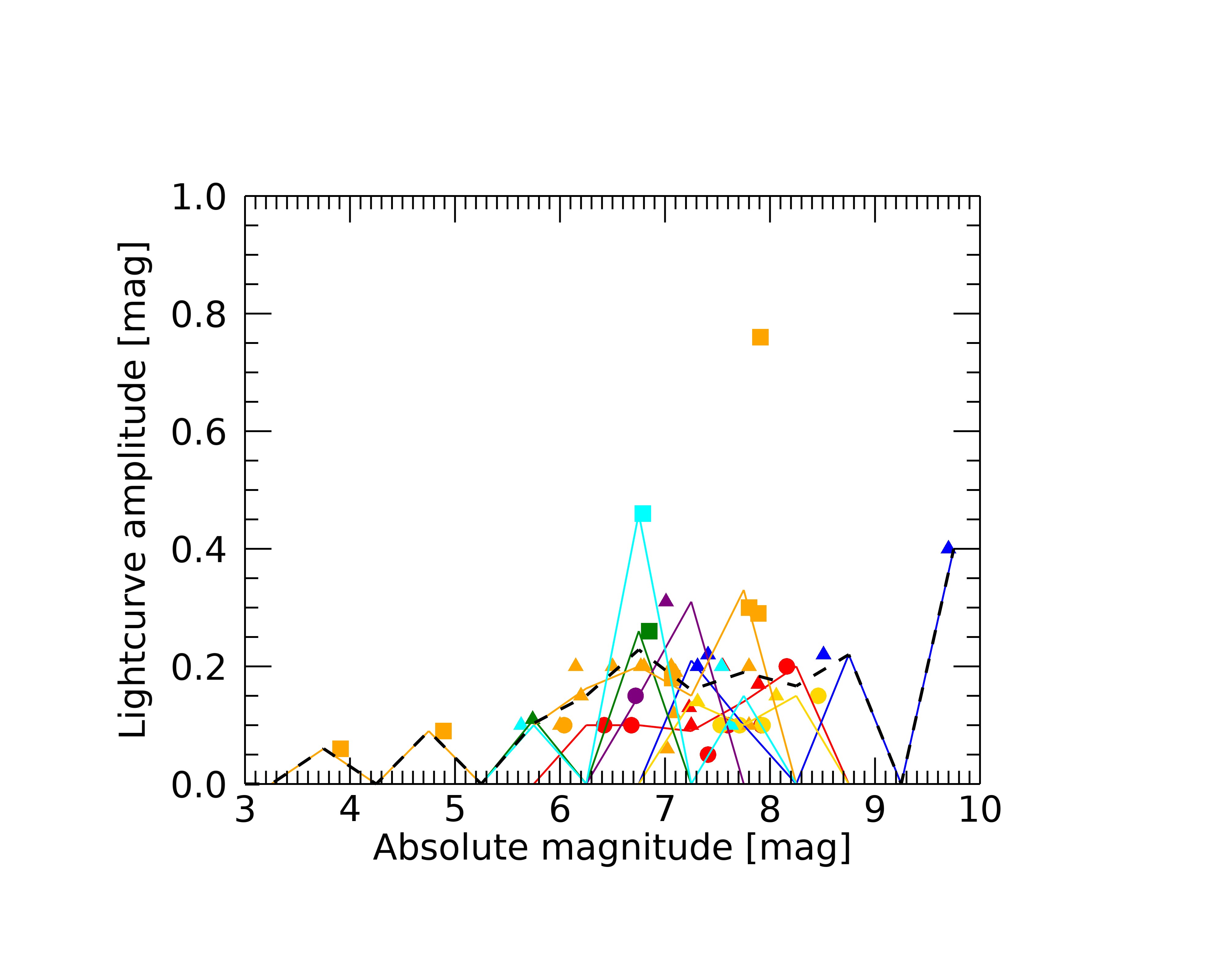} 
 \caption{Lightcurve amplitudes from complete, partial, and flat lightcurves are reported as a function of semi-major axis, eccentricity, and inclination as well as absolute magnitude. The same legend has been used for all the plots. Colored lines are running means based on partial, flat, and complete lightcurves from the literature and our survey for each resonance, and the black dashed line corresponds to the running mean across all resonances. The complete sample (literature + our survey) spans a wide range of orbital elements and absolute magnitudes. A weak trend is noticed between lightcurve amplitudes and absolute magnitudes, suggesting that the smaller objects have higher amplitudes, but this tendency, already noticed in several Kuiper Belt sub-populations, is weak and has a low confidence level in this sample.    }
\label{fig:DistributionAmplitude}
\end{figure*}

Figure~\ref{fig:DistributionAmplitude} shows lightcurve amplitude as a function of orbital elements (a, e, and i) and absolute magnitude (H). By probing both inner and outer resonances, our sample spans semi-major axes from $\sim$30 to $\sim$55~au, eccentricities from $\sim$0 to 0.5, and inclinations from $\sim$0 to $\sim$40$^{\circ}$. Combining our survey with literature data yields a sample covering H = 3.9–9.7~mag, corresponding to diameters of $\sim$76–1100~km assuming an albedo of 0.04.
To search for trends between lightcurve amplitude and orbital or physical properties, we used the \texttt{ASURV} package, which handles datasets containing upper and lower limits as well as secured datapoints \citep{Isobe1986}. Considering all resonances together, we find a weak correlation between lightcurve amplitude and absolute magnitude, with a correlation$\footnote{A (anti-)correlation is strong if the absolute value of the correlation coefficient, also known as the Spearman coefficient, is larger than 0.3. If the absolute value is higher than 0.6, the trend is very strong, but below 0.3, there is no trend. The tendency is very strong if the significance level is higher than 99\%, while the trend is strong/reasonably strong for the threshold of 97.5\%/95\%.}$ coefficient of 0.3 and a significance level of 91\% using complete and partial lightcurves (i.e., flat lightcurves are not considered). Similar trends have been reported for several Kuiper Belt subpopulations and for the overall KBO population, with the interpretation being that smaller objects are more irregular in shape, probably because self-gravity is less important for the smaller objects \citep{Sheppard2008, Thirouin2013, Benecchi2013, ThirouinSheppard2019, Alexandersen2019, ThirouinSheppard2024}. However, the correlation found here is weaker than in other populations and falls below the 95\% significance threshold.

In Figure~\ref{fig:DistributionAmplitude}, we overplot running means for each resonance and for the combined sample. Unlike other KBO populations, the commonly observed increase in lightcurve amplitude toward smaller sizes is not clearly apparent in these resonant populations. No significant correlations or anti-correlations are found between lightcurve amplitude and semi-major axis or inclination. However, we note that the KBOs with large amplitudes may tend to be at low inclinations; however, this potential anti-correlation has only a coefficient of -0.224 for a confidence level of 79.5\%. A similar anti-correlation has already been noticed in the 5:3 and 7:4 mean motion resonances with Neptune \citep{ThirouinSheppard2024}.
   
\begin{figure}
   \includegraphics[width=9.5cm,angle=0]{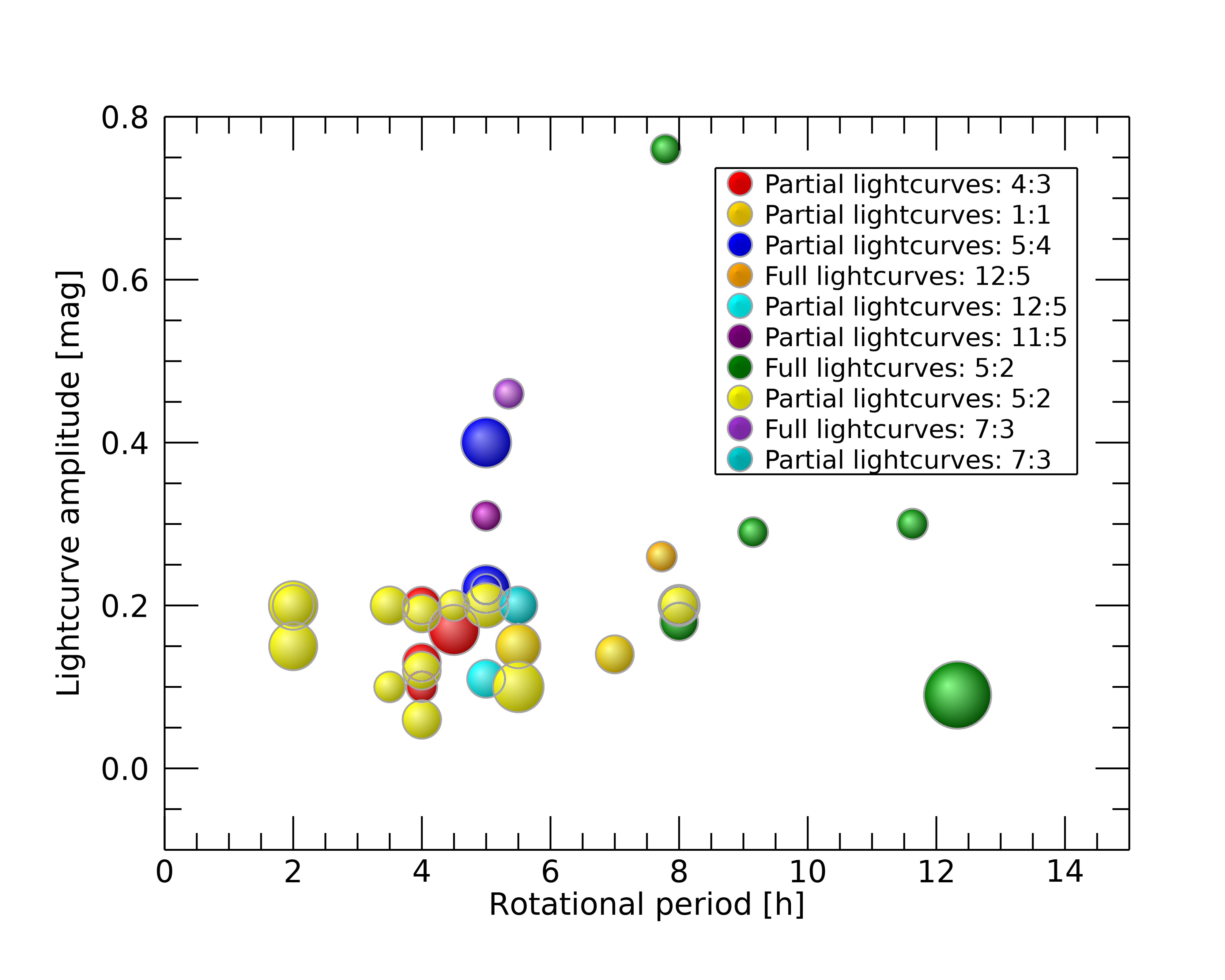} 
 \caption{Kuiper Belt Objects in the inner and outer resonances with a full or partial lightcurve are plotted in this lightcurve amplitude versus rotational period figure. The size of the object is indicated by the size of the point. The KBO 2002~TC$_{302}$ with a rotational period of about 56~h is out of the plotted range. There is no correlation between the lightcurve amplitude and the rotational period, as is the case for the dynamically Cold Classicals, as well as 5:3 and 7:4 resonant KBOs \citep{ThirouinSheppard2019, ThirouinSheppard2024}.}
\label{fig:Bubble}
\end{figure}

Figure~\ref{fig:Bubble} shows lightcurve amplitude as a function of rotational period for inner and outer resonant KBOs with partial or complete lightcurves. Our correlation analysis reveals no significant relationship between amplitude and rotational period in these resonant populations with a coefficient of -0.189 and a significance level of 72\%. In contrast, such a trend has been reported for at least three other Kuiper Belt subpopulations: the dynamically Cold Classical objects and the 5:3 and 7:4 resonant KBOs \citep{ThirouinSheppard2019, ThirouinSheppard2024}. 
           
   \subsubsection{Rotational period distributions}  
 
\begin{figure*}
   \includegraphics[width=10cm,angle=0]{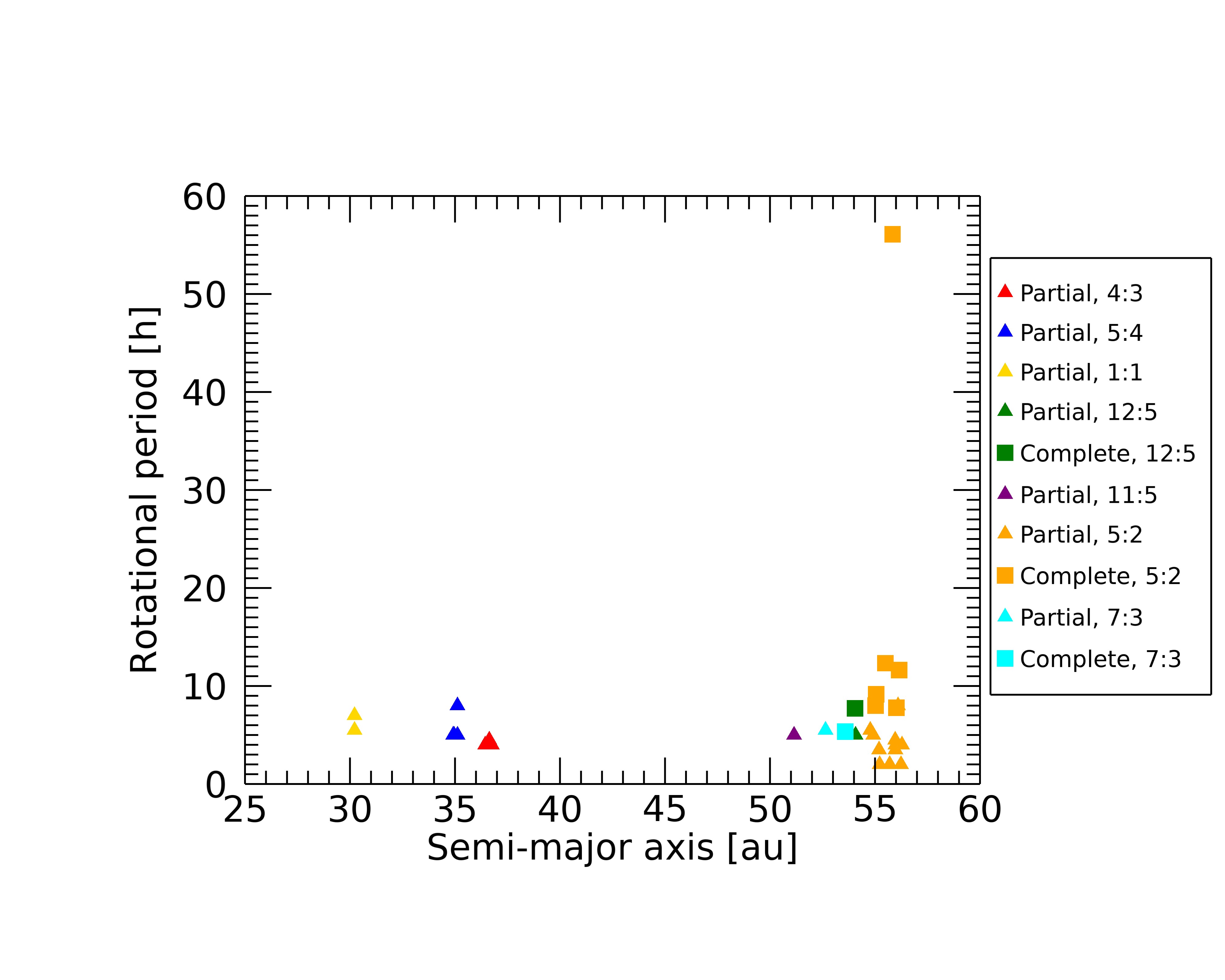} 
         \includegraphics[width=10cm,angle=0]{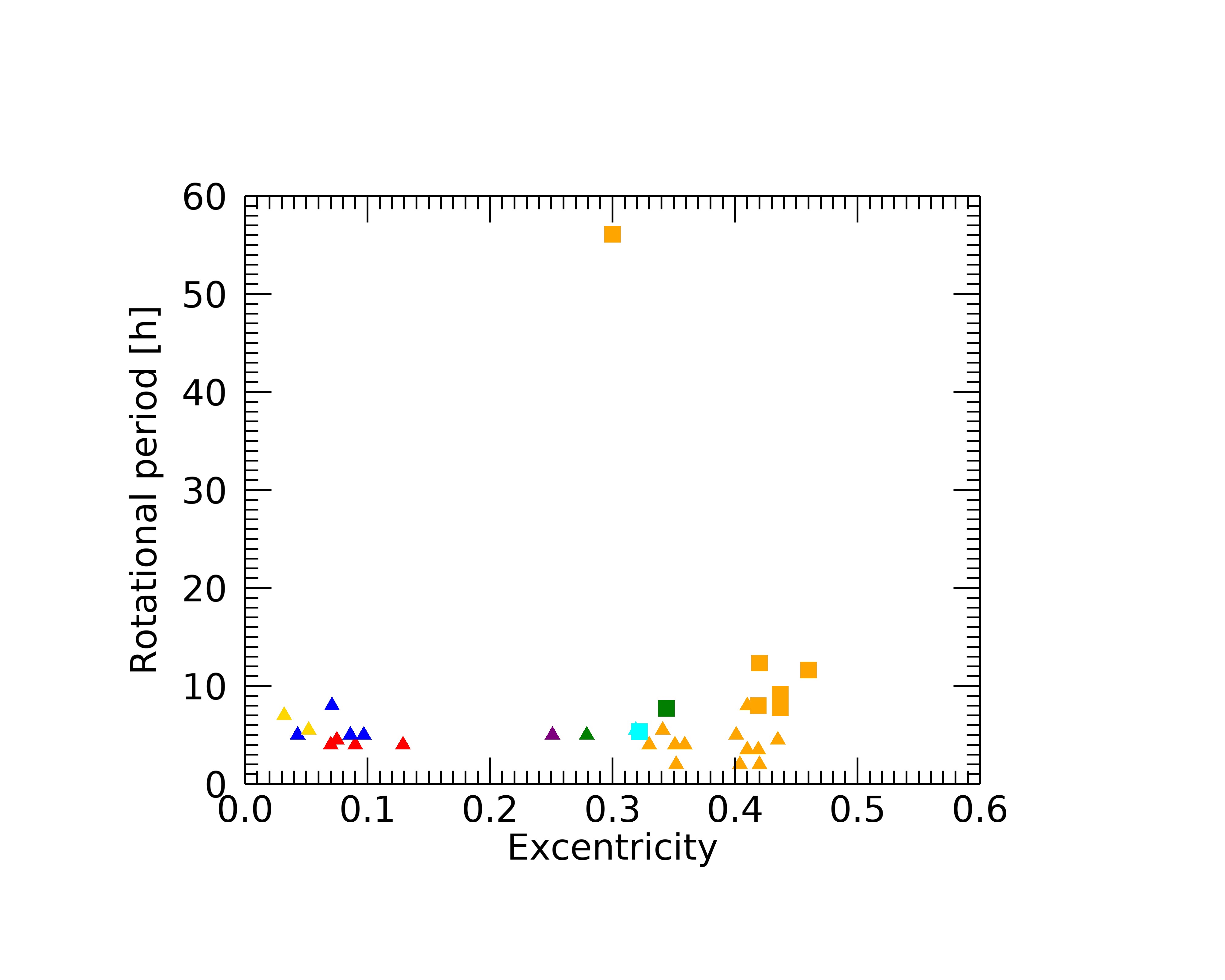}
     \includegraphics[width=10cm,angle=0]{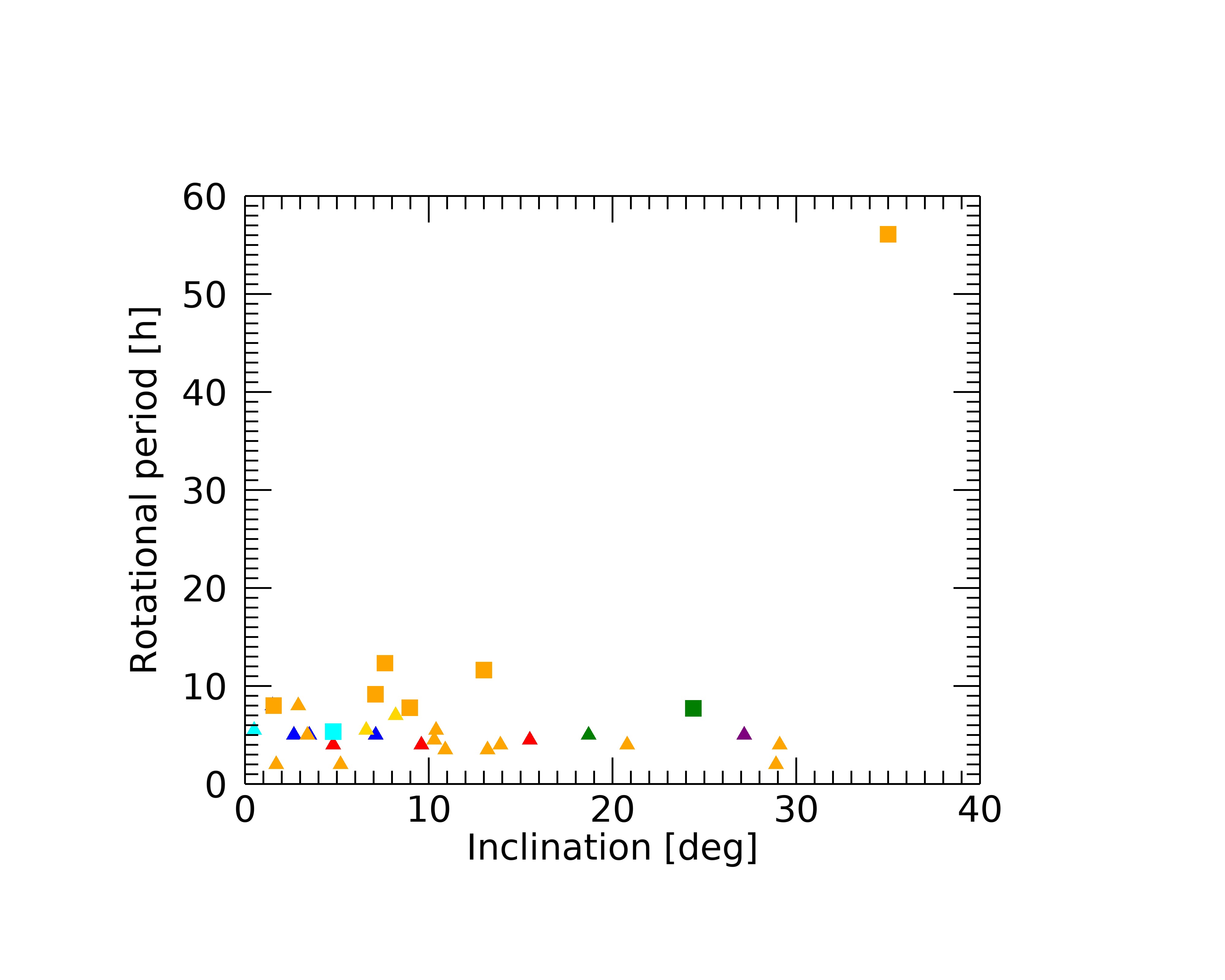}  
       \includegraphics[width=10cm,angle=0]{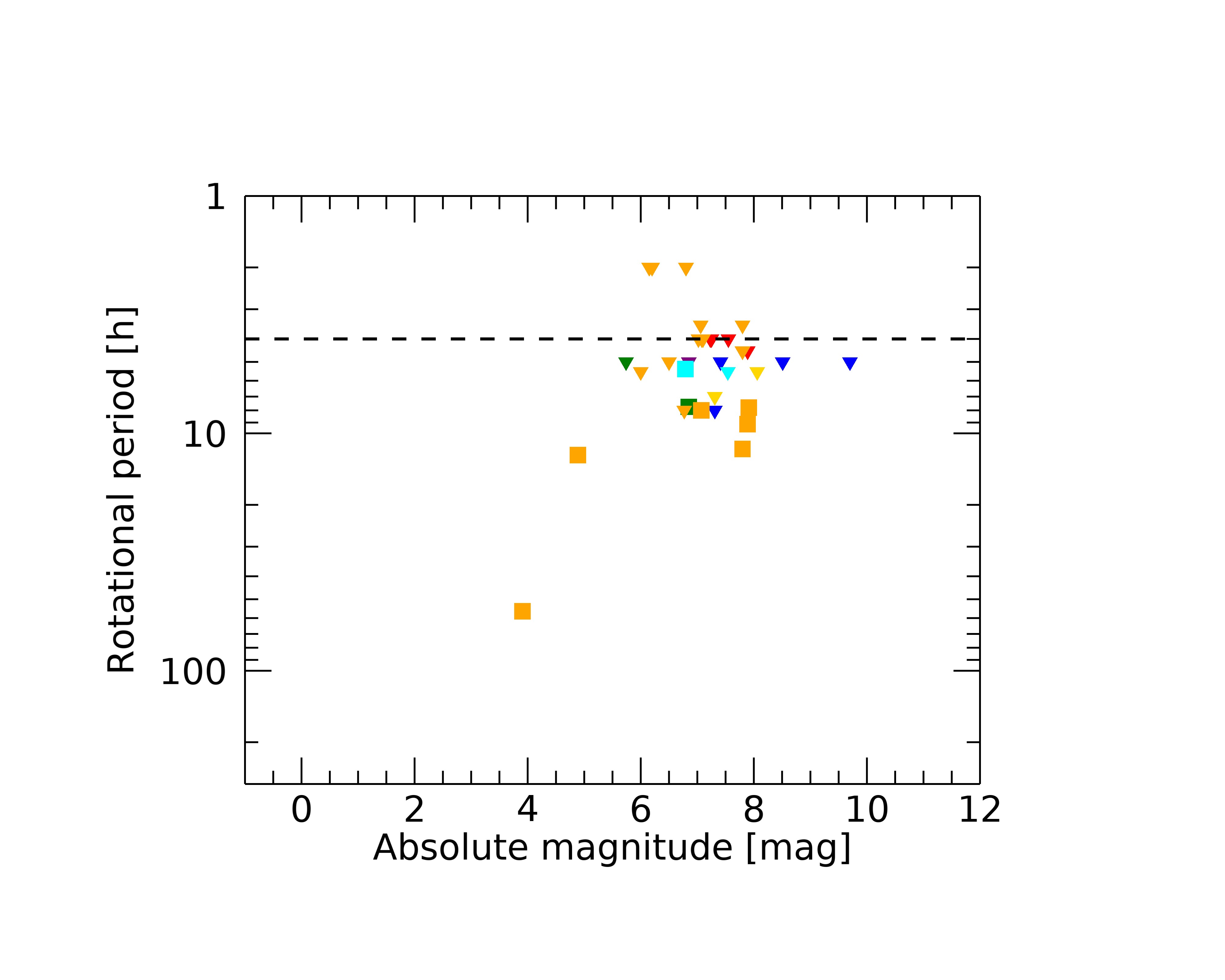} 
 \caption{Rotational periods from complete and partial lightcurves are reported as a function of semi-major axis, eccentricity, and inclination as well as absolute magnitude. The same legend has been used for all plots. The black dashed line in the period versus absolute magnitude plot is the KBO spin barrier at 4~h \citep{Thirouin2010}.  }
\label{fig:DistributionPeriod}
\end{figure*}

As for the lightcurve amplitudes discussed above, we also examined the rotational period as a function of absolute magnitude and orbital elements. Figure~\ref{fig:DistributionPeriod} presents the partial and complete lightcurves from both our survey and the literature; objects with flat lightcurves are excluded. Using the full sample, the correlation between rotational period and semi-major axis yields a coefficient of 0.263 with a significance level of 86\%. Restricting the analysis to the outer resonances gives a coefficient of 0.231 and a significance level of 72\%. In both cases, the weak coefficients ($<0.3$) indicate no significant correlation between rotational period and semi-major axis. A possible correlation is found between rotational period and eccentricity, with a coefficient of 0.331 and a significance level of 94\% for the full sample. However, because the significance remains below the 95\% threshold and the correlation is weak, additional data are needed to confirm this trend. For the outer resonances alone, the coefficient decreases to 0.226 with a significance level of 71\%, indicating no detectable correlation. There is no trend between period and inclination, as well as period and absolute magnitude, in our dataset. 
         
   \subsubsection{Density derived from lightcurves}  

 As presented above, we derived lower limits to the densities of the four resonant KBOs that have a large amplitude and a complete lightcurve. Three of the objects have relatively low density limits of 0.30 to 0.66~g~cm$^{-3}$, but 2013~TJ$_{159}$ requires a substantially higher lower limit of $\rho \geq 1.46$~g~cm$^{-3}$ to sustain its rapid 5.35~h rotation. These density ranges are consistent with other measurements of small- to medium-sized KBOs \citep{Grundy2019, Thirouin2010, Sheppard2008}. \citet{Thirouin2013} derived a lower limit of 0.26~g~cm$^{-3}$ for 1999~DE$_{9}$ while \citet{Ortiz2020} assumed a density of 0.8~g~cm$^{-3}$ for their work. Based on the lightcurve of 2002~GP$_{32}$, we calculate a lower limit of 0.63~g~cm$^{-3}$ for its density using the same assumptions presented in this work. \citet{Thirouin2017} estimated that the density of the likely contact binary 2004~TT$_{357}$ is larger than 2~g~cm$^{-3}$, but this result is based on only one lightcurve and using the \citet{Leone1984} modeling. We emphasize, however, that the densities derived from lightcurves are lower limits and rely on the assumptions of hydrostatic equilibrium and an equatorial viewing geometry. The low bulk densities are fairly common for small to medium-sized KBOs based on densities derived from mutual orbits of binary and multiple systems. In fact, \citet{Grundy2019, Noll2020} showed that objects with a diameter of less than $\sim$300~km tend to have densities well below that of water ice, which is around 1~g~cm$^{-3}$. Such low densities are generally interpreted as evidence for highly porous interiors composed primarily of volatile-rich material \citep{Grundy2019, Bierson2019}. Comet nuclei also show similarly low densities \citep{AHearn2011, Jorda2016, Groussin2019}.  
           
   \subsubsection{Contact binaries}  
   
One goal of our survey is to identify and characterize contact binaries. Following the criteria summarized in \citet{ThirouinSheppard2024}, we infer the contact binary nature of an object from its lightcurve morphology and amplitude. Despite observing 41 resonant KBOs, we did not identify any new likely contact binaries. To date, the only likely contact binary in the surveyed inner and outer resonances is 2004~TT$_{357}$, located in the outer 5:2 mean motion resonance with Neptune \citep{Thirouin2017}.

Using the method described in \citet{SheppardJewitt2004, ThirouinSheppard2022}, we estimate a lower limit to the nearly equal-sized contact binary fraction in the 5:2 resonance. Briefly, this method is used to debias the object's pole orientations to estimate the lower limit for the fraction of contact binaries. As an example reported in \citet{SheppardJewitt2004}, an object with an a/b axis ratio of 3 will present a lightcurve amplitude of 0.9~mag if the angle of the object's pole relative to the perpendicular of the line of sight is $\theta$=10$^{\circ}$. The probability to observe an object from a random distribution within 10$^{\circ}$ of the sight line is P($\theta$$\leq$10$^{\circ}$)=0.17. Therefore, the lower fraction of objects with an amplitude larger than 0.9mag is f($\Delta m$$\geq$0.9~mag)= (number of objects with $\Delta m$$\geq$0.9~mag)/(number of observed objects $\times$ P($\theta$$\leq$10$^{\circ}$)). In our case, there is 1 object with a $\Delta m$$\geq$0.7~mag) in a sample of 19 KBOs in the 5:2 resonance, and using Equations 3 and 4 of \citet{SheppardJewitt2004}, we derive a lower limit to the fraction of contact binaries of $\sim$12 and 15\%, respectively. 
With one likely contact binary among 19 KBOs studied for short-term variability, the inferred lower limit is $\sim$12–15\%. This value is consistent with the fraction derived for the 2:1 resonance \citep{ThirouinSheppard2022} and with the lower end of the models by \citet{Nesvorny2019} (see Section 6.1 of \citet{Nesvorny2019} for more details), which predict contact binary fractions of 10-30\%. The fraction of contact binaries in the 5:2 resonance is lower than in the Cold Classical population, 3:2, 7:4, and 5:3 resonances \citep{ThirouinSheppard2018, ThirouinSheppard2019, ThirouinSheppard2024}. However, we emphasize that using lightcurves to derive the fraction of contact binaries only gives a lower limit as we are dealing with several observational biases and limitations, and thus the true fraction can be higher \citet{Porter2024}. We note that evidence of contact binaries in several populations has also been suggested by different ground-based lightcurve surveys, such as \citet{Ashton2023, Strauss2024}, among others. The existence of close/contact binaries has also been studied by \citet{Porter2024} using the \textit{Hubble Space Telescope}. 

   \subsubsection{Origin of the inner and outer resonances?}  

 Because of the limited number of complete lightcurves, we cannot apply a 2D Kolmogorov–Smirnov (K-S) test to quantitatively compare subpopulations. Nevertheless, the period and amplitude distributions suggest that the inner and outer resonances studied here likely do not share a common origin with the middle 7:4 and 5:3 resonances, and dynamically cold Classical populations. In particular, those three subpopulations associated with the main Kuiper belt, which likely formed in-situ, contain many large-amplitude objects, unlike the inner and outer resonant KBOs analyzed in this work. Likewise, the correlation between rotational period and lightcurve amplitude reported for some other KBO populations is absent here. These differences are also consistent with color studies by \citet{Sheppard2012}, who showed that the inner and outer resonant populations have color distributions more similar to Detached and Scattered Disk objects than to dynamically cold Classicals. To date some resonances have a low number of objects well observed for colors, and thus more color measurements are needed to better determine possible statistical correlations. An additional difference between the inner and outer resonances with respect to the middle 5:3 and 7:4 resonances, and Cold Classical subpopulations is the low fraction of likely contact binaries. Therefore, our work is suggestive of the fact that the inner and outer resonances do not share a common origin and/or had significantly different dynamical histories compared with the low inclination Cold Classicals and middle 5:3 and 7:4 Neptune resonances. We also emphasize that additional observational data would be useful to confirm such a conclusion, as the sample size of well observed objects in the inner and outer resonances are still limited. 
 

 \section{Conclusion}
 
 Our photometric survey of the inner and outer mean motion resonances with Neptune can be summarized as: 
 \begin{itemize}
 \item We report the flat, partial, or complete lightcurves of 41 KBOs trapped into inner and outer mean motion resonances to increase by about 600\% the number of objects with a photometric study.
 \item Three KBOs with complete lightcurves out of the four derived for this work show an asymmetric lightcurve: 2001~XQ$_{254}$, 2013~TJ$_{159}$, and 2013~RZ$_{108}$ while the lightcurve of 2015~AR$_{293}$ is symmetric. Such asymmetry can be attributed to an albedo spot on the object's surface or an irregular shape. 
 \item The lightcurve of 2013~TJ$_{159}$ has the fastest rotation, highest lightcurve amplitude, and largest asymmetry in our sample. The 5.35 hour double-peaked rotation suggests this object might be elongated from its large angular momentum, giving it the large amplitude of $0.46 \pm 0.03$ magnitudes. Future observations to monitor the change in amplitude versus observing geometry of this object will better determine the nature of the short-term lightcurve and its possible pole orientation.
 \item Except for two KBOs, most of the objects trapped in the inner and outer resonances have a low to moderate lightcurve amplitude ($<$0.4~mag). Low amplitudes can be attributed to a spheroidal object and/or a nearly pole-on orientation.
 \item Based on the literature and our survey, we estimate that the fraction of nearly equal-sized contact binaries in the 5:2 resonance is low at about 12 to 15\%, but is consistent with the lower estimate of the modeling by \citet{Nesvorny2019} regarding binary survival. Such a low fraction of contact binaries is also similar to the one derived for the 2:1 resonance, but differs from that of the 3:2, 5:3, 7:4, and Cold Classicals \citep{ThirouinSheppard2018, ThirouinSheppard2019, ThirouinSheppard2022, ThirouinSheppard2024}. 
 \item The lightcurve amplitude, rotational period, and surface color distributions of the inner and outer resonances are different from the distributions of the middle 5:3 and 7:4 resonances, as well as the dynamically Cold Classical KBOs. We suggest that the origin and/or past dynamics of the inner and outer resonances differ significantly from the origin of the middle 5:3 and 7:4 resonances and Cold Classical KBOs \citep{Sheppard2012, ThirouinSheppard2024}. 
 \end{itemize}



\begin{acknowledgments}
Authors thank the reviewer for useful comments and careful reading of the paper. 
This paper includes data gathered with the 6.5~m Magellan-Baade Telescope located at Las Campanas Observatory, Chile. This research is based on data obtained at the Lowell Discovery Telescope (LDT). Lowell Observatory is a private, non-profit institution dedicated to astrophysical research and public appreciation of astronomy, and operates the LDT in partnership with Boston University, the University of Maryland, the University of Toledo, Northern Arizona University, and Yale University. Partial support of the LDT was provided by Discovery Communications. LMI was built by Lowell Observatory using funds from the National Science Foundation (AST-1005313). We are grateful to the Magellan and LDT staffs.  \\
Authors acknowledge support from the National Science Foundation with grant \#1734484 awarded to the ``Comprehensive Study of the Most Pristine Objects Known in the Outer Solar System'' and grant \#2109207 awarded to the ``Resonant Contact Binaries in the Trans-Neptunian Belt''.  
\end{acknowledgments}

\begin{contribution}

All authors contributed to this paper.


\end{contribution}

%
\facilities{Magellan:Baade, LDT}



 \clearpage
\appendix
 \section{Appendix information}
\label{sec:appA}

In this Appendix, we report the flat and partial lightcurves from our survey. 

 
      \begin{figure}
   \includegraphics[width=10cm,angle=0]{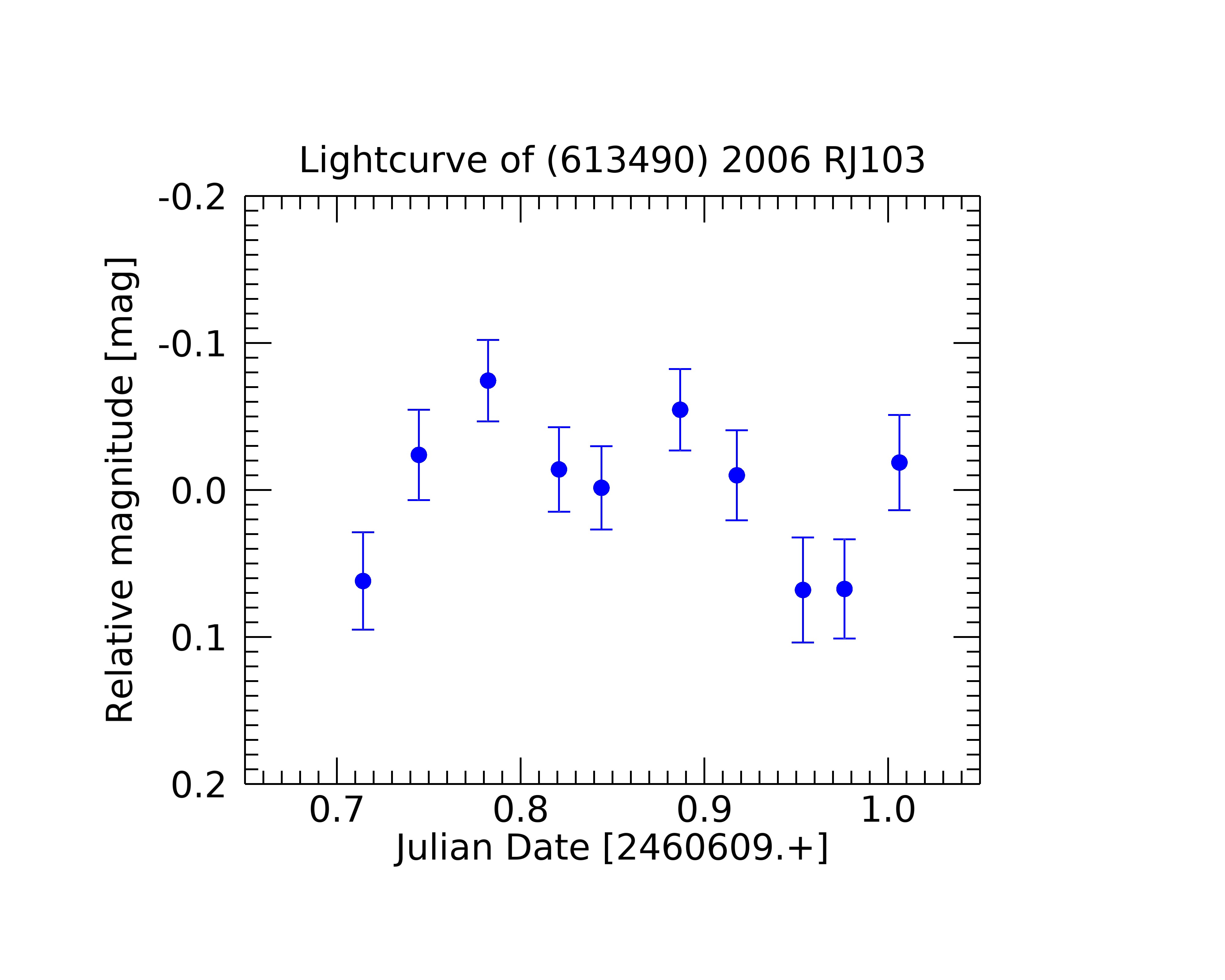} 
      \includegraphics[width=10cm,angle=0]{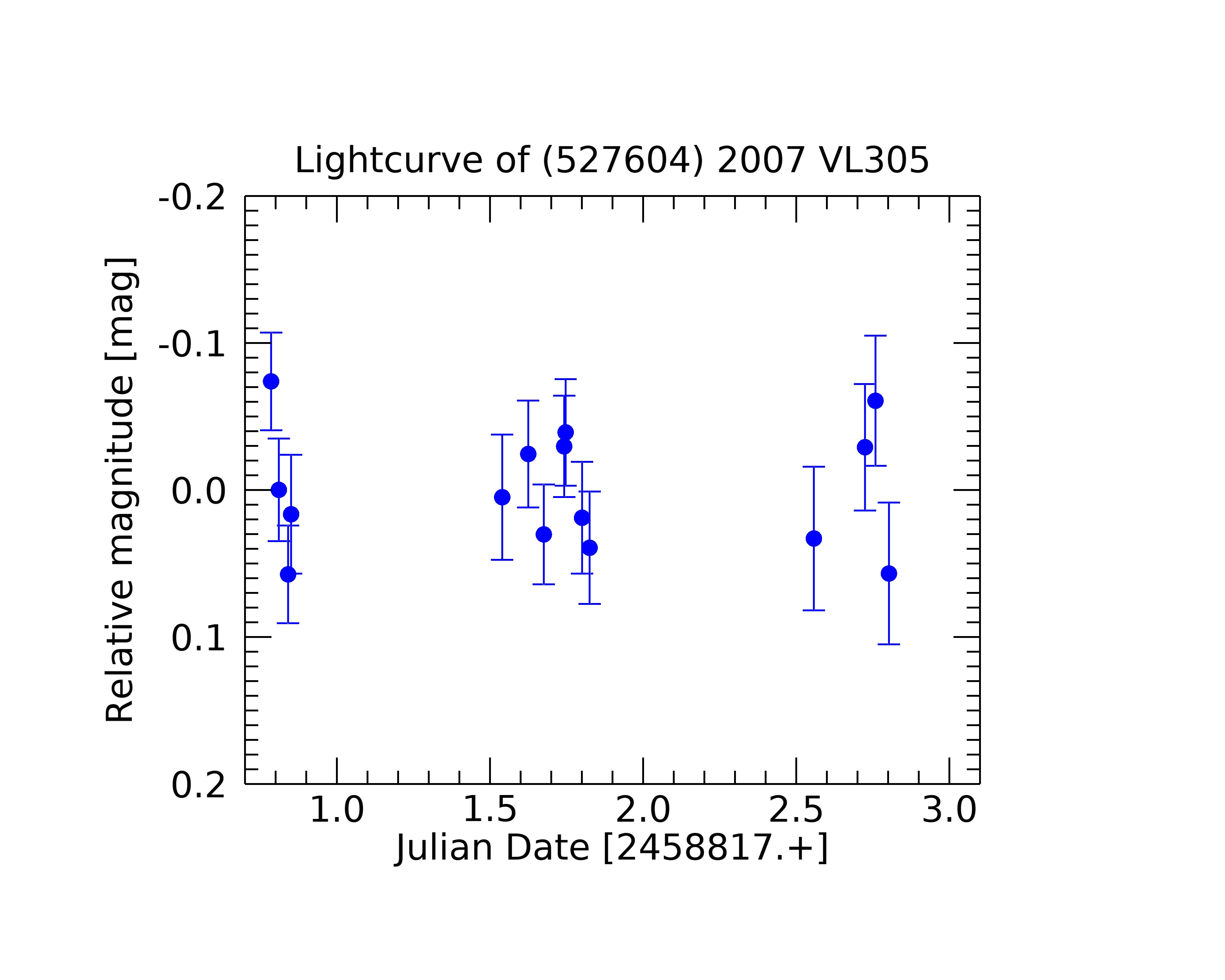} 
    \includegraphics[width=10cm,angle=0]{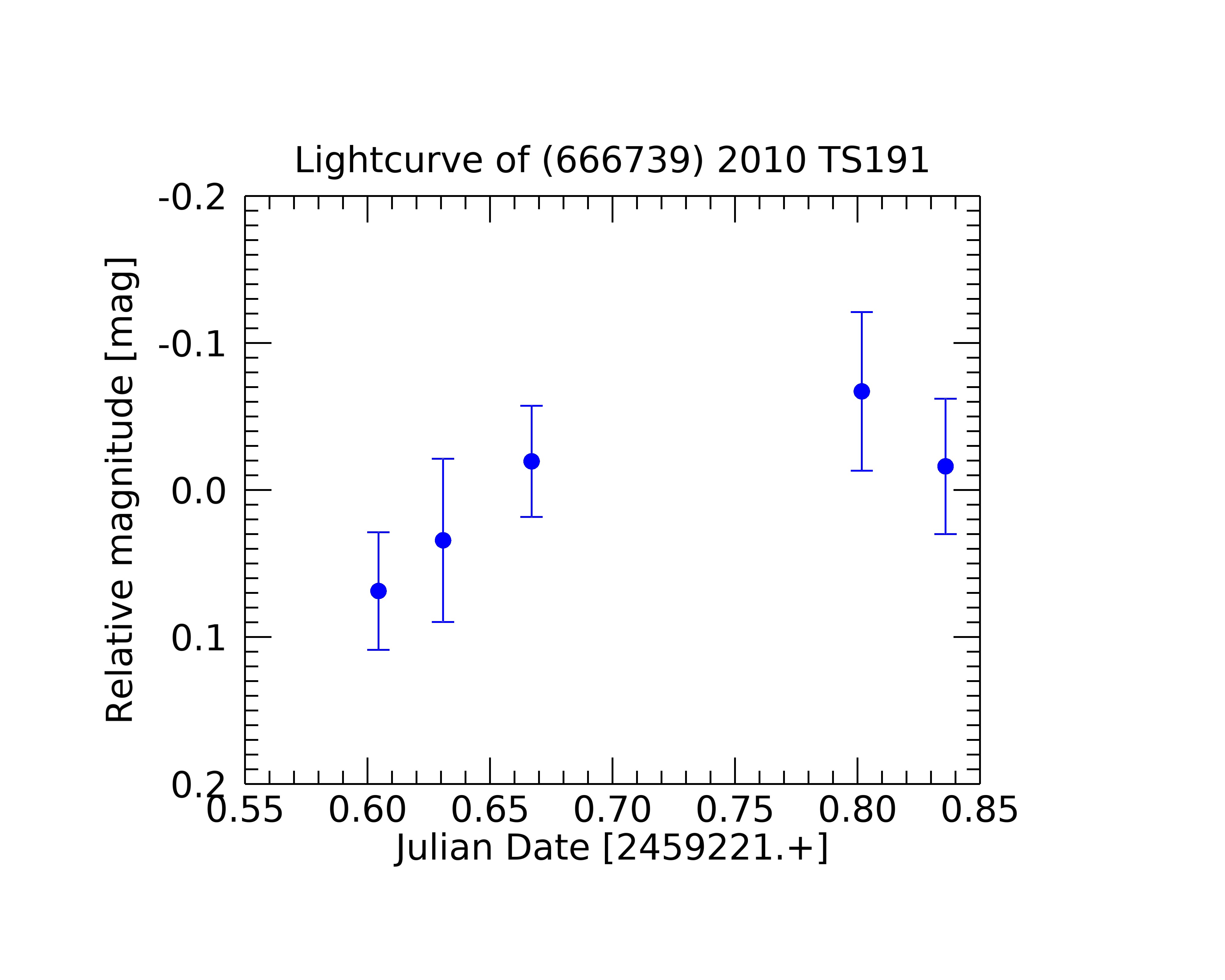} 
  \includegraphics[width=10cm,angle=0]{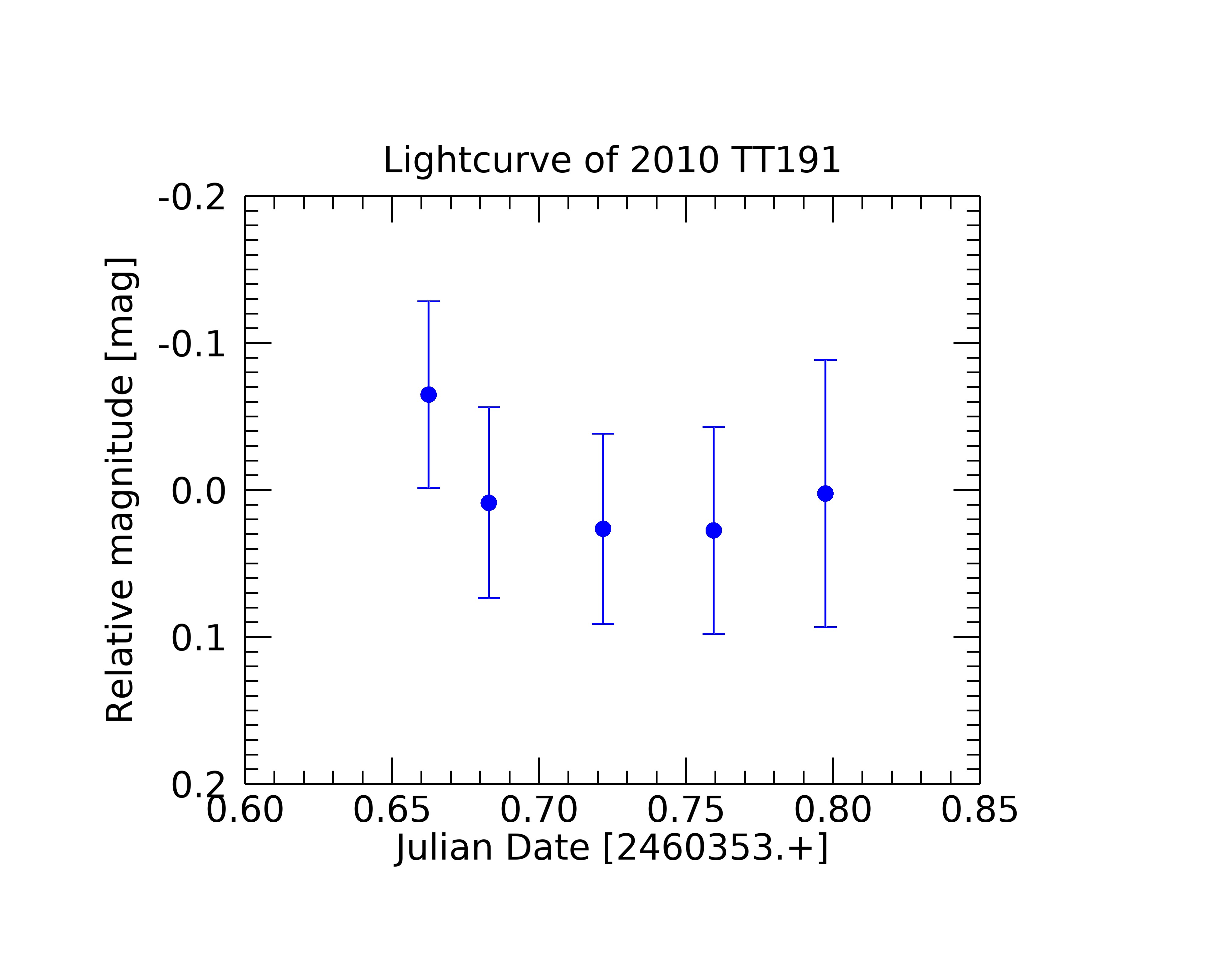} 
     \includegraphics[width=10cm,angle=0]{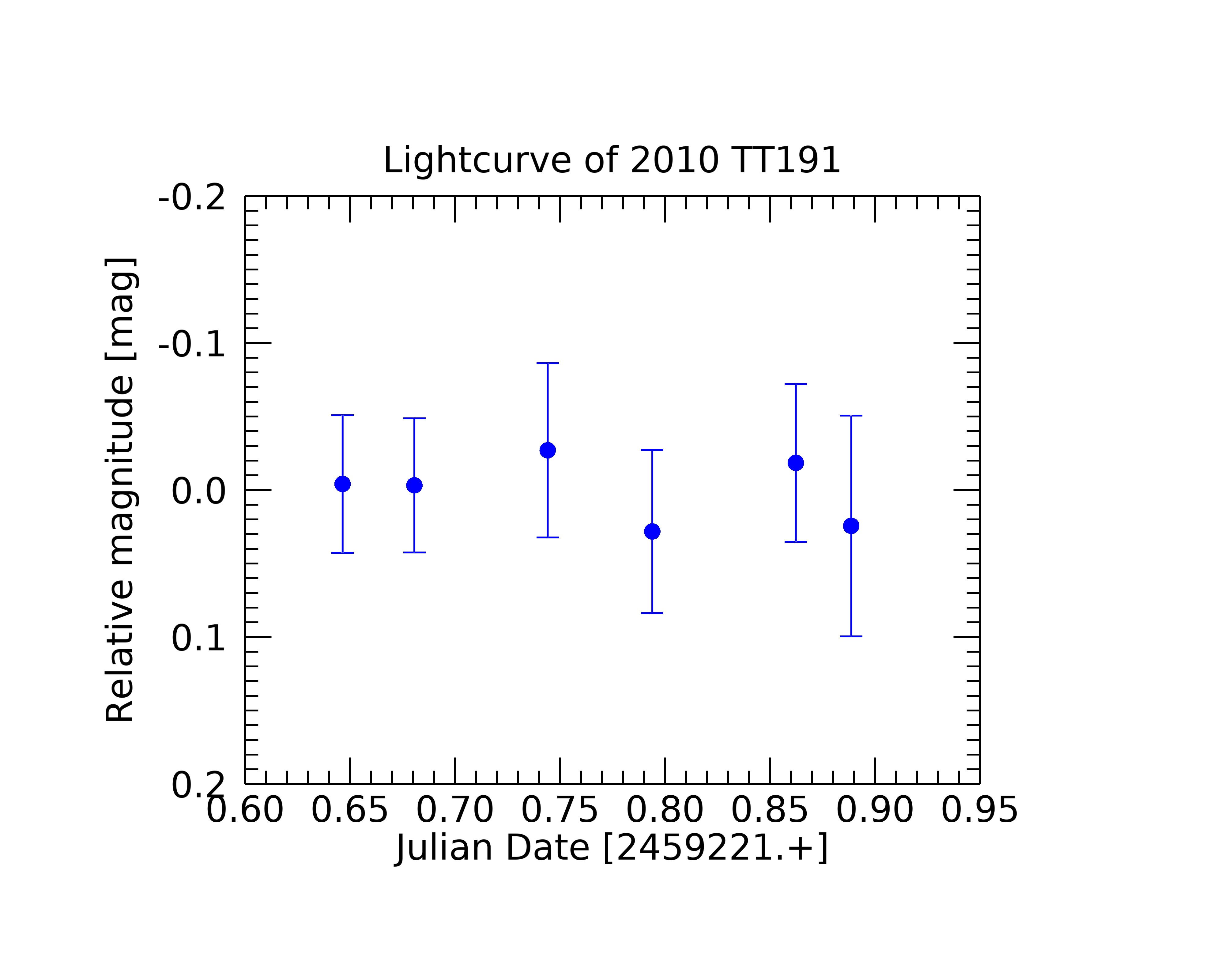} 
   \includegraphics[width=10cm,angle=0]{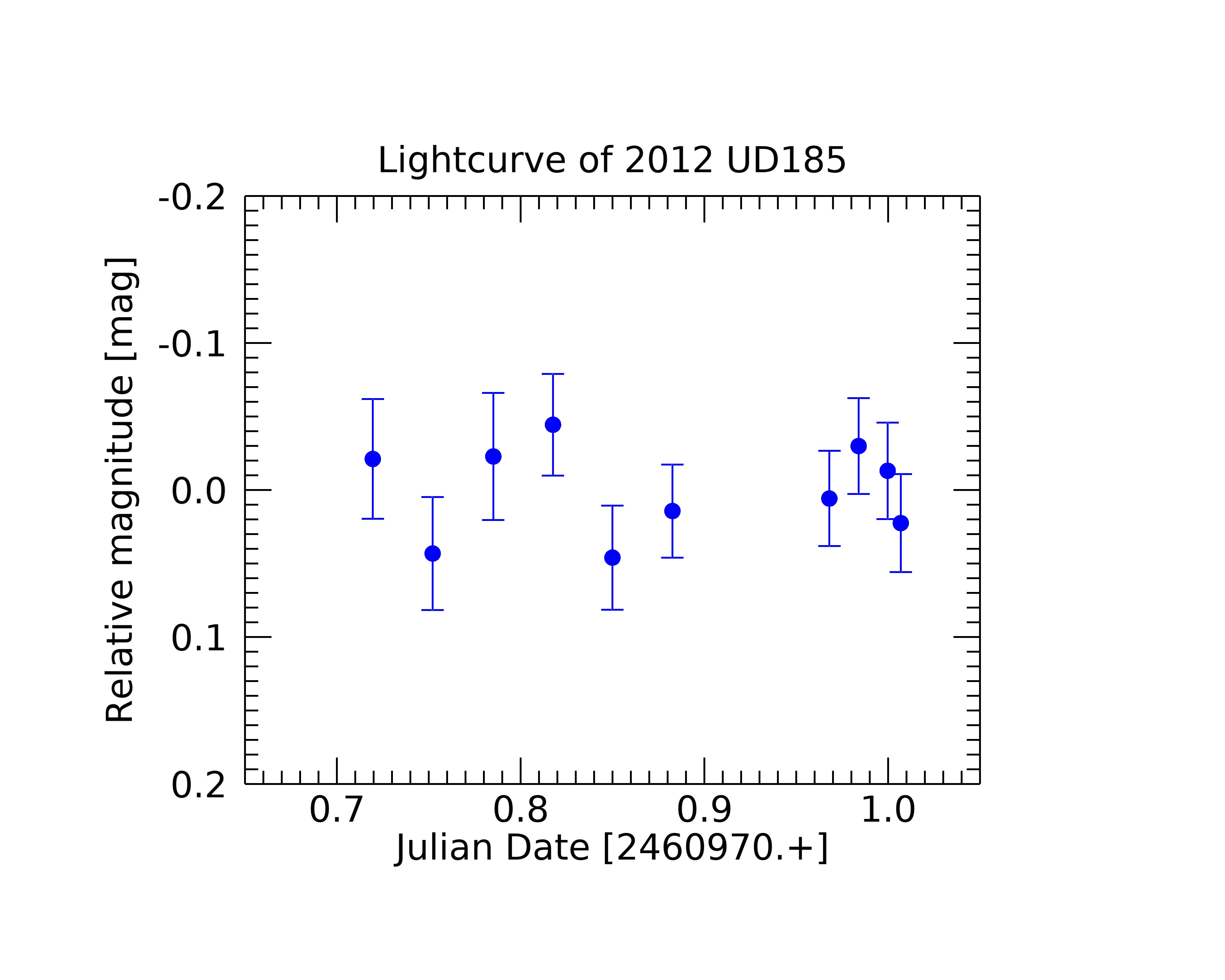} 
 \caption{Partial lightcurves of KBOs trapped in the 1:1 resonance }
\label{fig:LC11}
\end{figure}
      \begin{figure}
  \includegraphics[width=10cm,angle=0]{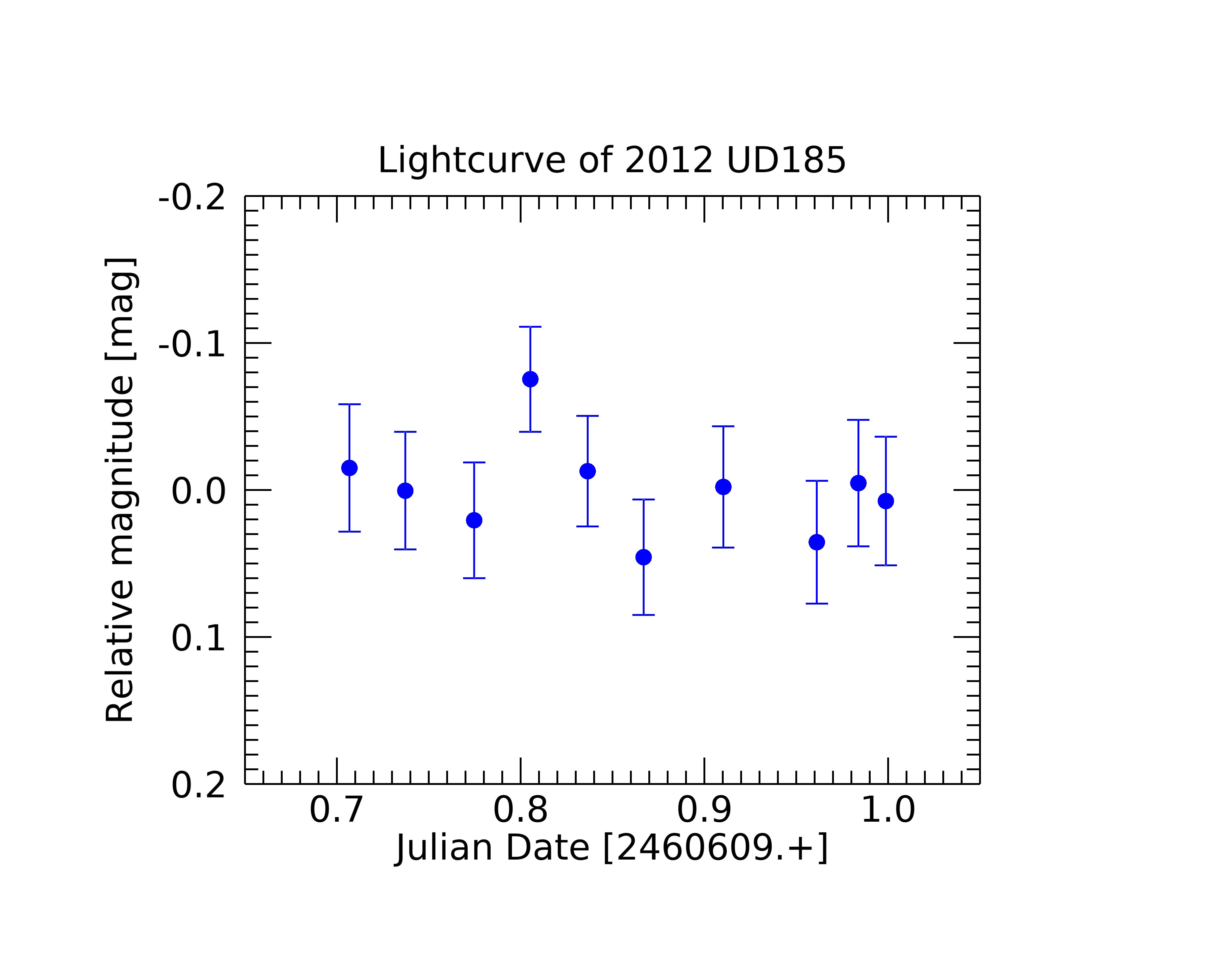} 
     \includegraphics[width=10cm,angle=0]{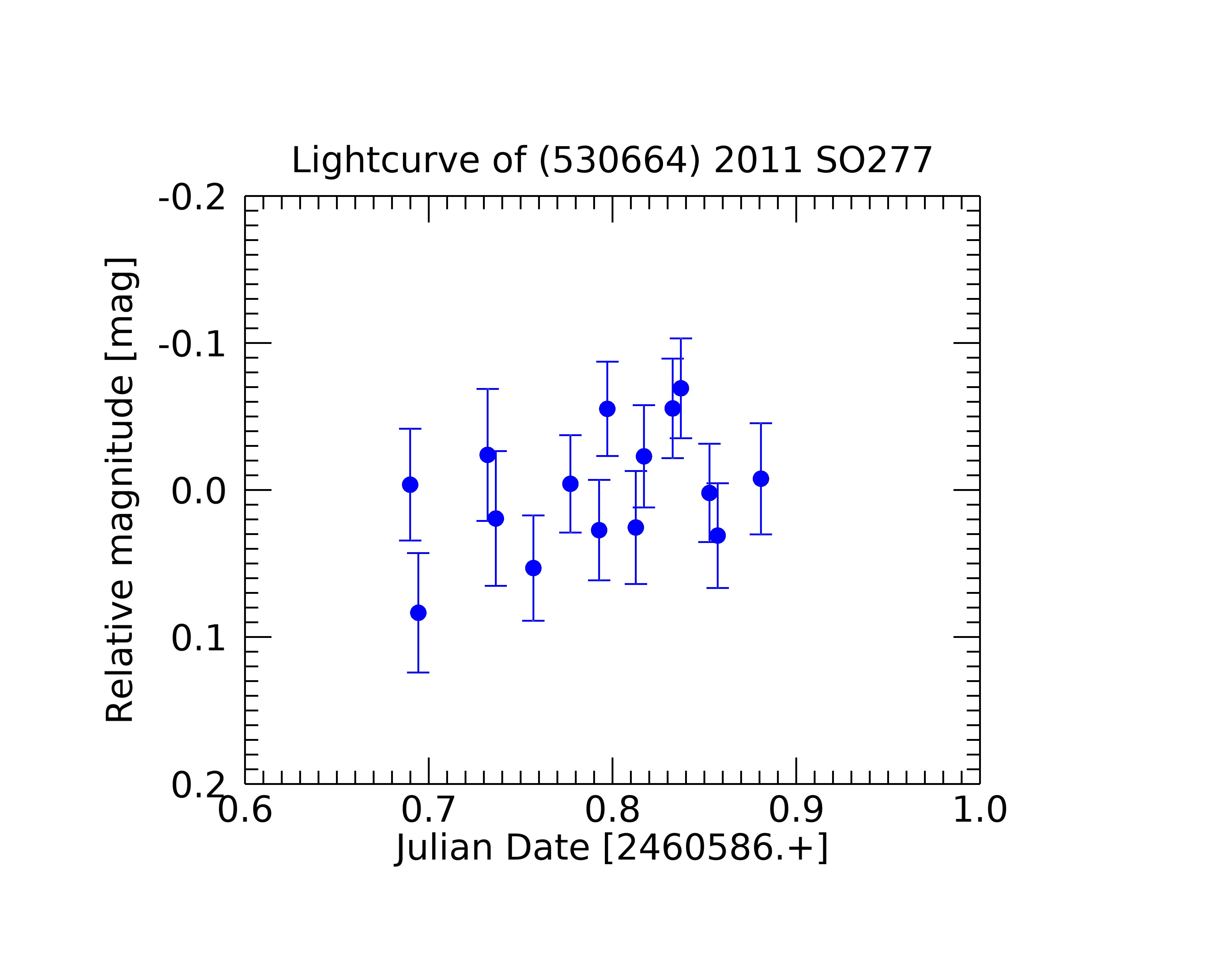} 
\caption{ continued}
\label{fig:LC112}
\end{figure}
 
 
 \clearpage
      \begin{figure}
  \includegraphics[width=10cm,angle=0]{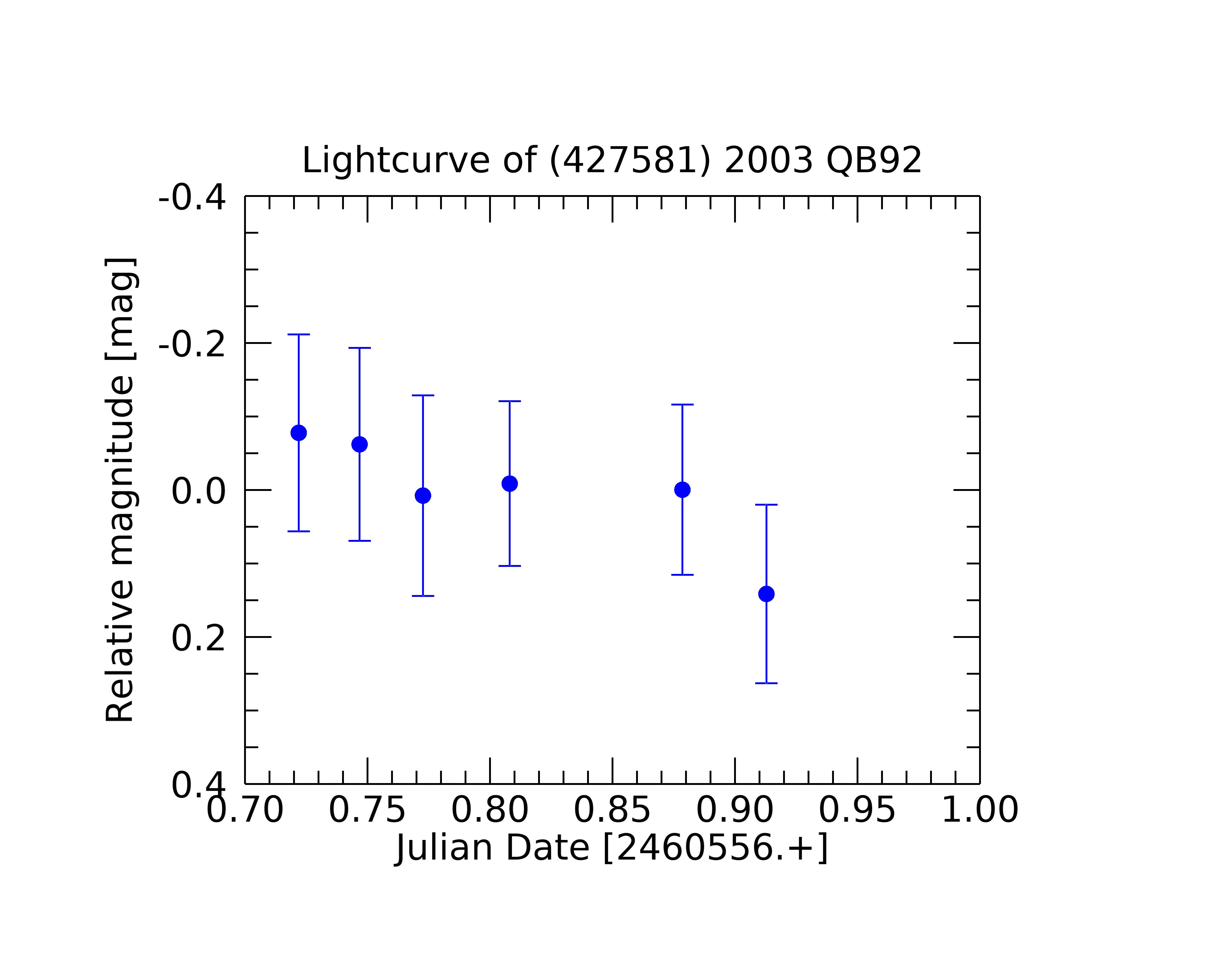} 
   \includegraphics[width=10cm,angle=0]{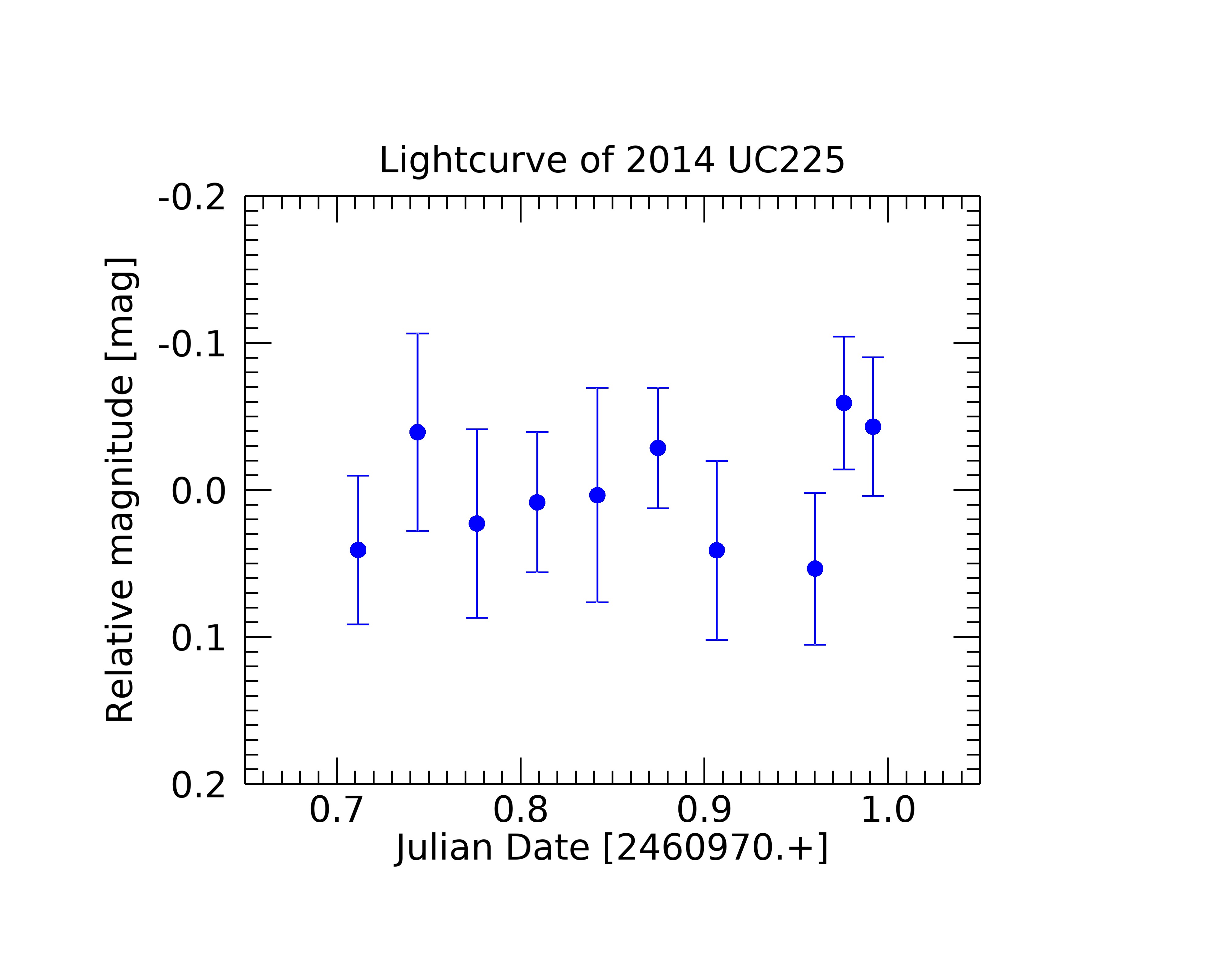} 
    \includegraphics[width=10cm,angle=0]{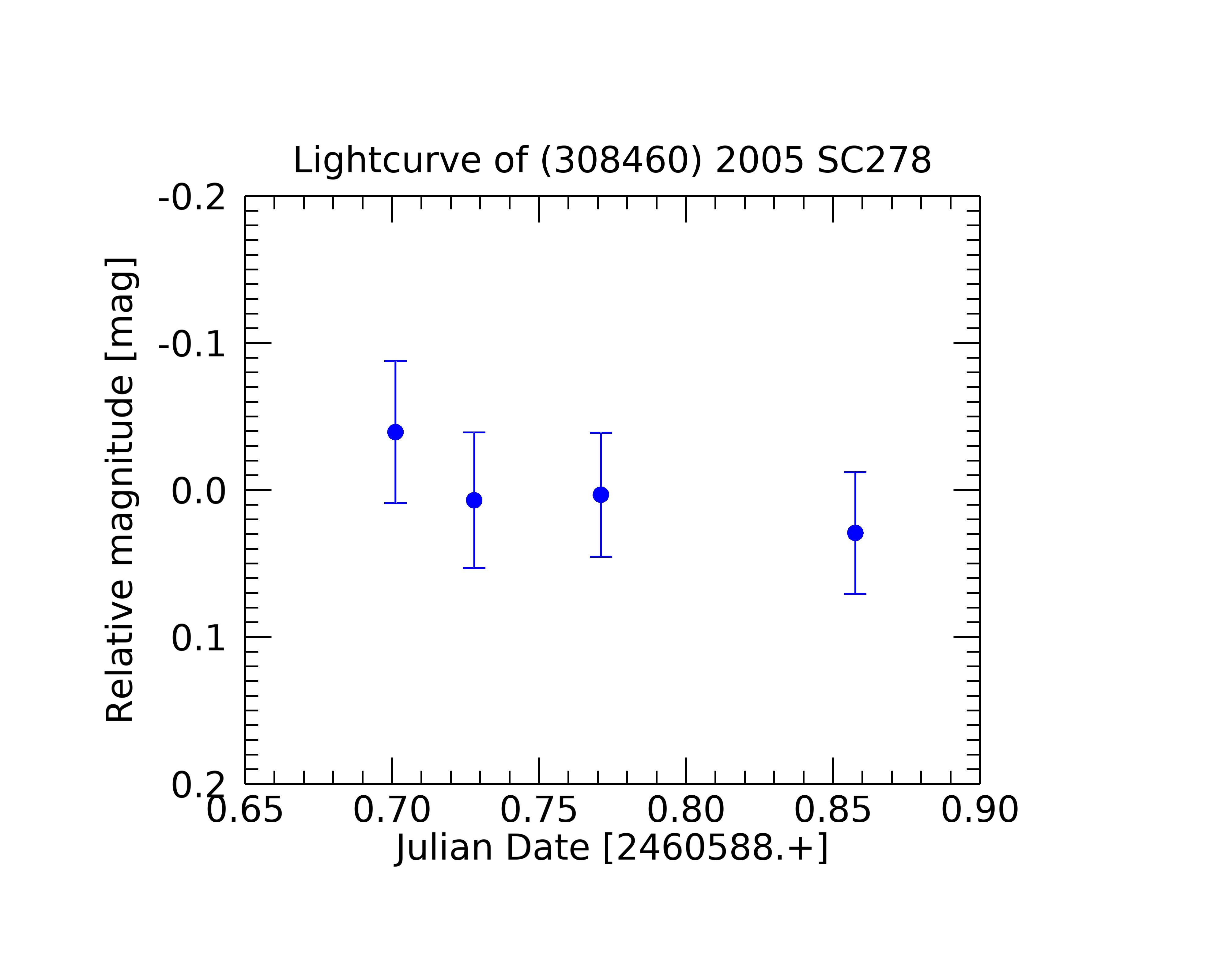} 
 \includegraphics[width=10cm,angle=0]{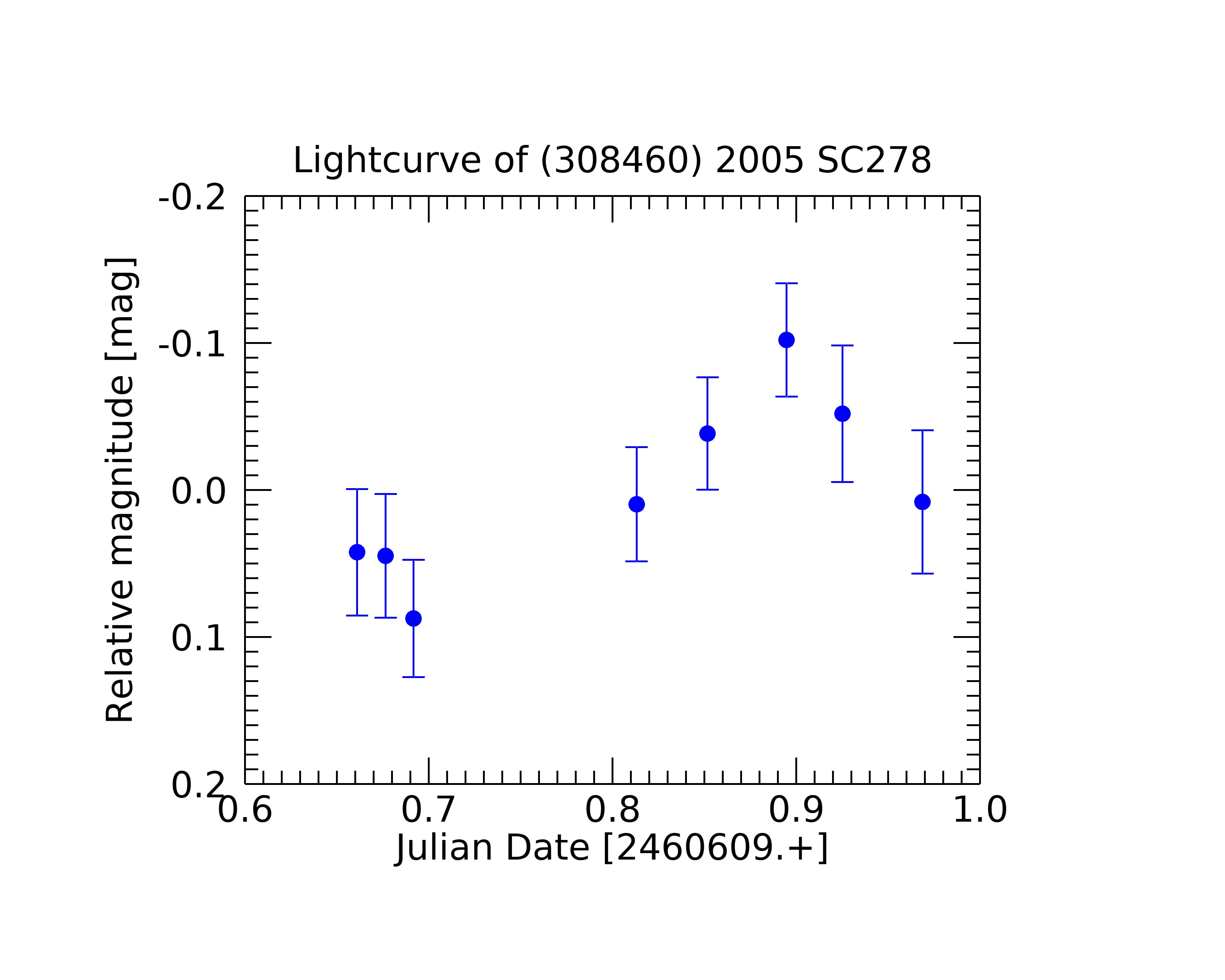} 
 \includegraphics[width=10cm,angle=0]{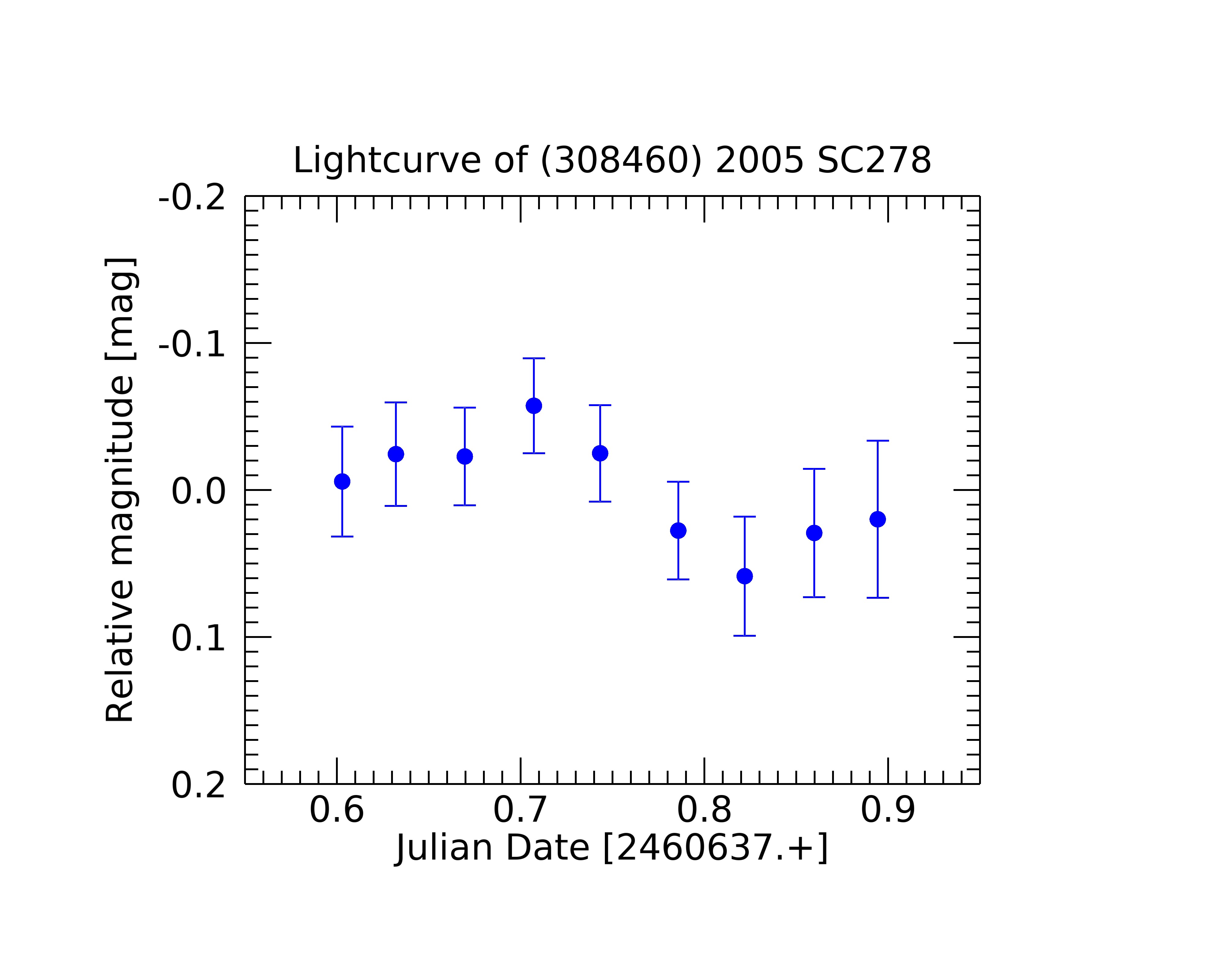} 
     \includegraphics[width=10cm,angle=0]{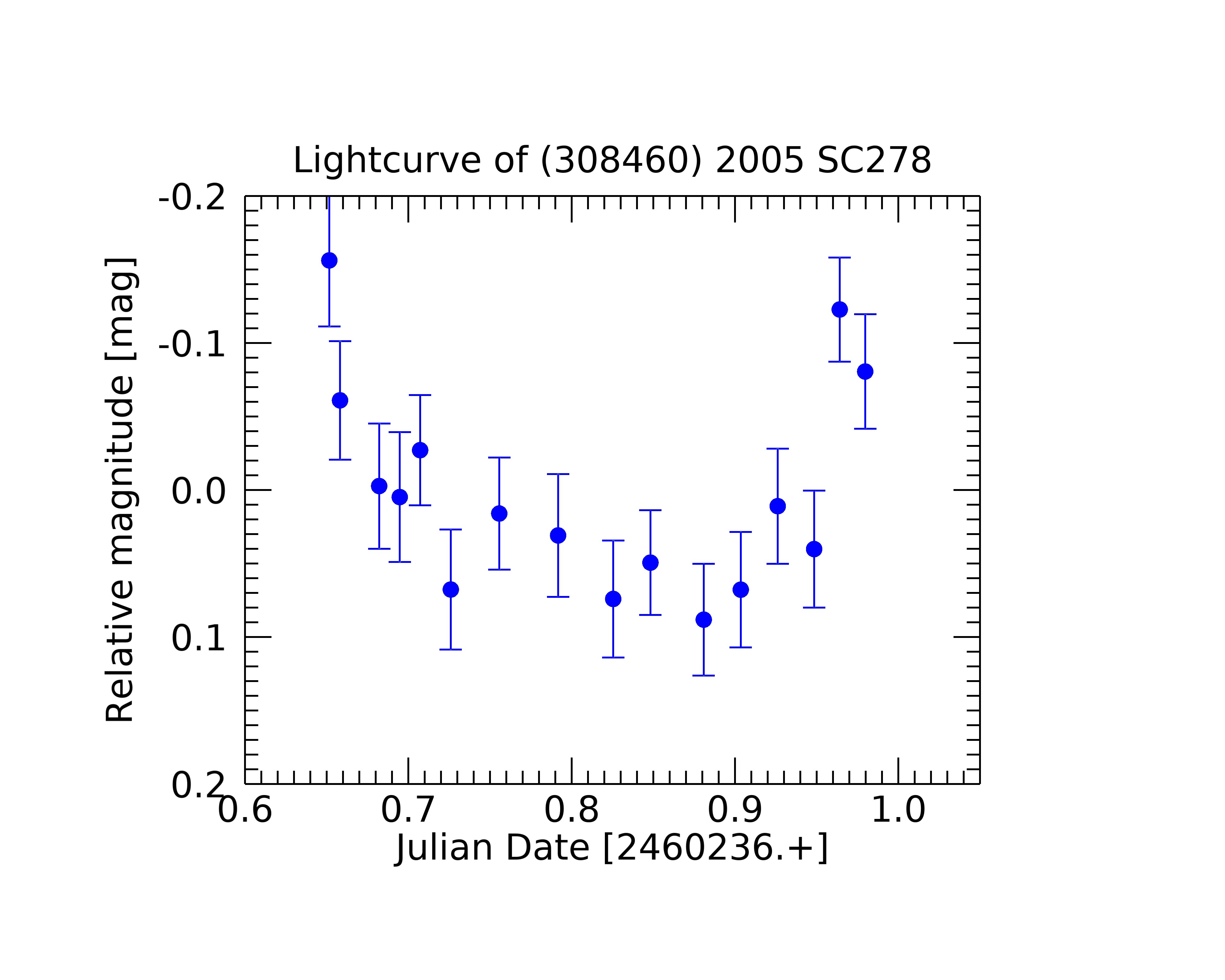} 
 \caption{Partial lightcurves of KBOs trapped in the 5:4 resonance }
\label{fig:LC54}
\end{figure}

   
 \clearpage
        \begin{figure}
 \includegraphics[width=10cm,angle=0]{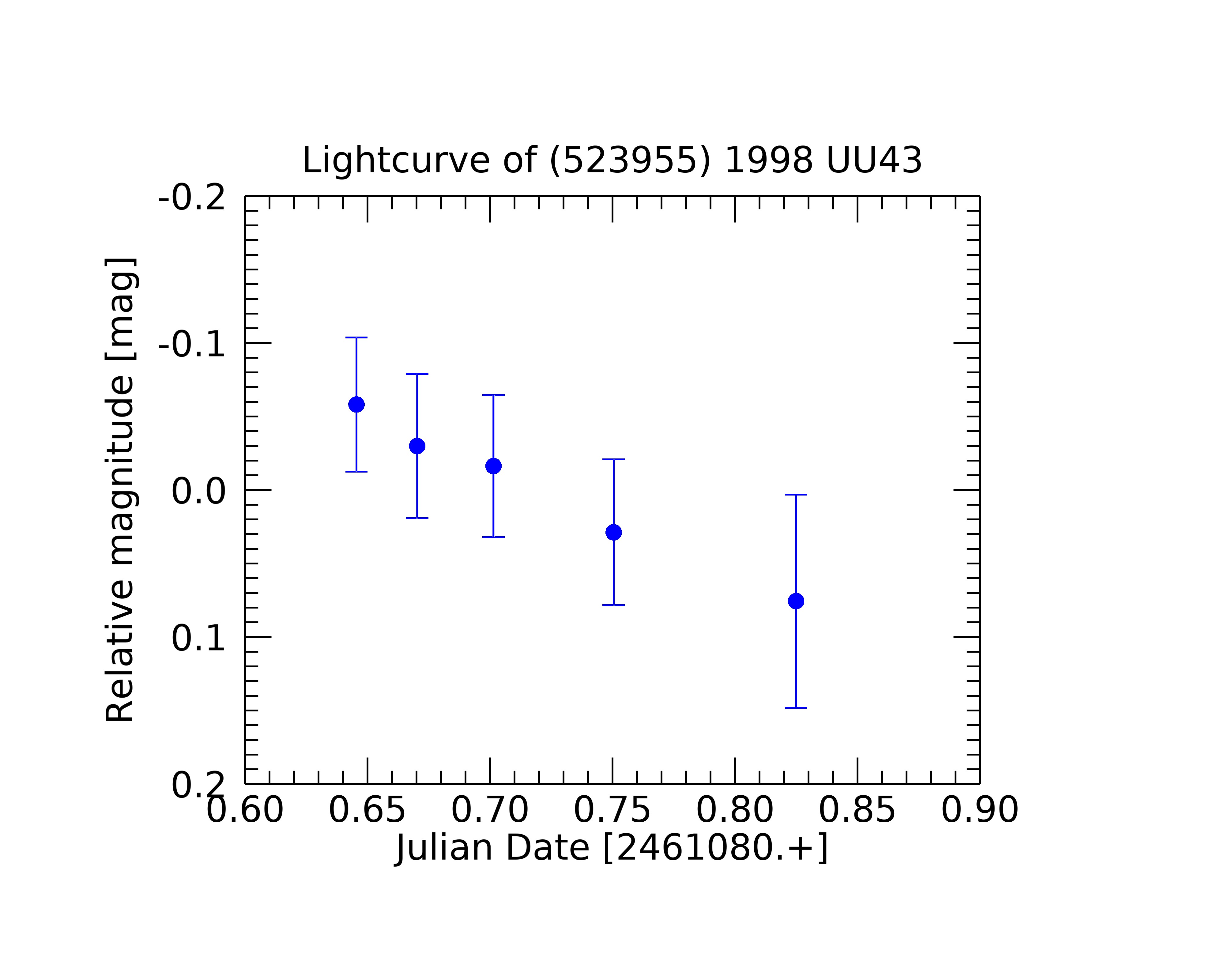} 
  \includegraphics[width=10cm,angle=0]{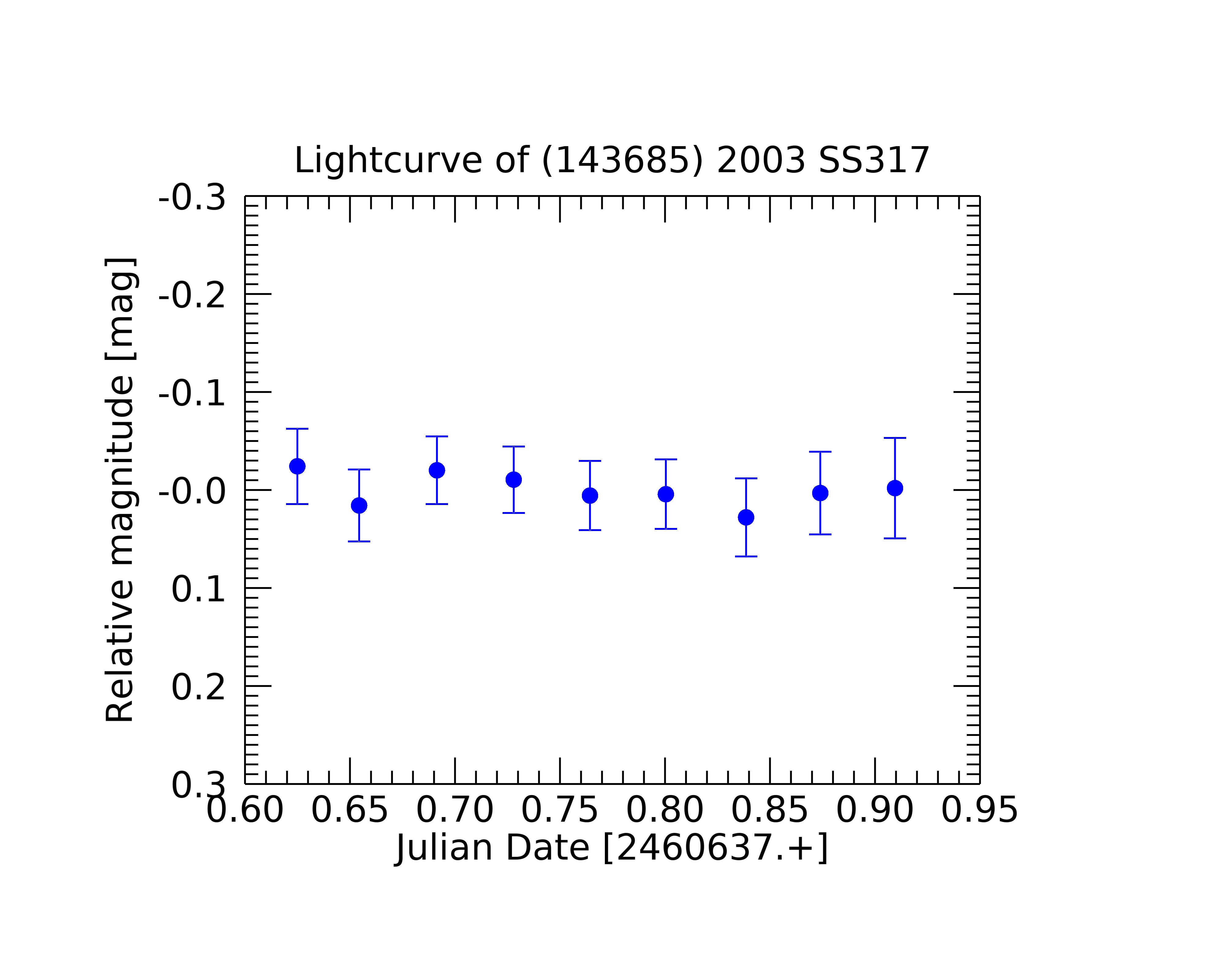} 
   \includegraphics[width=10cm,angle=0]{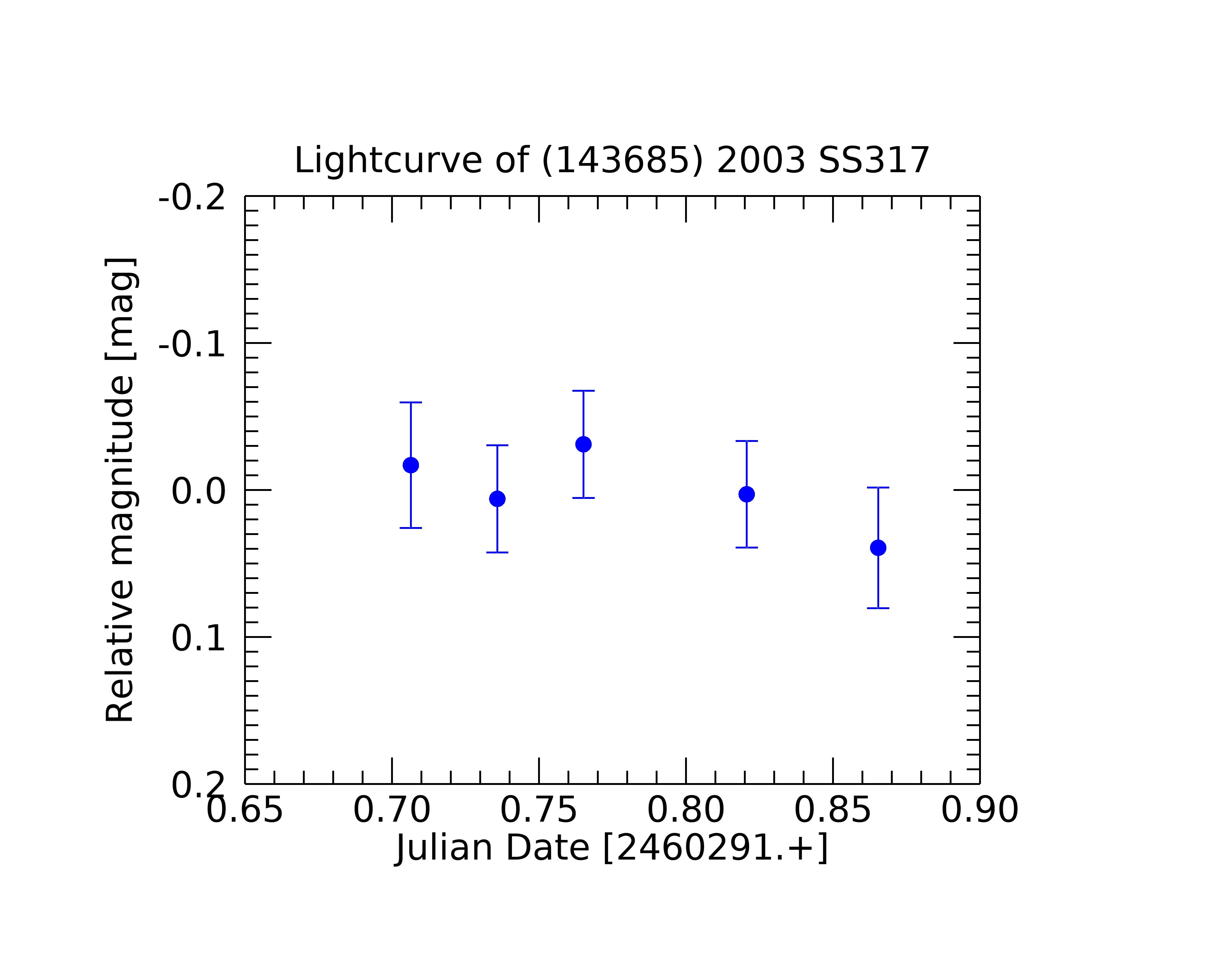} 
     \includegraphics[width=10cm,angle=0]{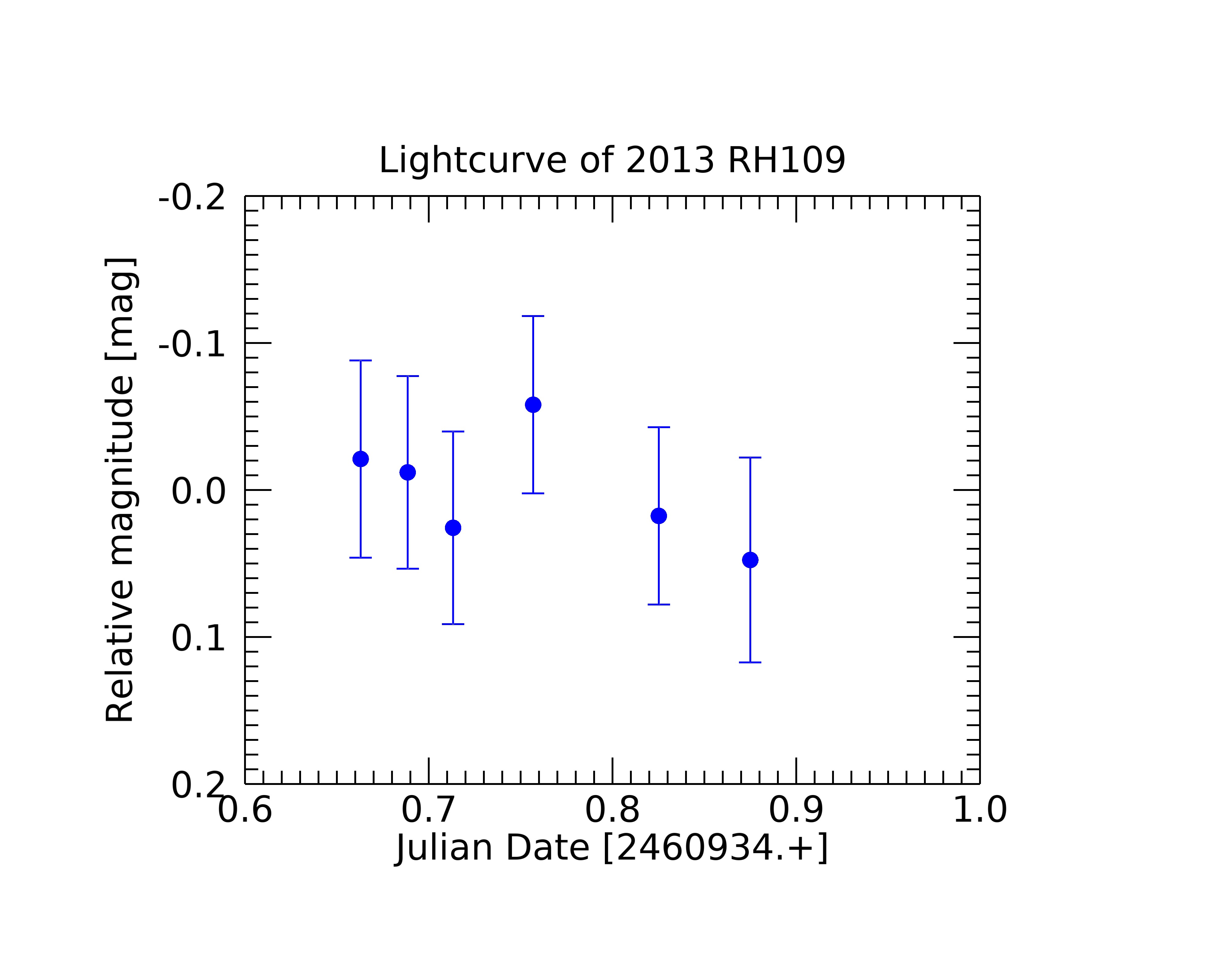} 
    \includegraphics[width=10cm,angle=0]{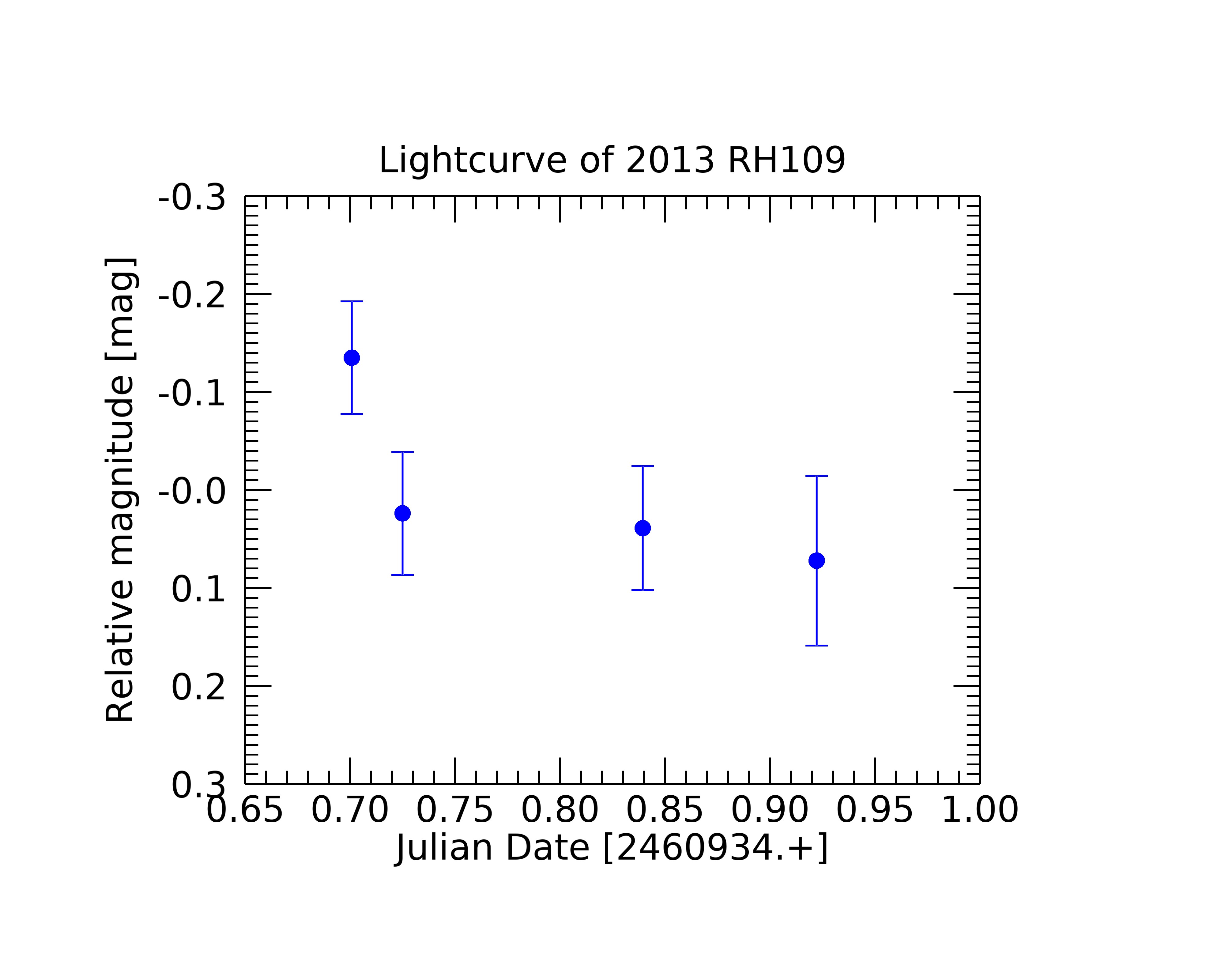} 
    \includegraphics[width=10cm,angle=0]{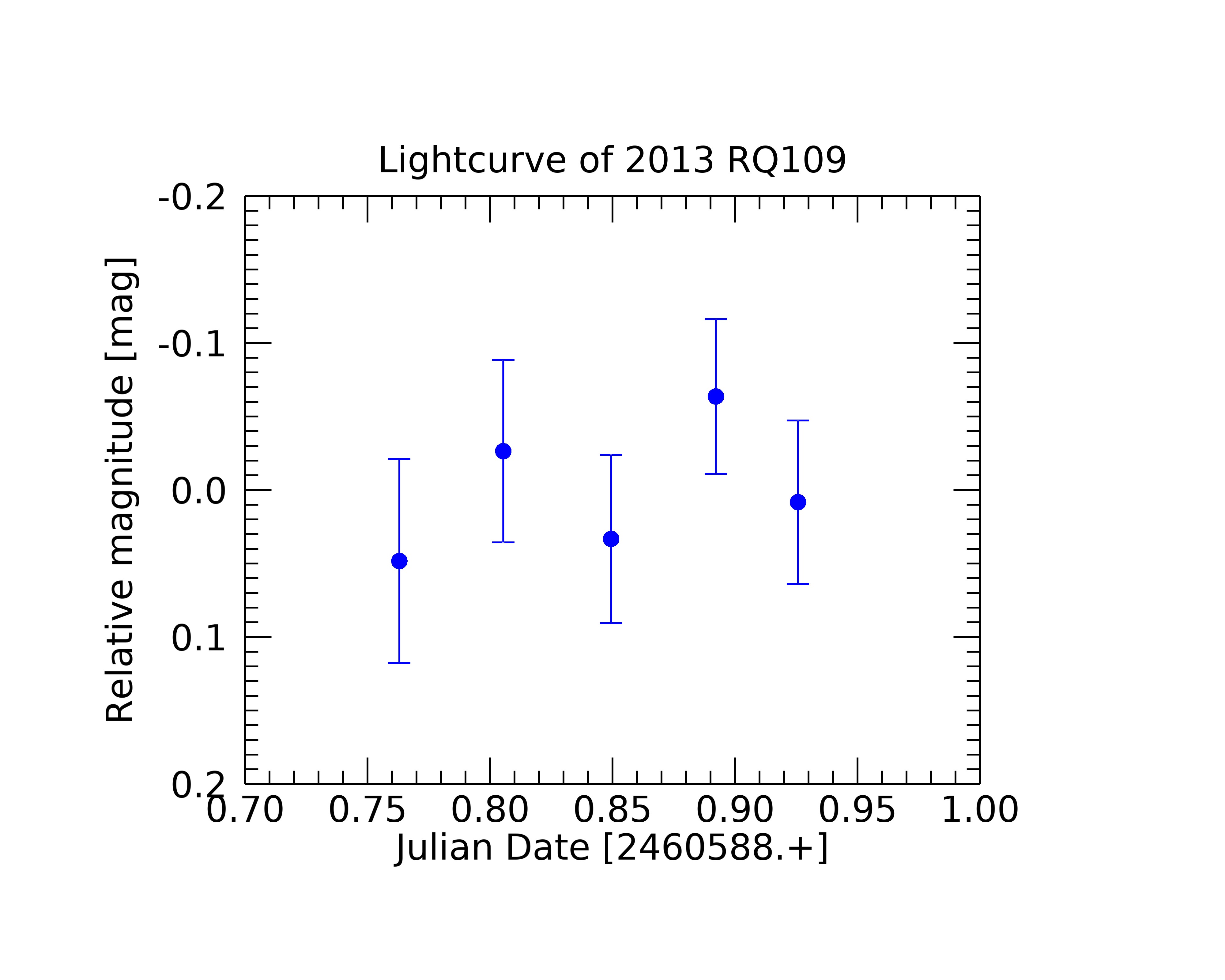} 
 \caption{Partial lightcurves of KBOs trapped in the 4:3 resonance }
\label{fig:LC431}
\end{figure}

     \begin{figure}
        \includegraphics[width=10cm,angle=0]{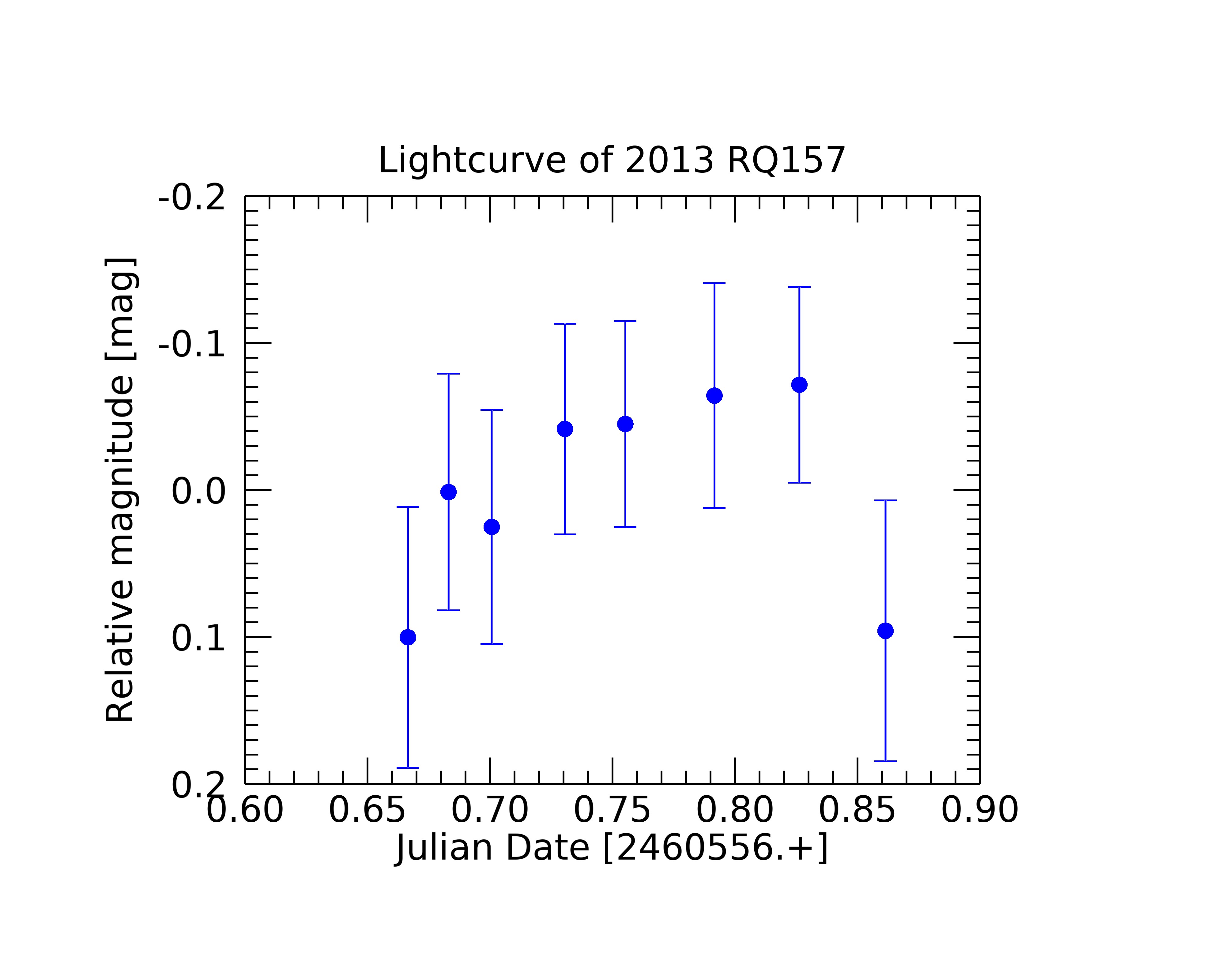} 
 \includegraphics[width=10cm,angle=0]{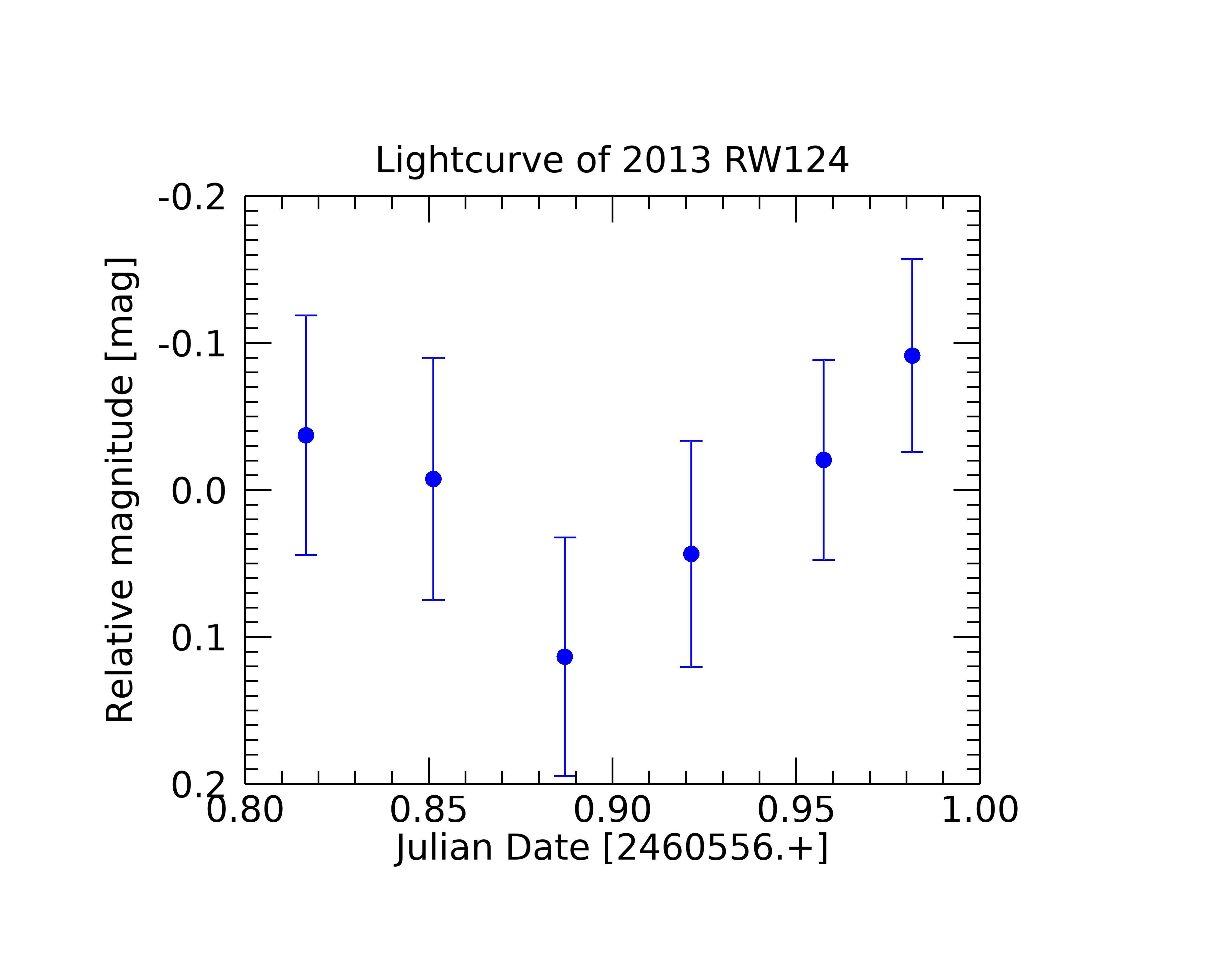} 
         \includegraphics[width=10cm,angle=0]{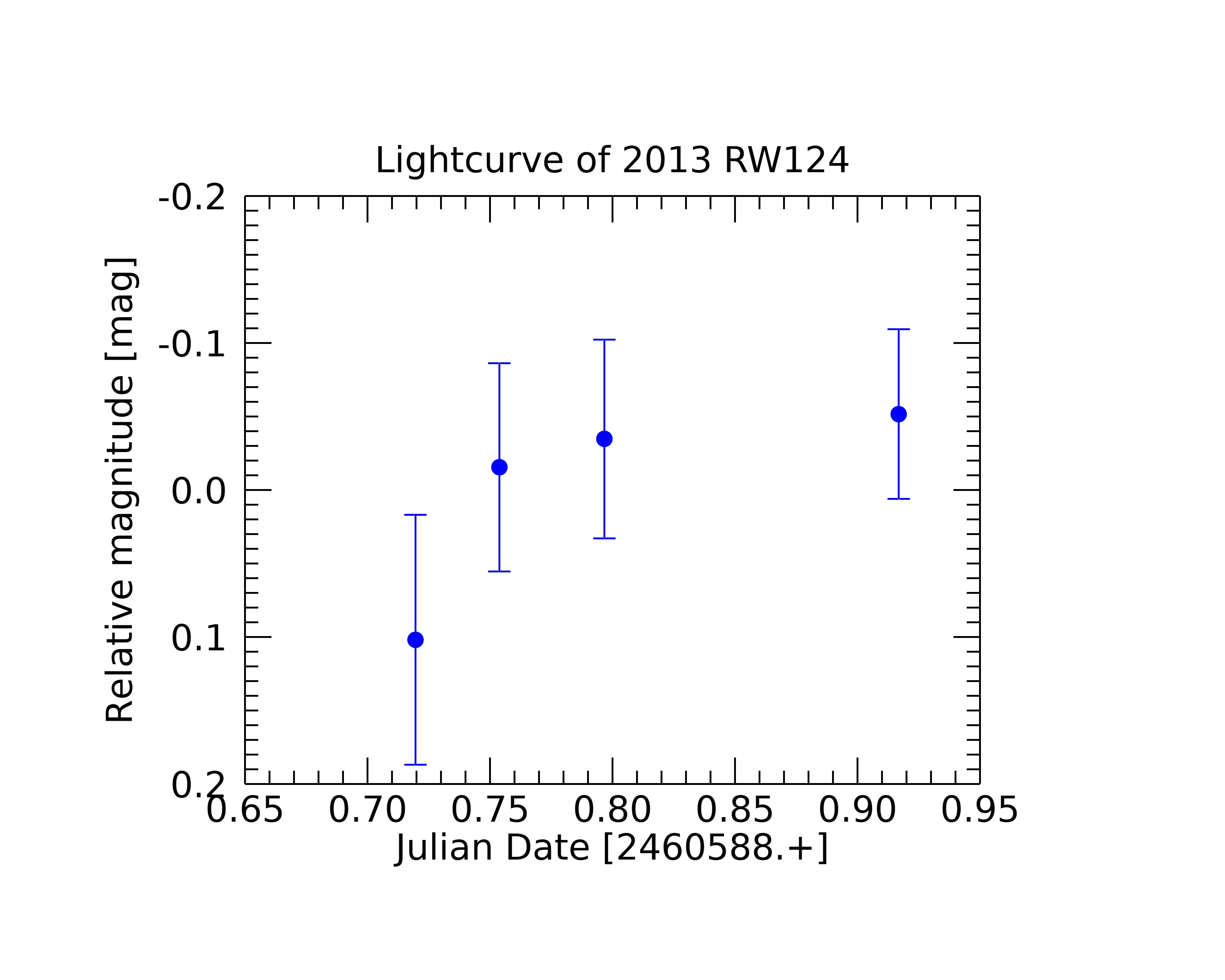} 
  \includegraphics[width=10cm,angle=0]{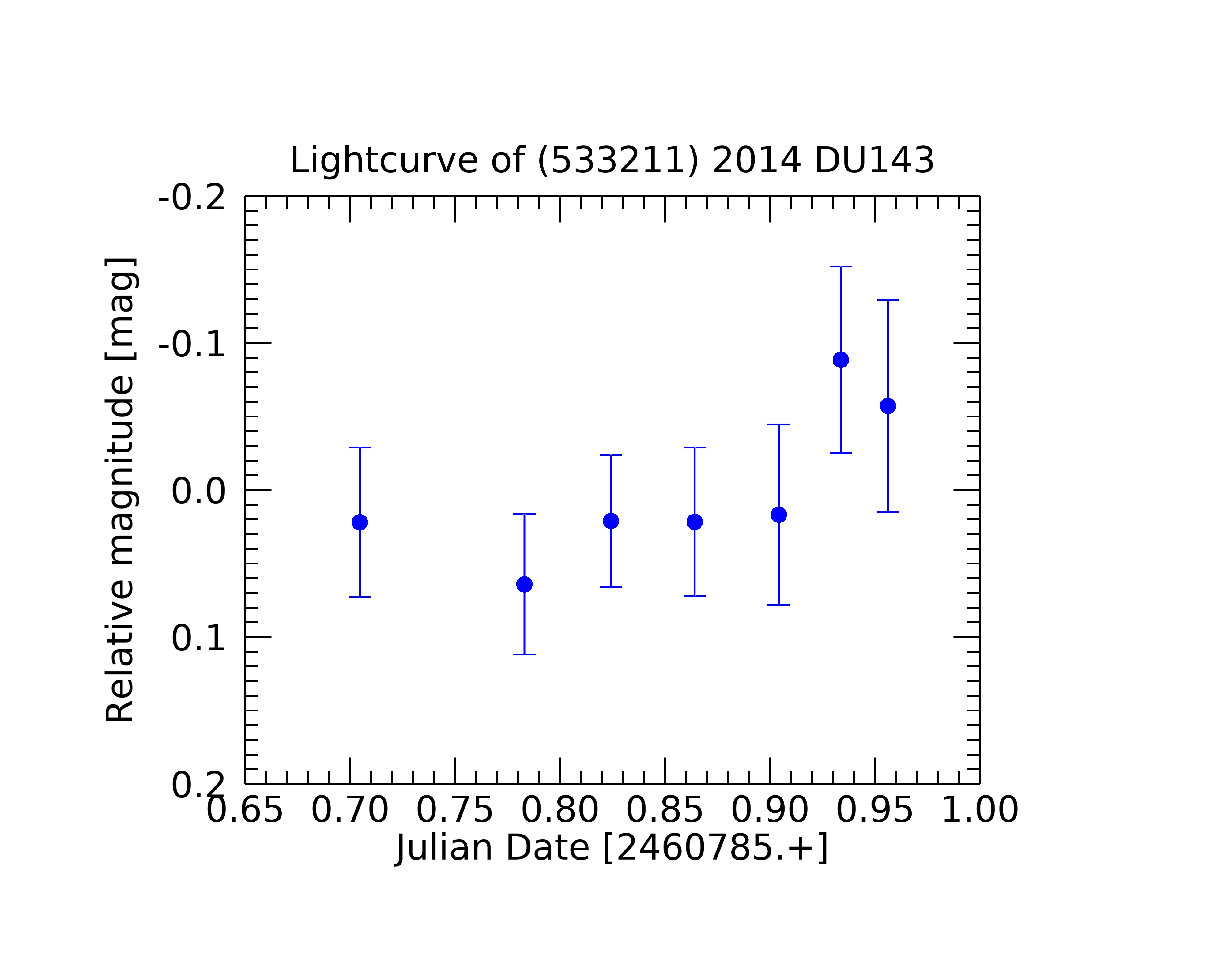} 
    \includegraphics[width=10cm,angle=0]{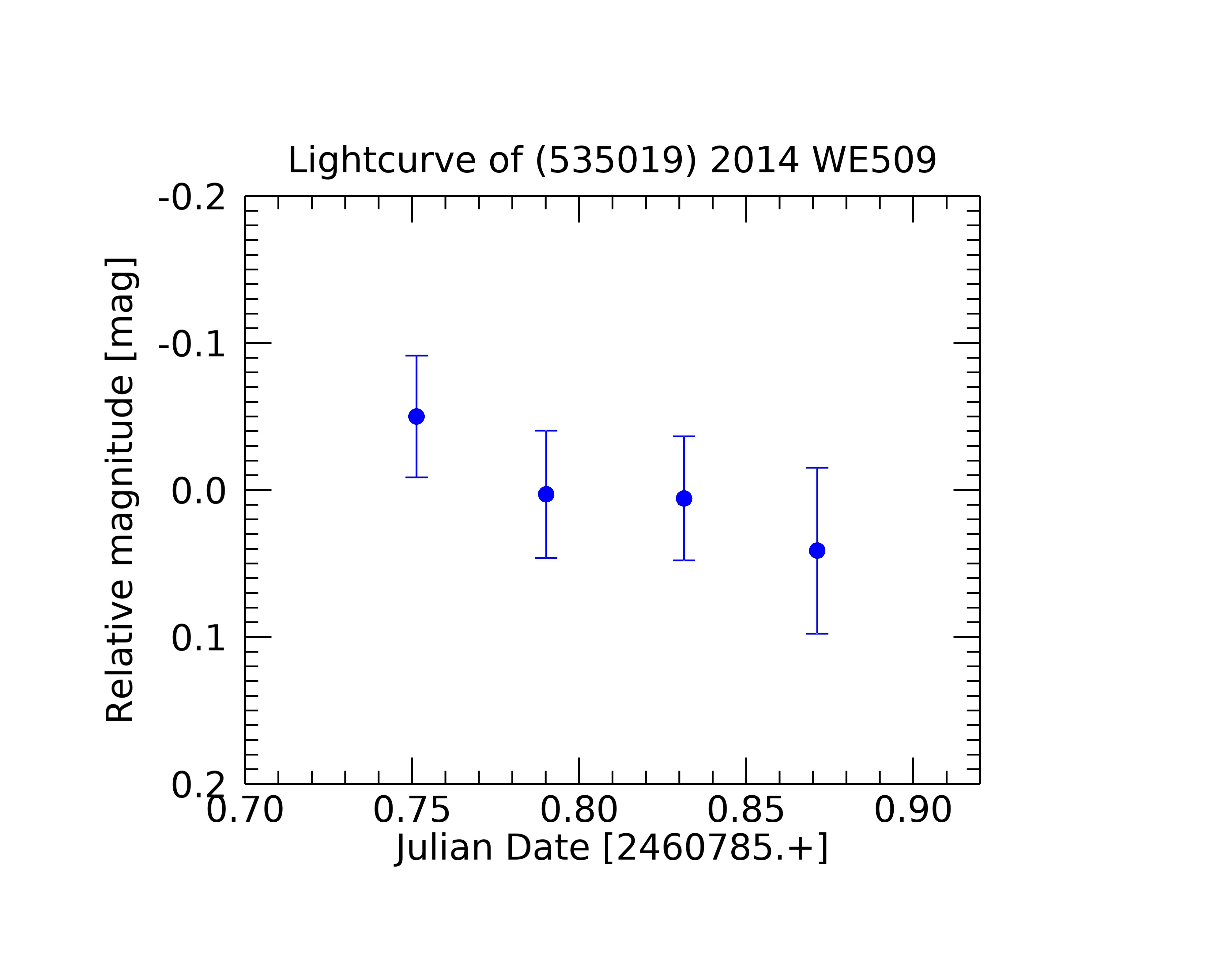} 
    \includegraphics[width=10cm,angle=0]{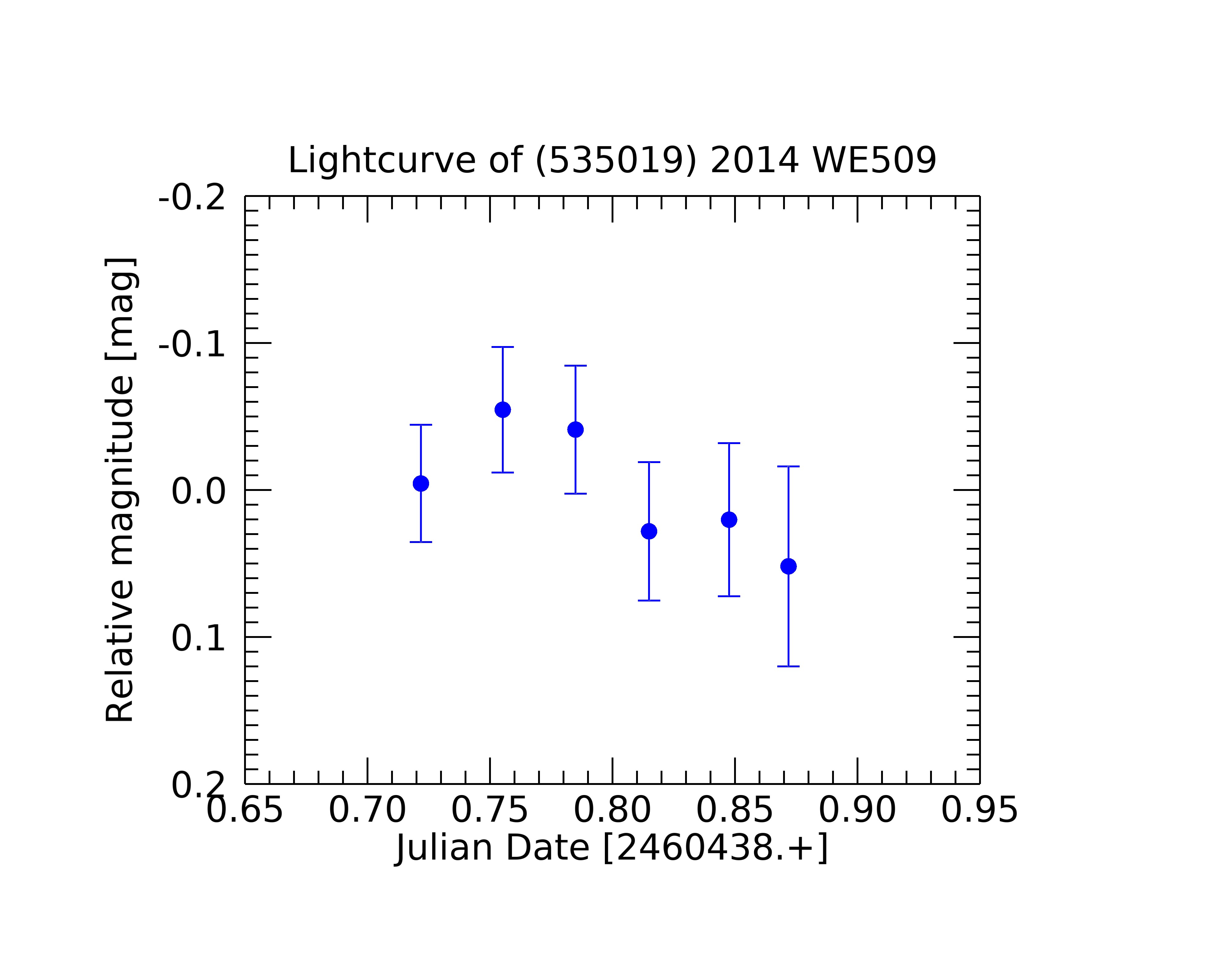} 
 \caption{ Continued}
\label{fig:LC432}
\end{figure}

      \begin{figure} 
        \includegraphics[width=10cm,angle=0]{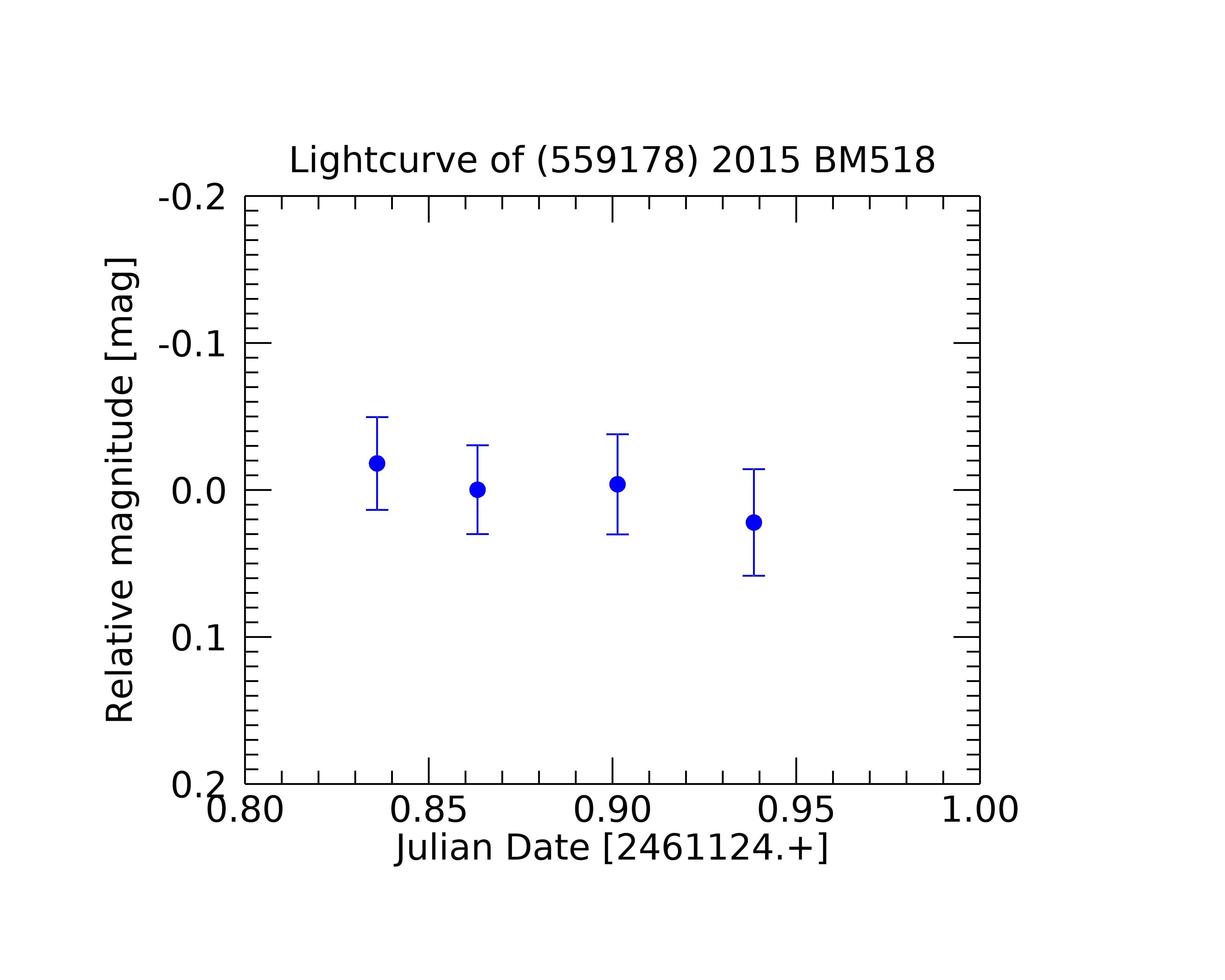} 
   \includegraphics[width=10cm,angle=0]{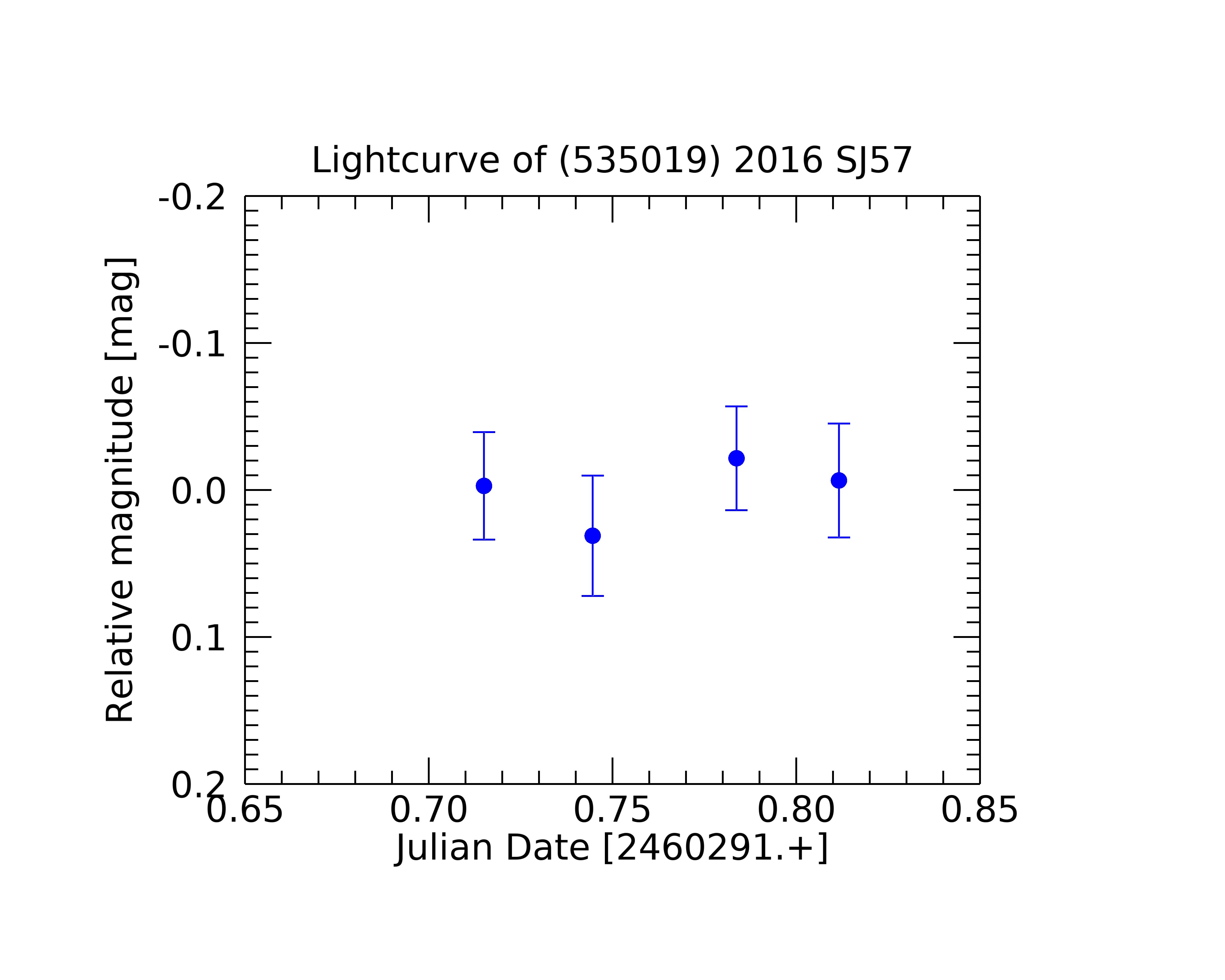} 
 \caption{ Continued}
\label{fig:LC433}
\end{figure}

   
 \clearpage
      \begin{figure}
 \includegraphics[width=10cm,angle=0]{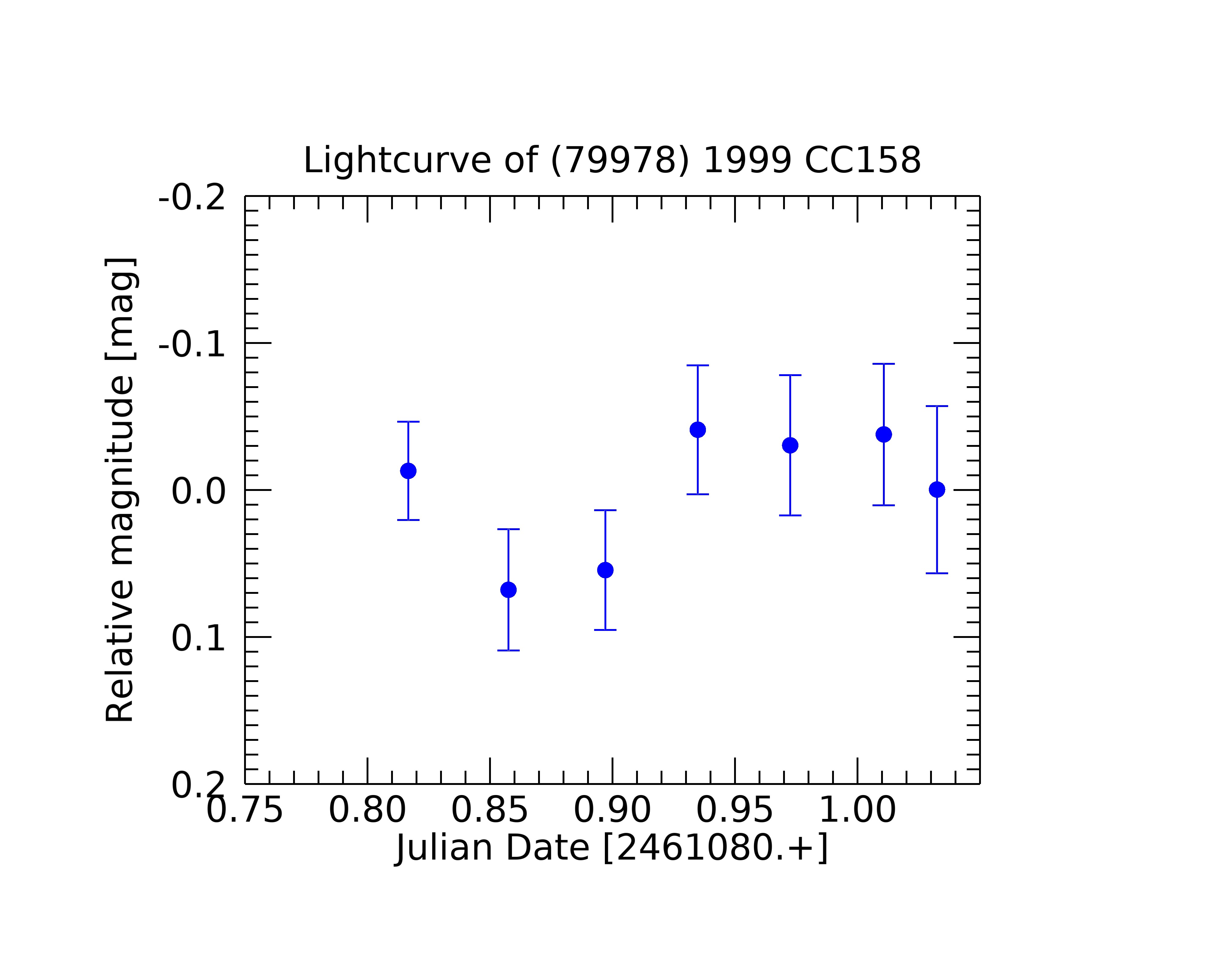} 
 \caption{Partial lightcurves of KBOs trapped in the 12:5 resonance }
\label{fig:LC125}
\end{figure}


 \clearpage
      \begin{figure}
 \includegraphics[width=10cm,angle=0]{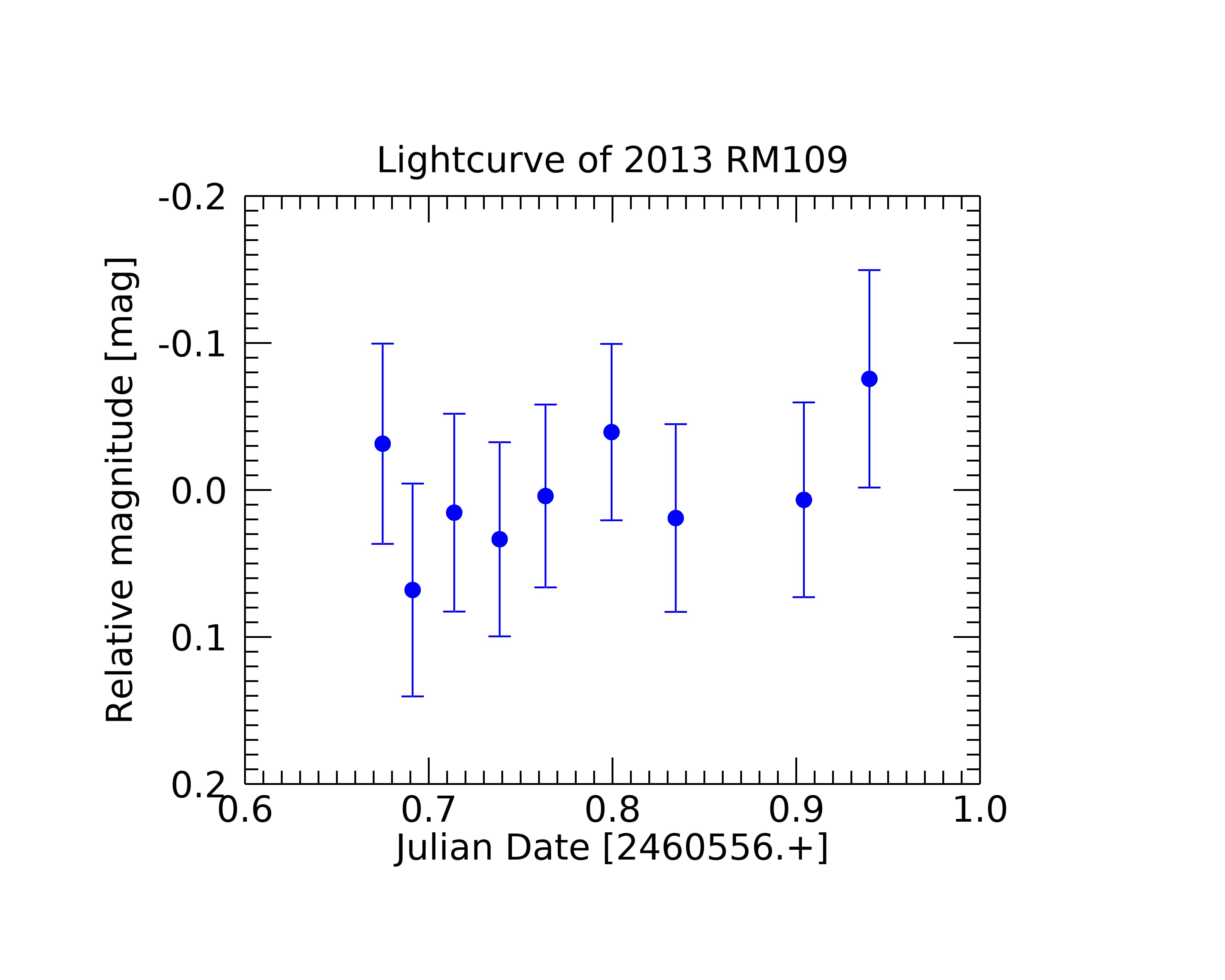} 
 \caption{Partial lightcurves of KBOs trapped in the 11:5 resonance }
\label{fig:LC115}
\end{figure}

 \clearpage
      \begin{figure}
 \includegraphics[width=10cm,angle=0]{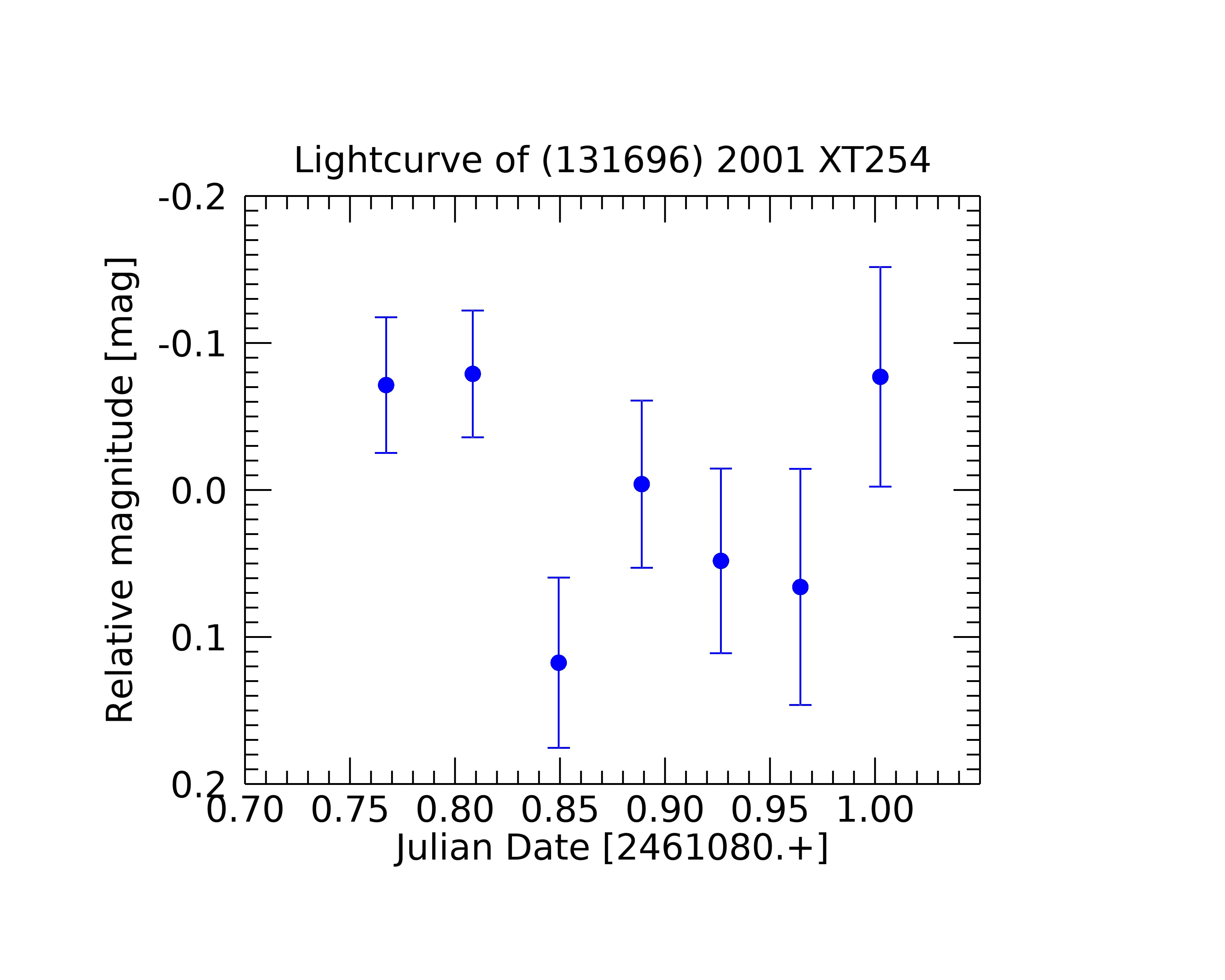} 
  \includegraphics[width=10cm,angle=0]{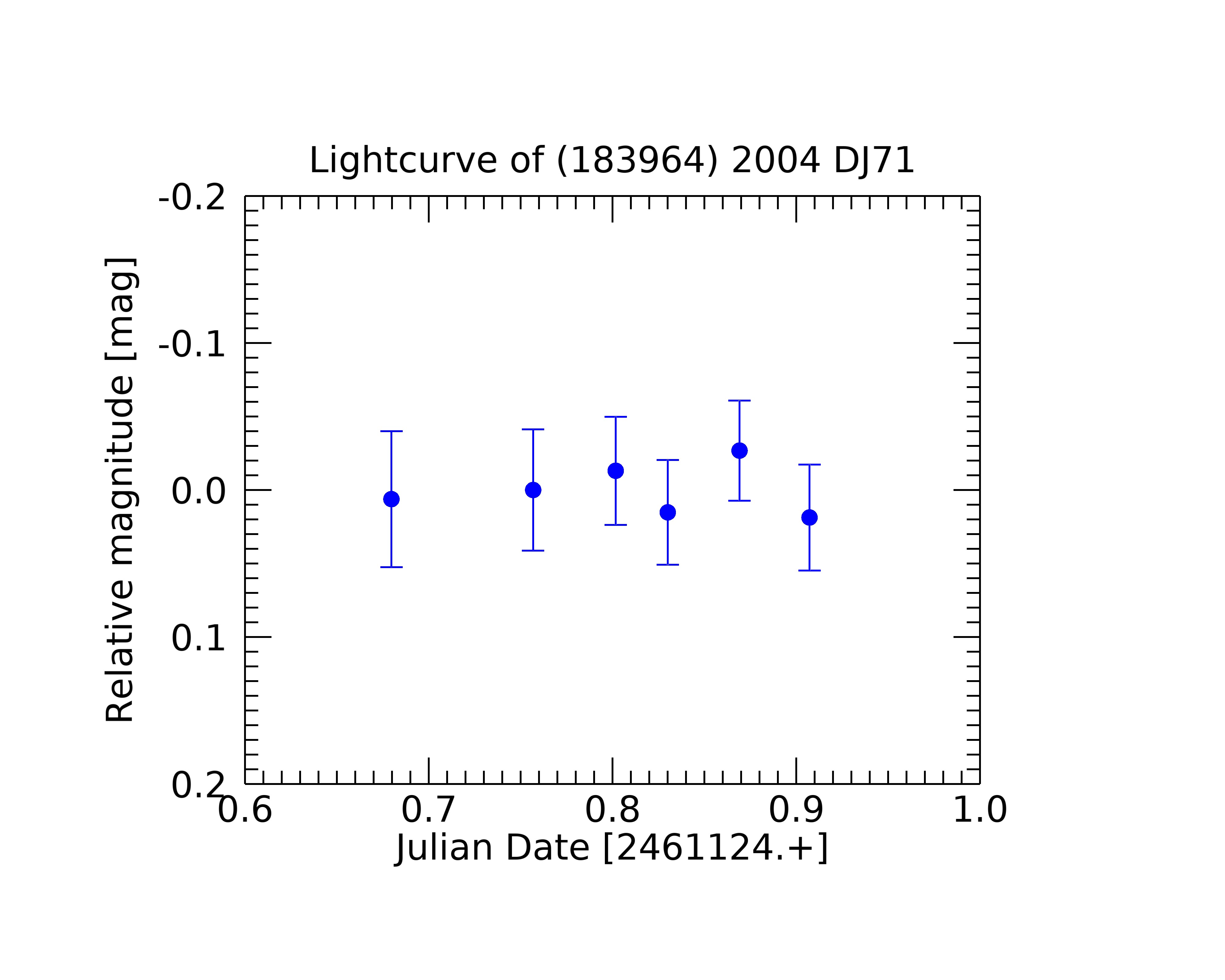} 
  \includegraphics[width=10cm,angle=0]{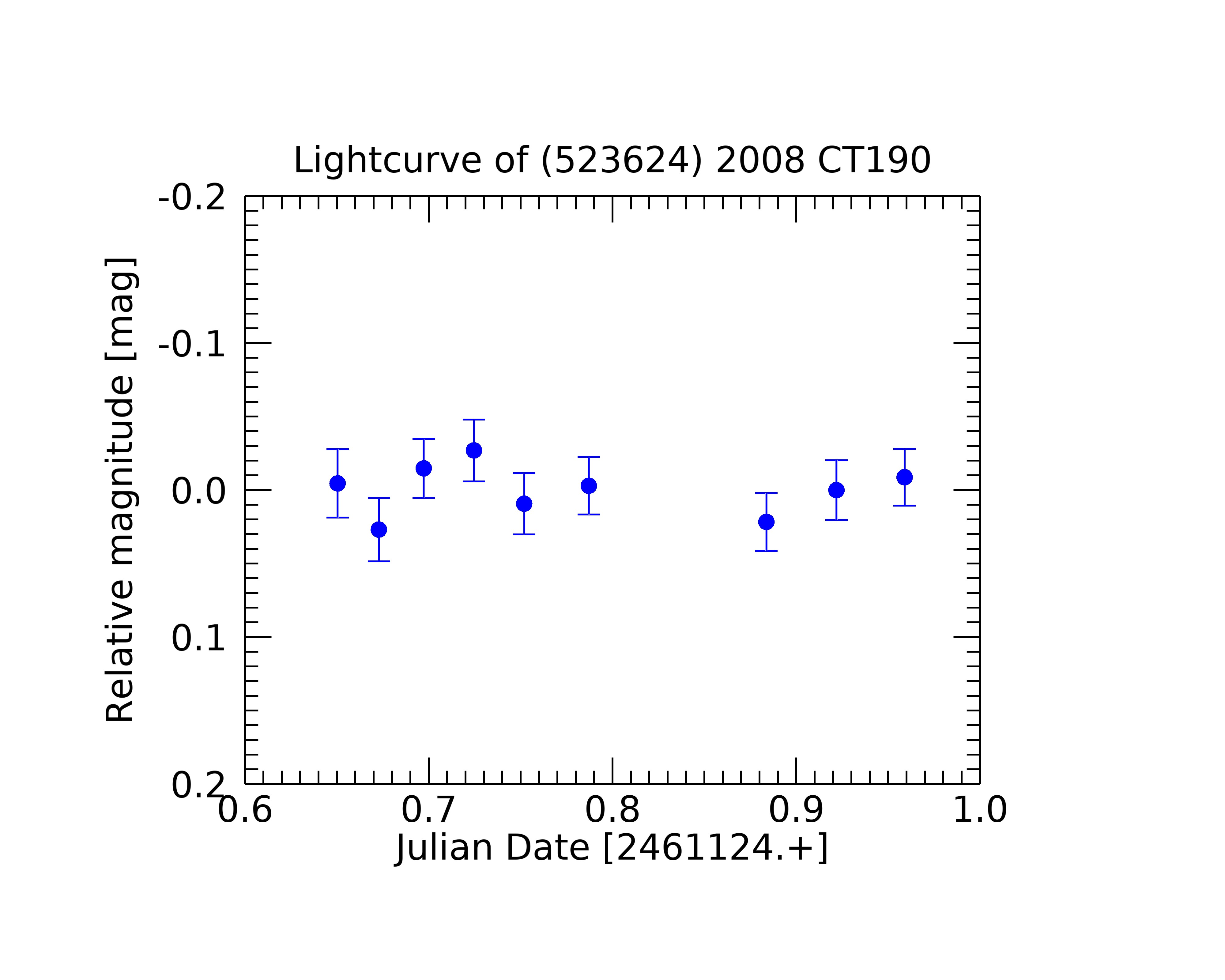} 
 \caption{Partial lightcurves of KBOs trapped in the 7:3 resonance }
\label{fig:LC73}
\end{figure}

 \clearpage
 
      \begin{figure}
  \includegraphics[width=10cm,angle=0]{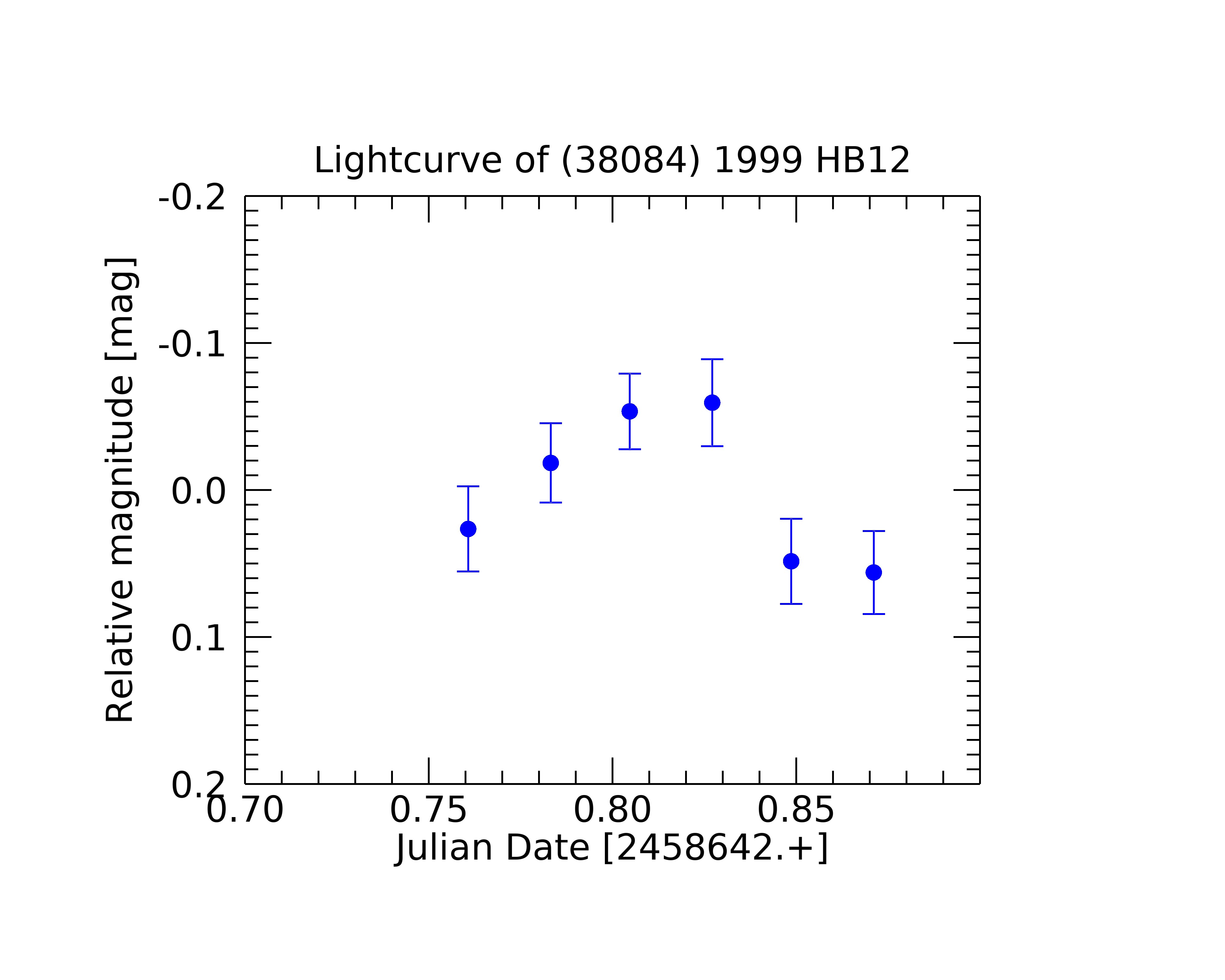} 
    \includegraphics[width=10cm,angle=0]{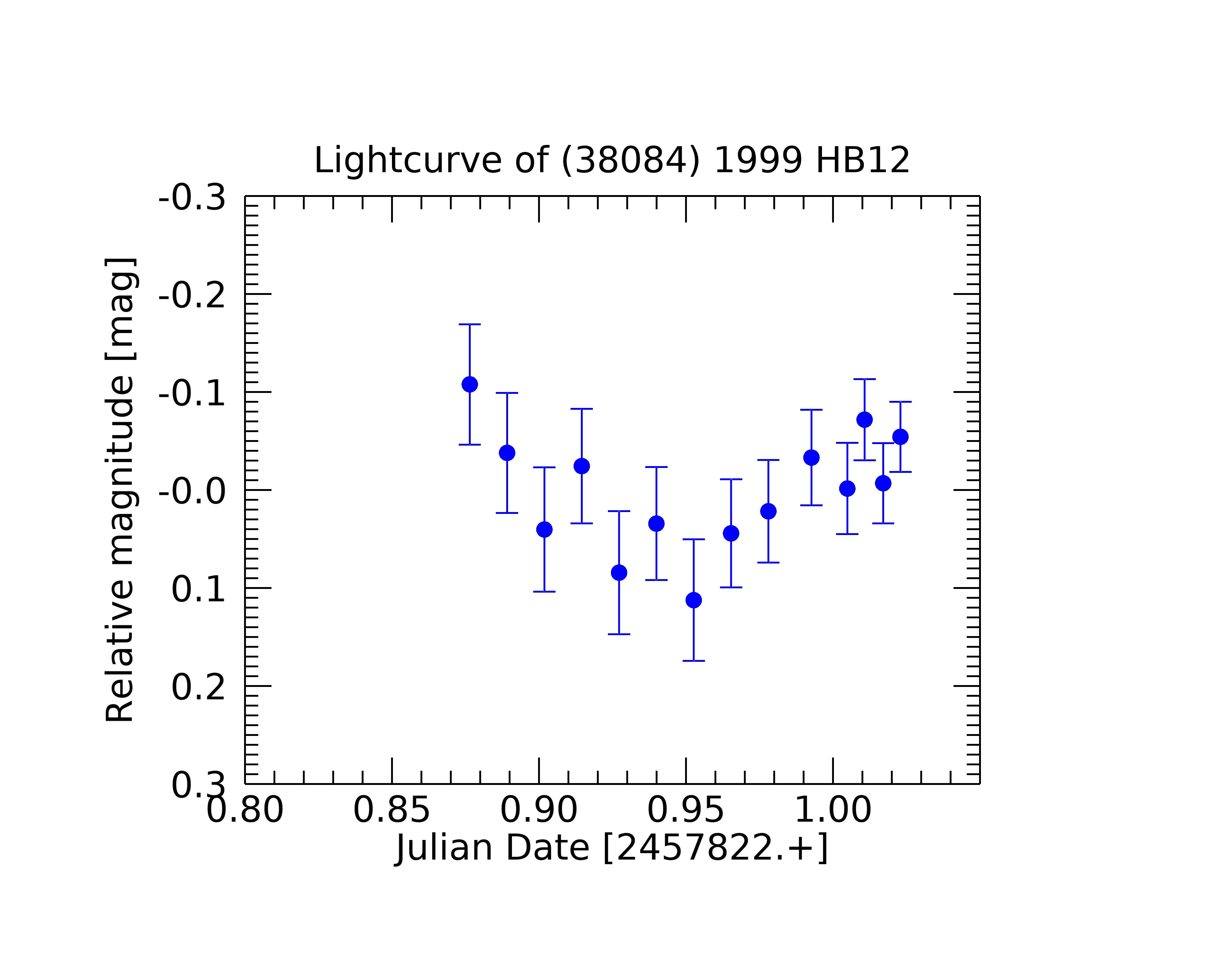} 
 \includegraphics[width=10cm,angle=0]{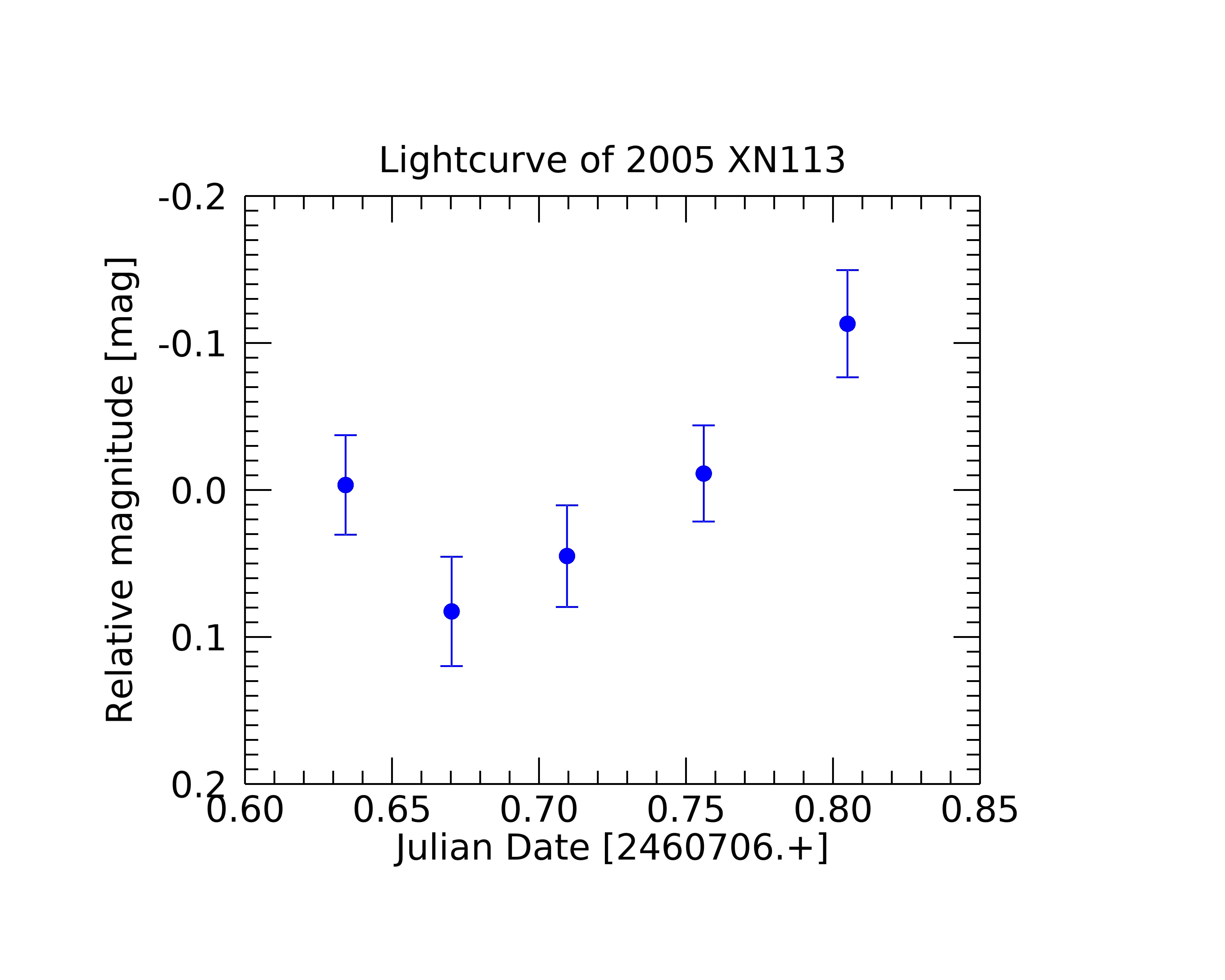} 
  \includegraphics[width=10cm,angle=0]{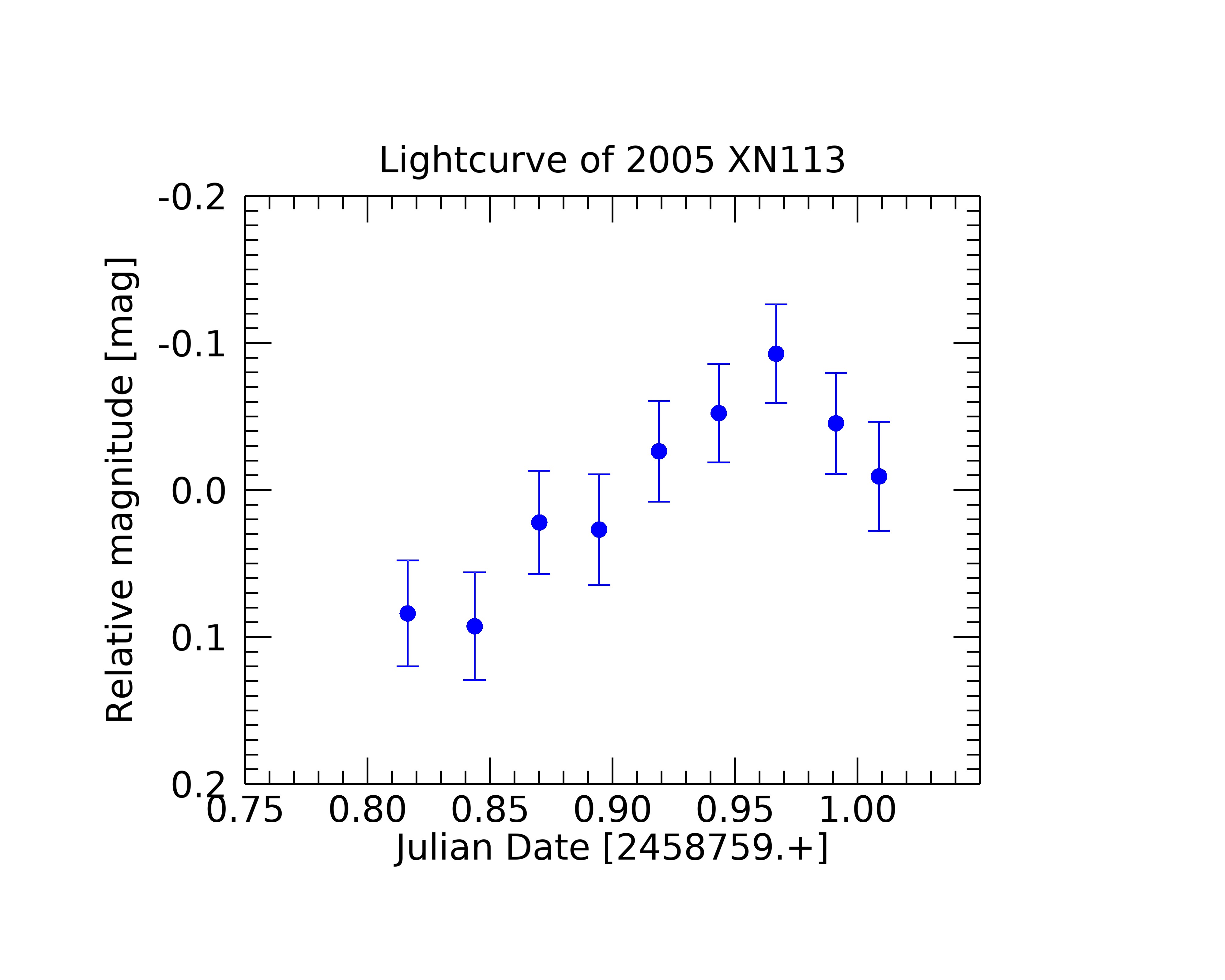} 
    \includegraphics[width=10cm,angle=0]{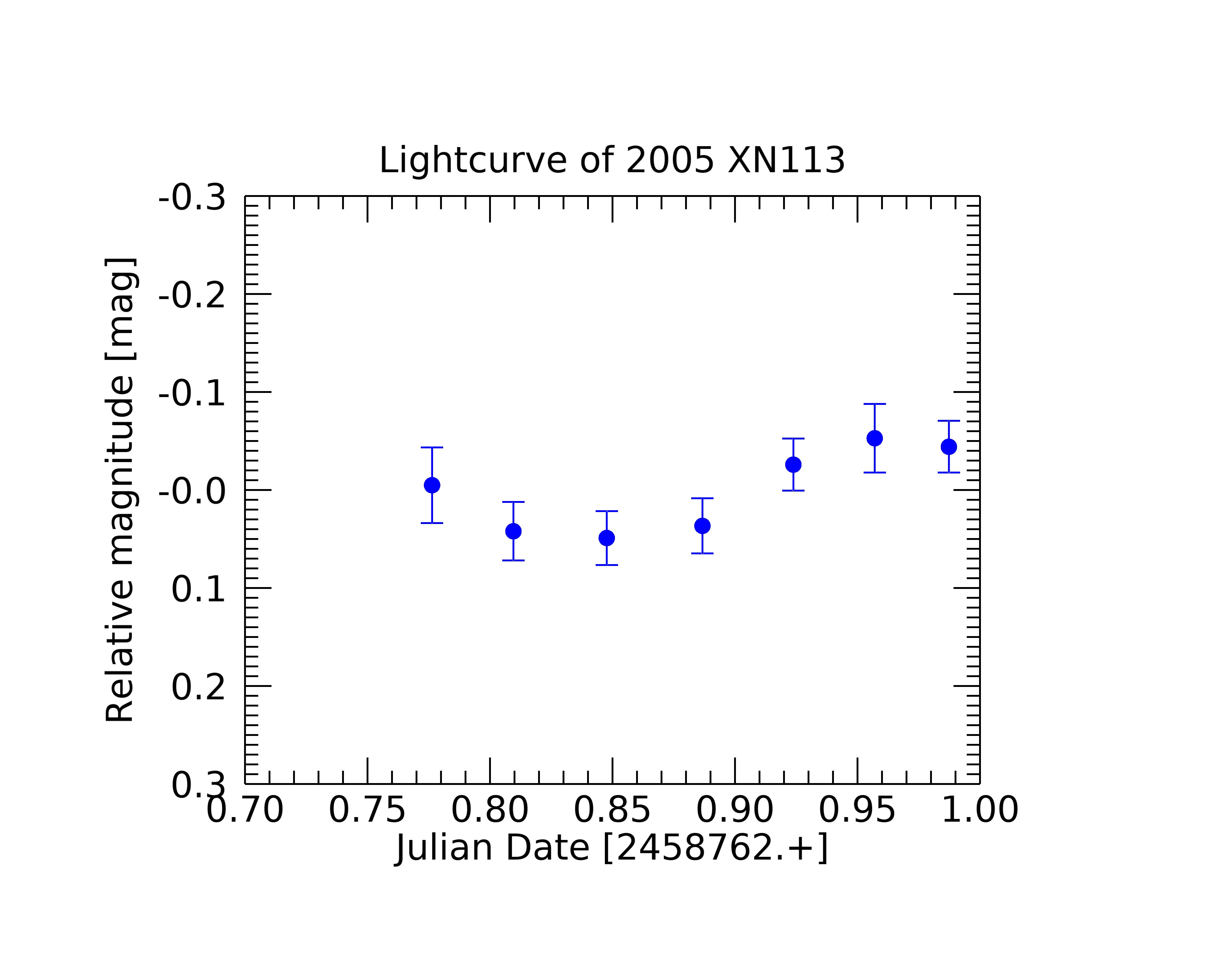} 
   \includegraphics[width=10cm,angle=0]{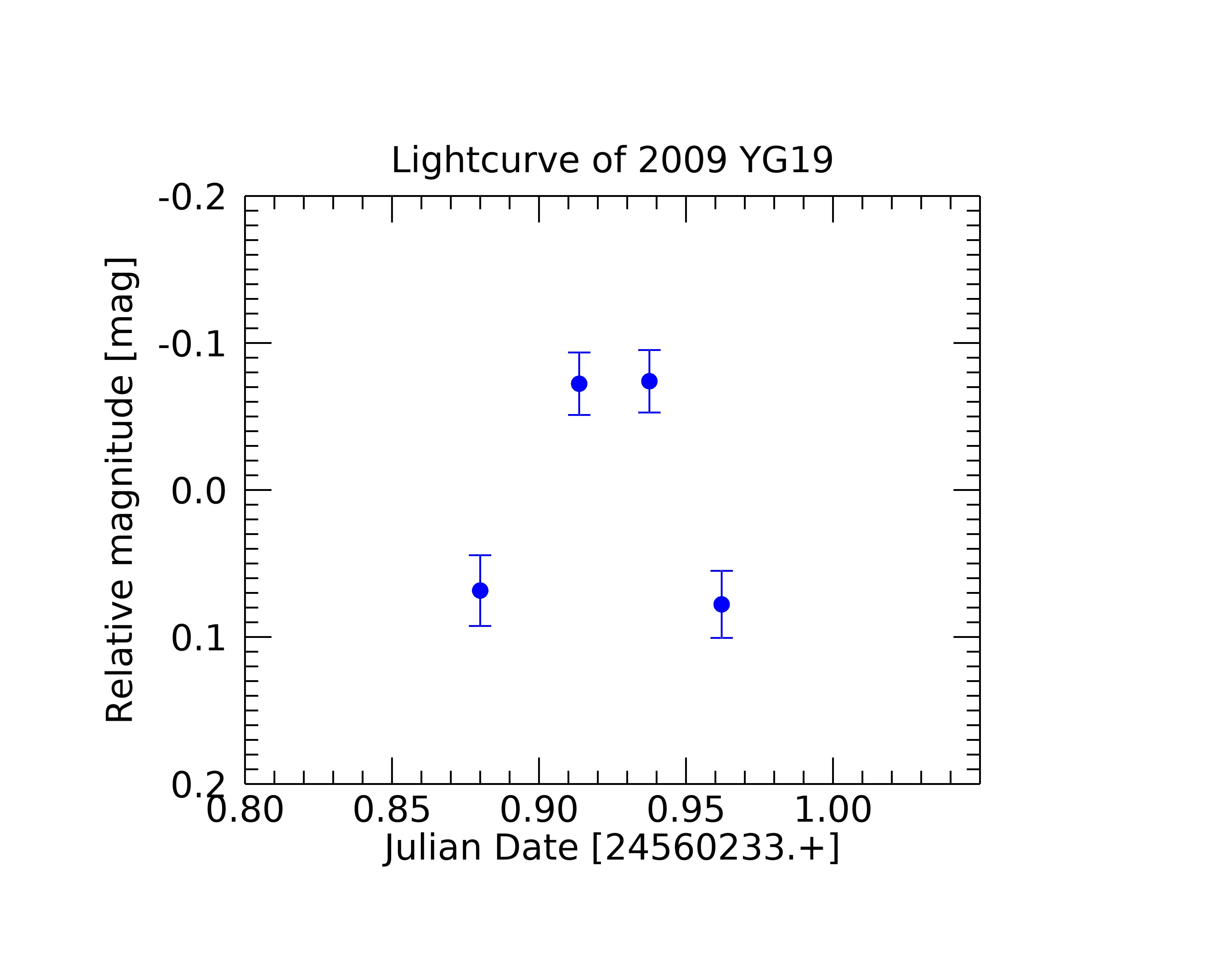}  
 \caption{Partial lightcurves of KBOs trapped in the 5:2 resonance }
\label{fig:LC521}
\end{figure}

      \begin{figure}
            \includegraphics[width=10cm,angle=0]{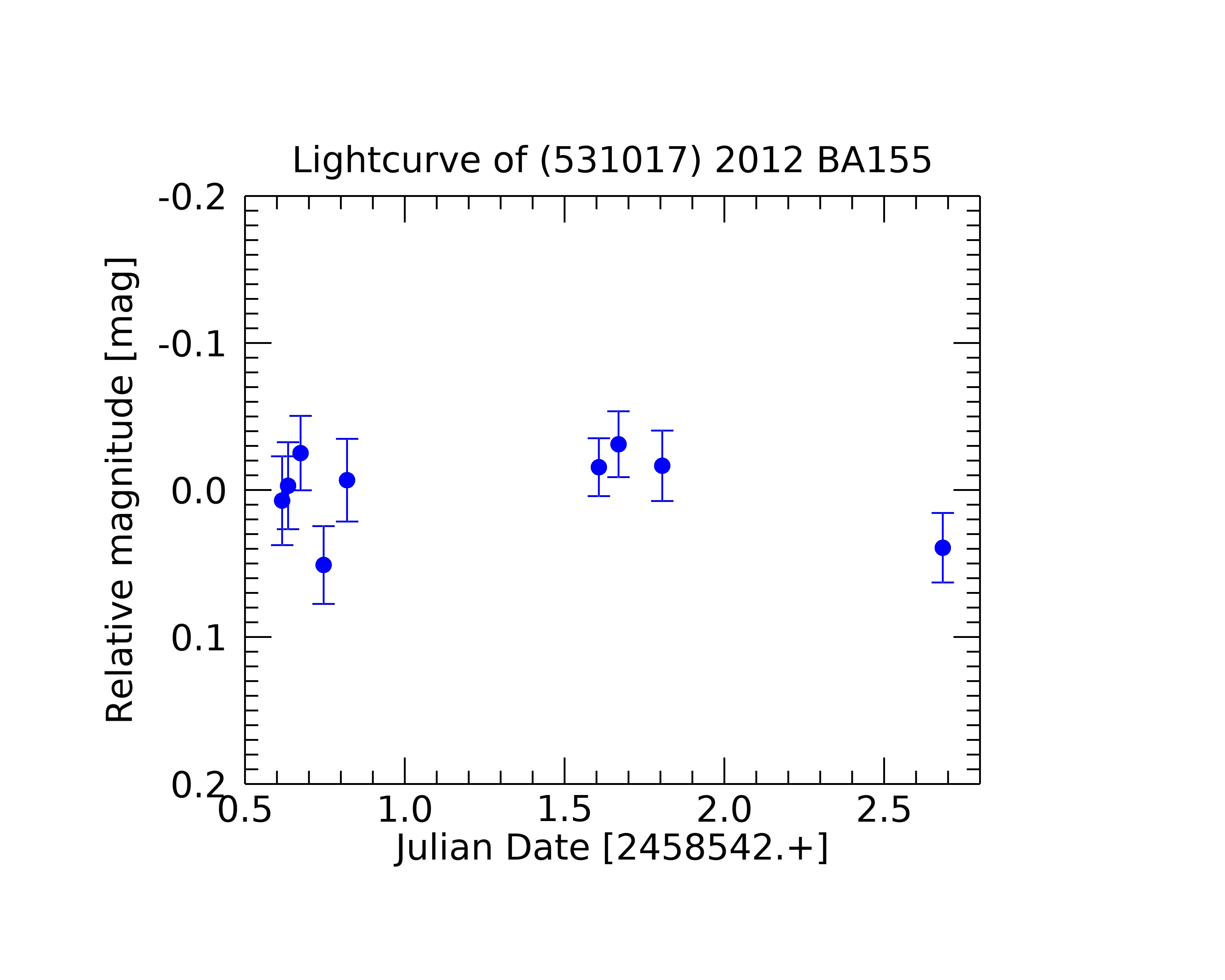}   
               \includegraphics[width=10cm,angle=0]{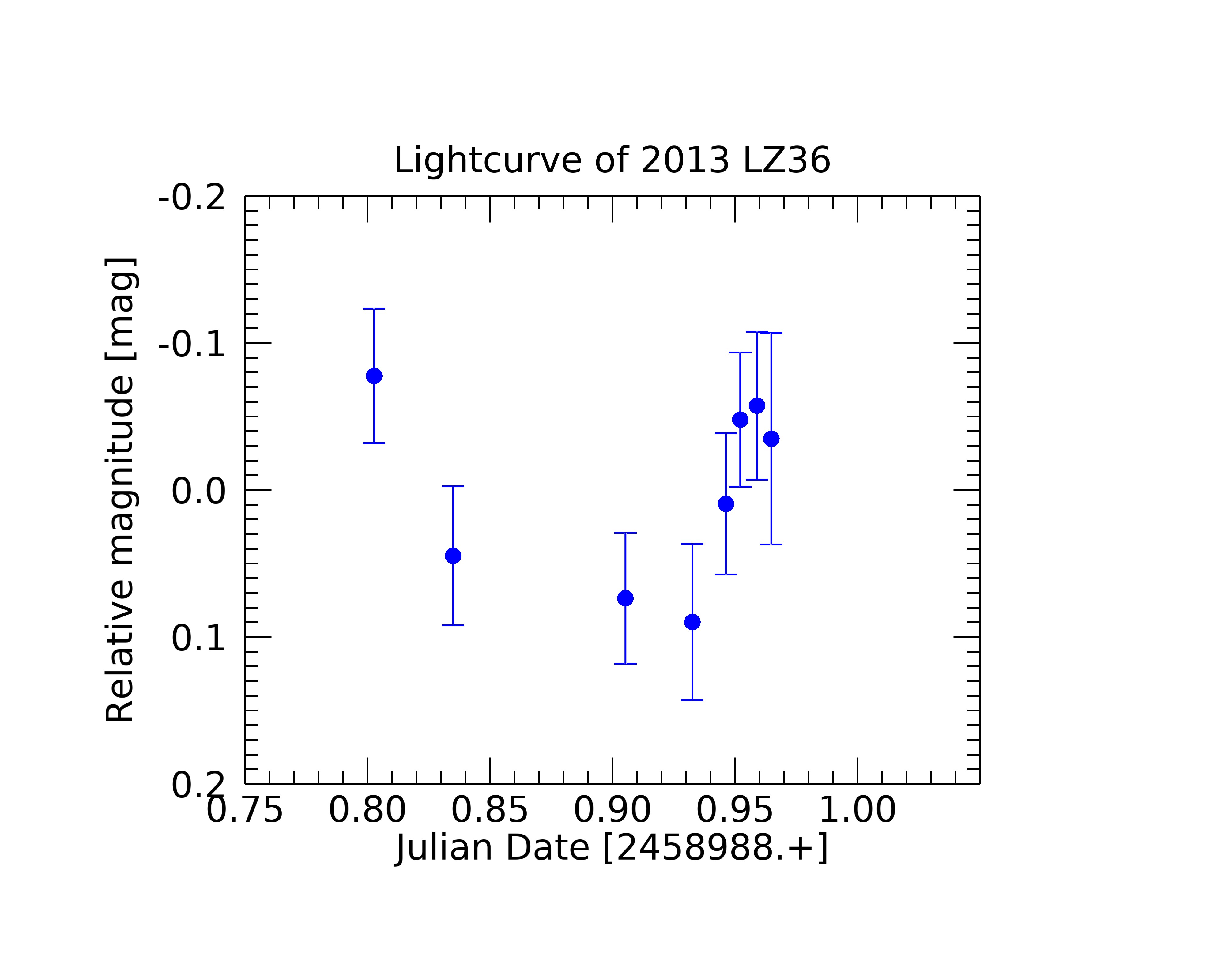}
  \includegraphics[width=10cm,angle=0]{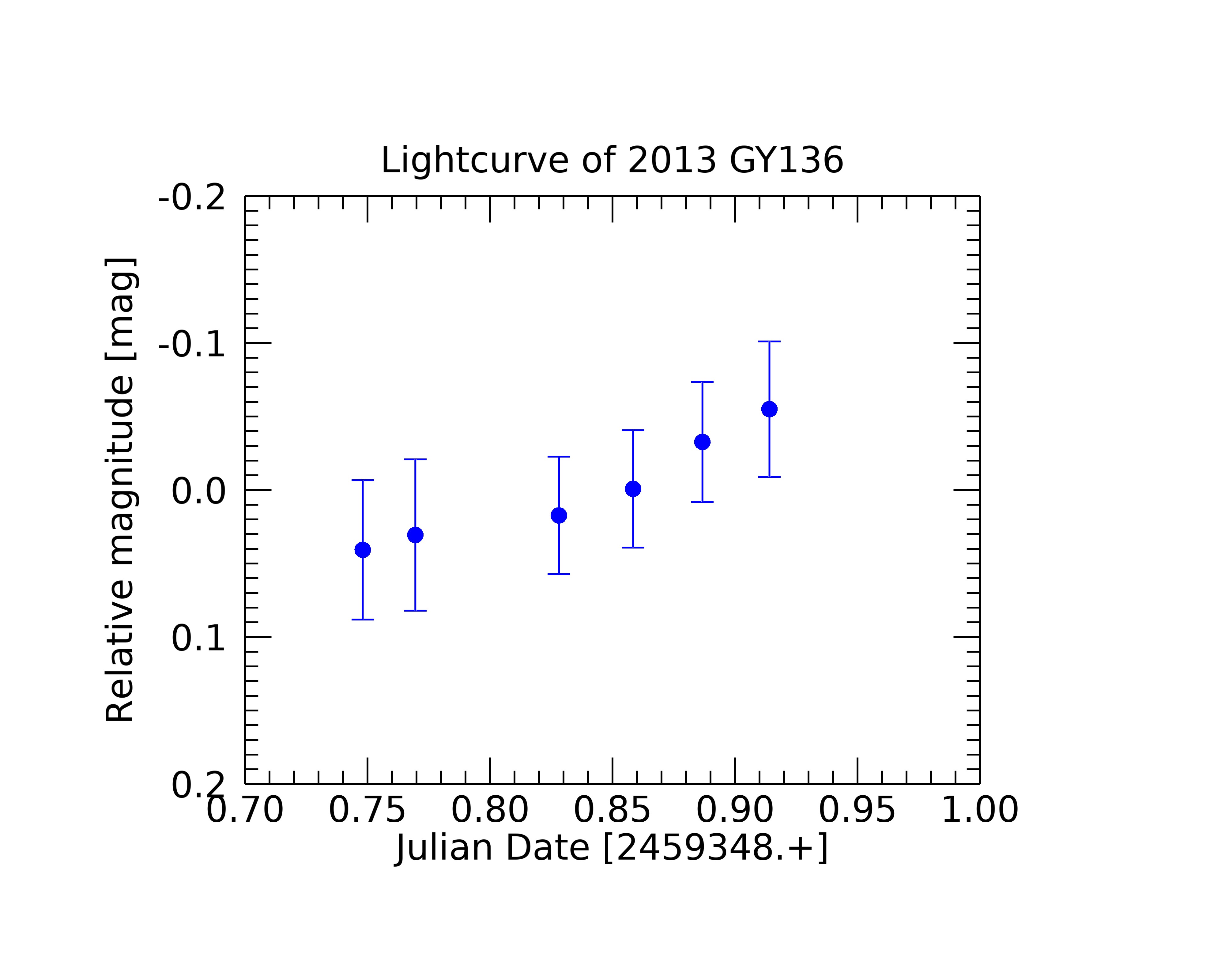}  
      \includegraphics[width=10cm,angle=0]{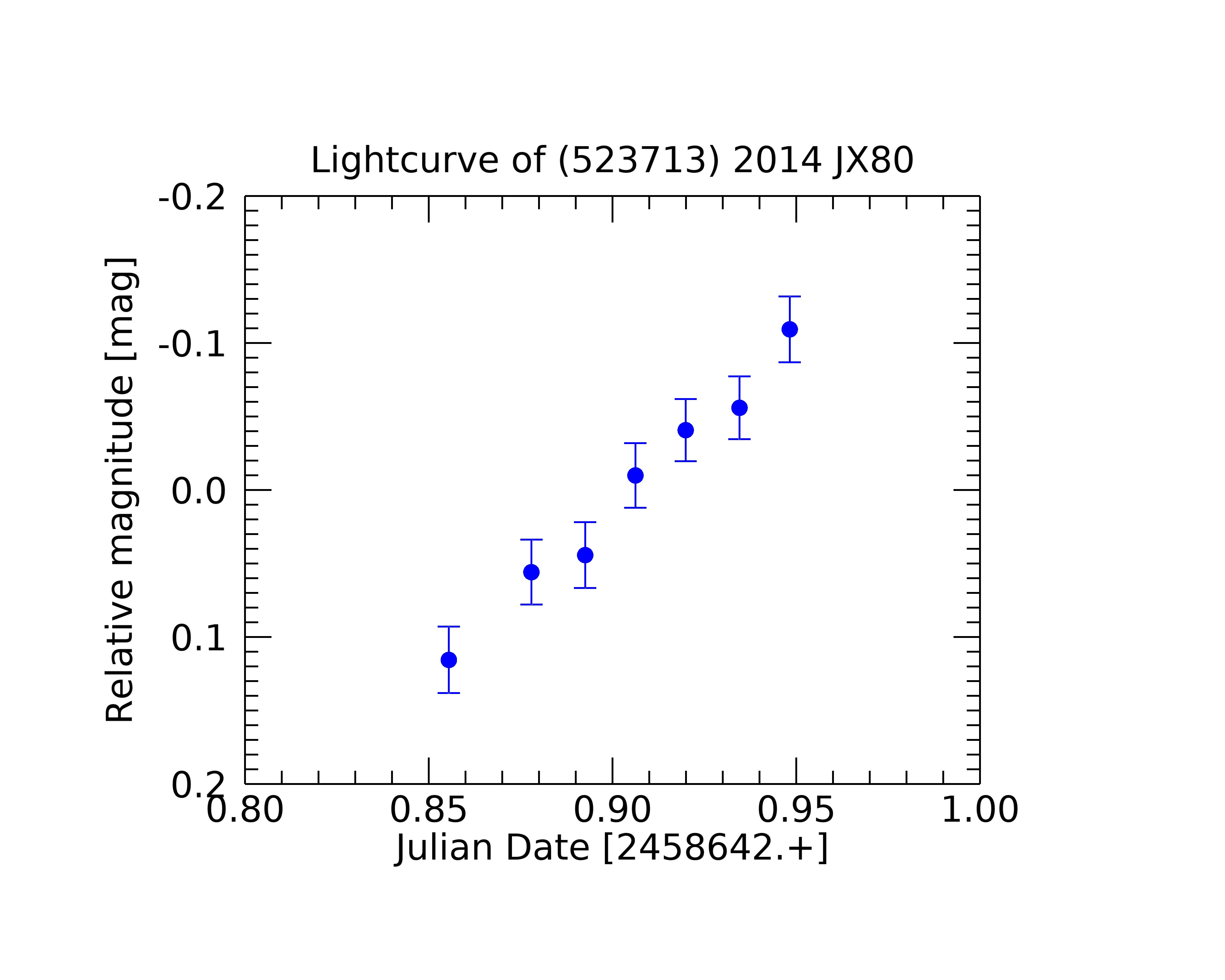} 
 \includegraphics[width=10cm,angle=0]{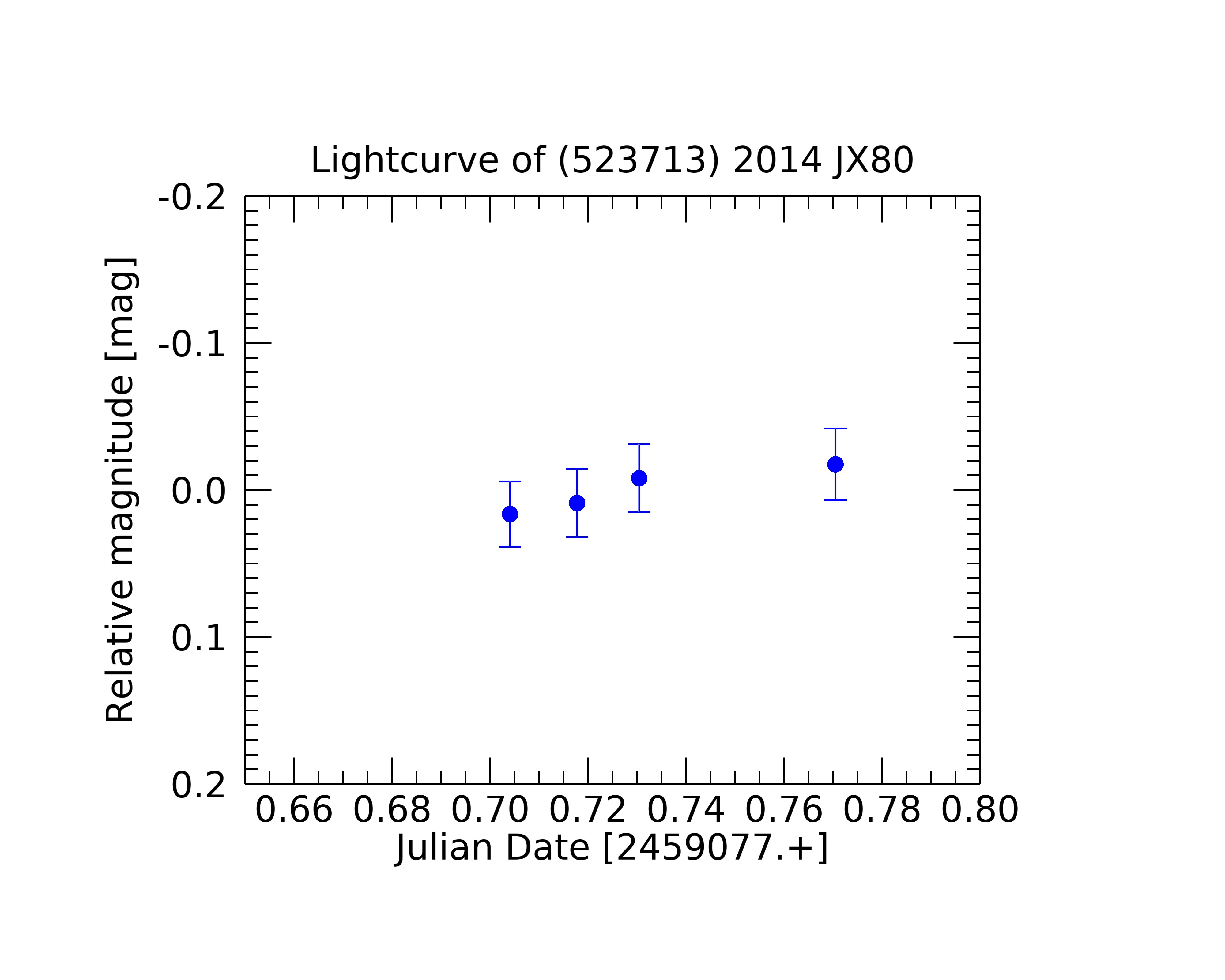} 
  \includegraphics[width=10cm,angle=0]{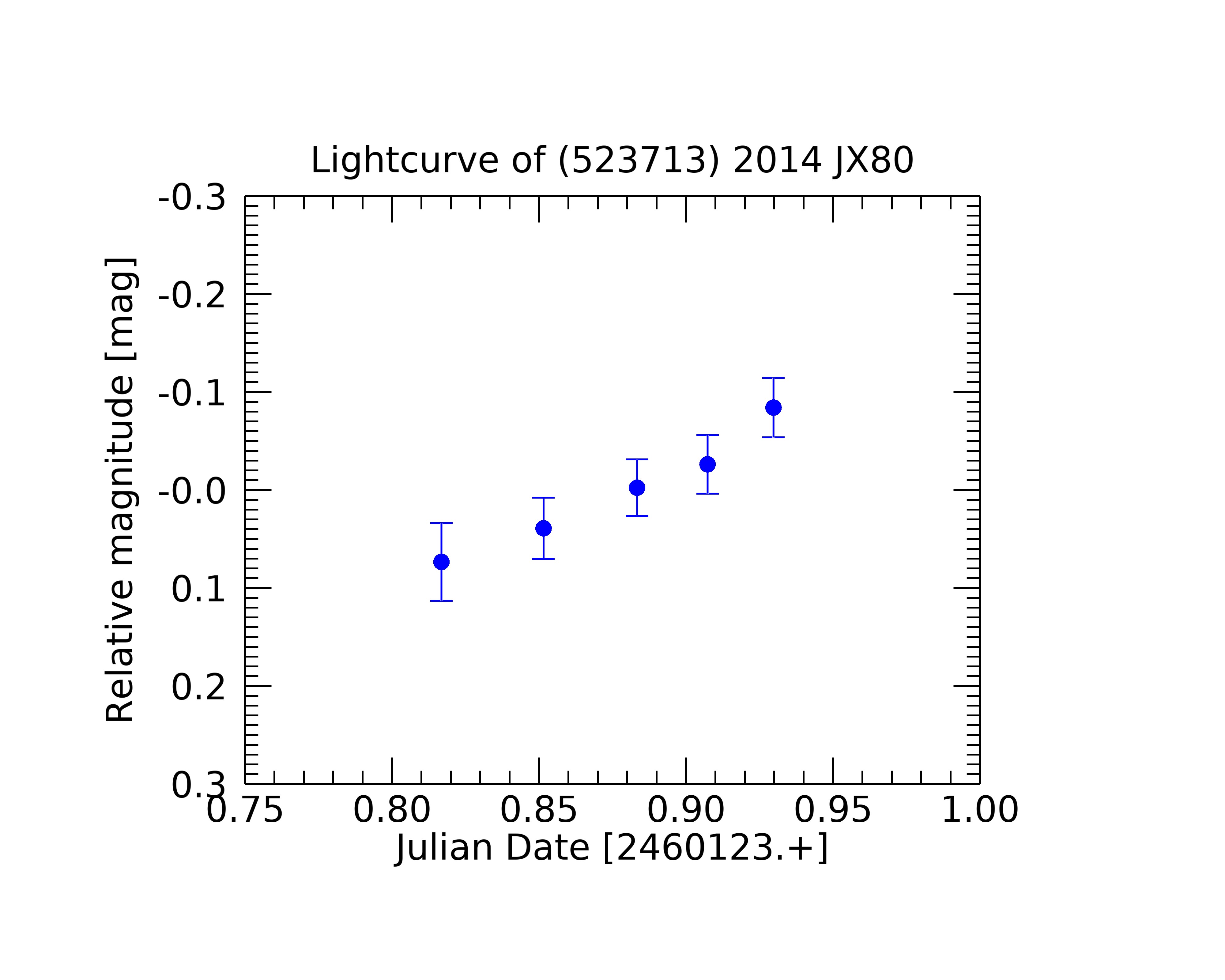} 
      \caption{ Continued }
\label{fig:LC522}
\end{figure}

      \begin{figure}
    \includegraphics[width=10cm,angle=0]{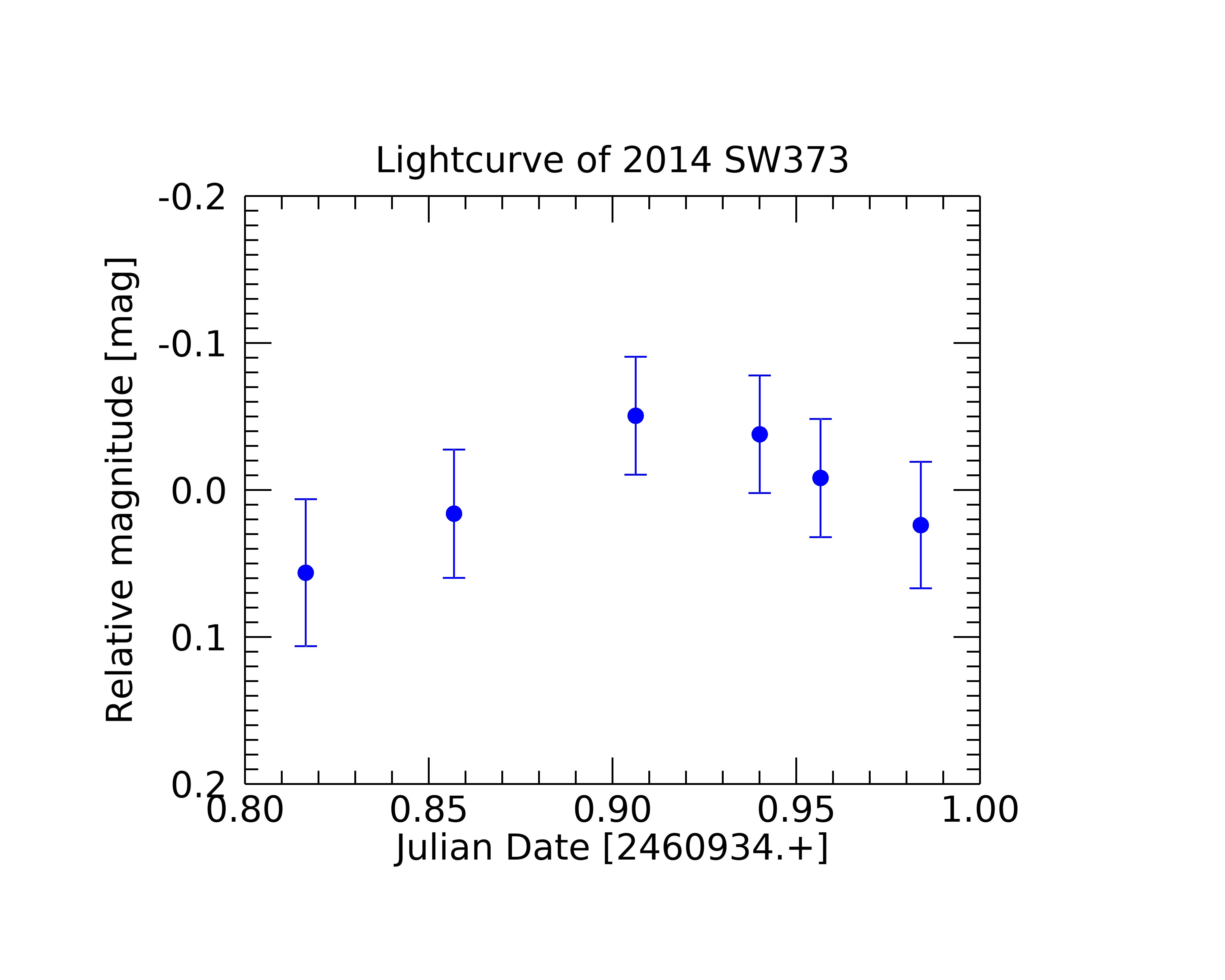} 
        \includegraphics[width=10cm,angle=0]{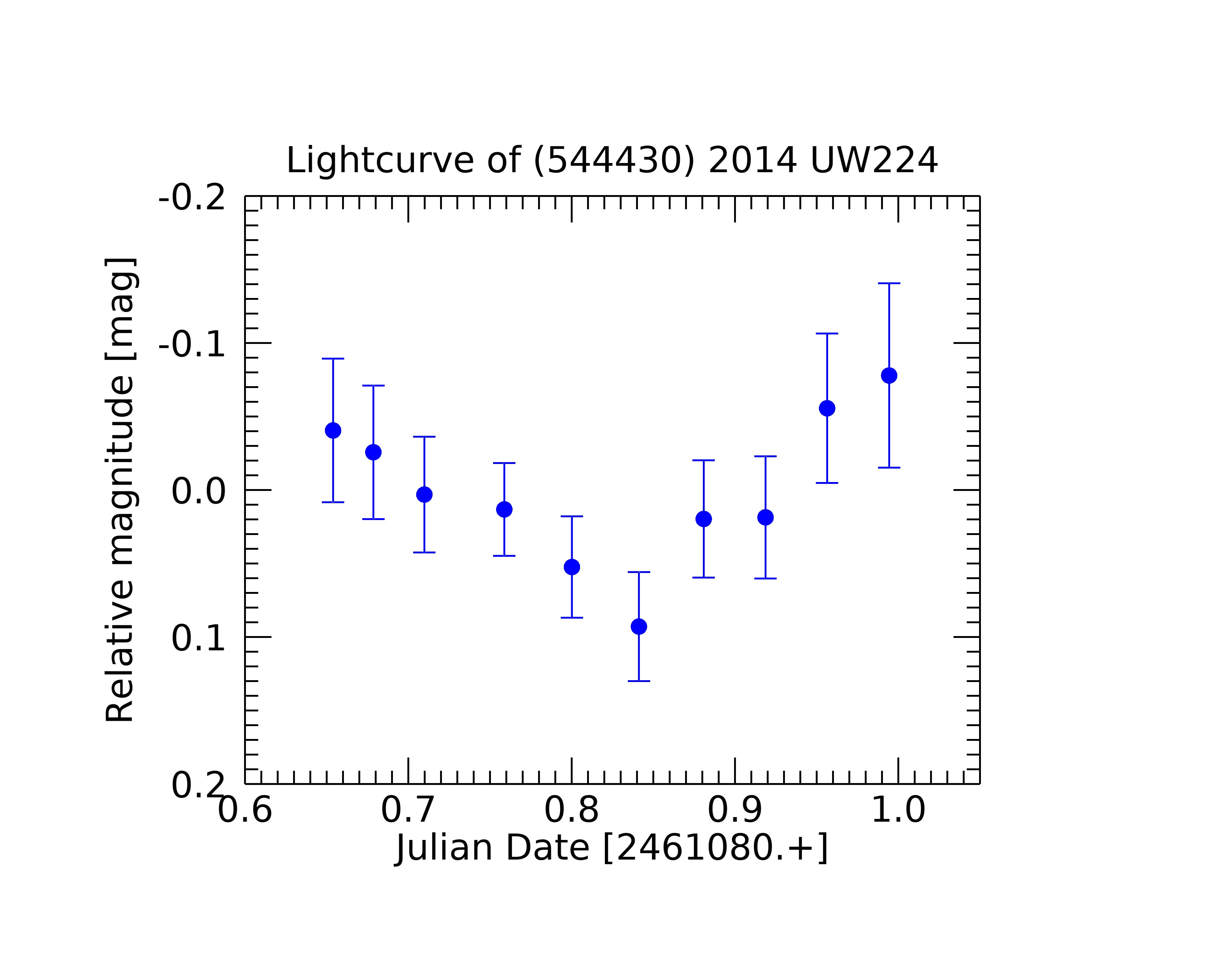} 
  \includegraphics[width=10cm,angle=0]{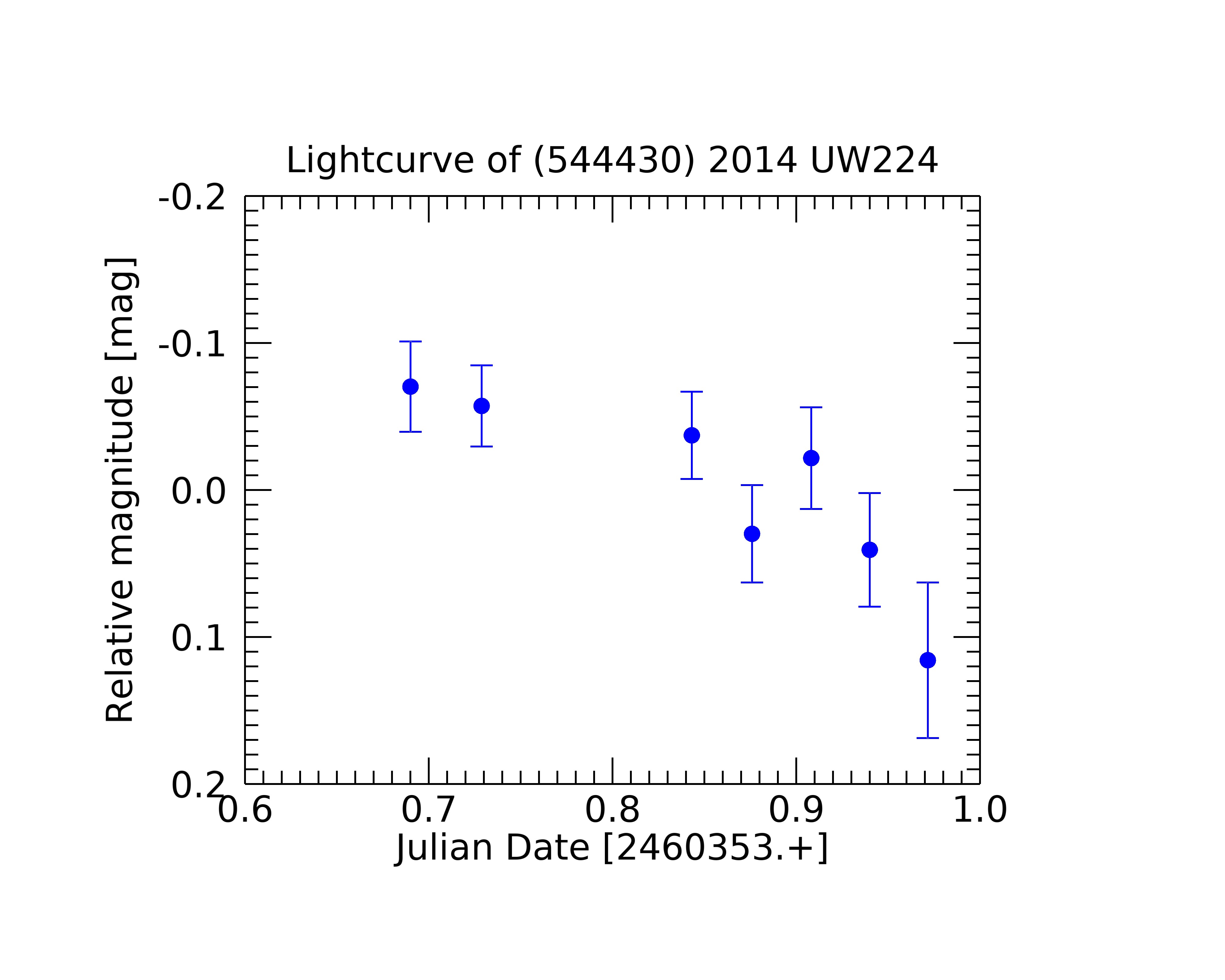} 
   \includegraphics[width=10cm,angle=0]{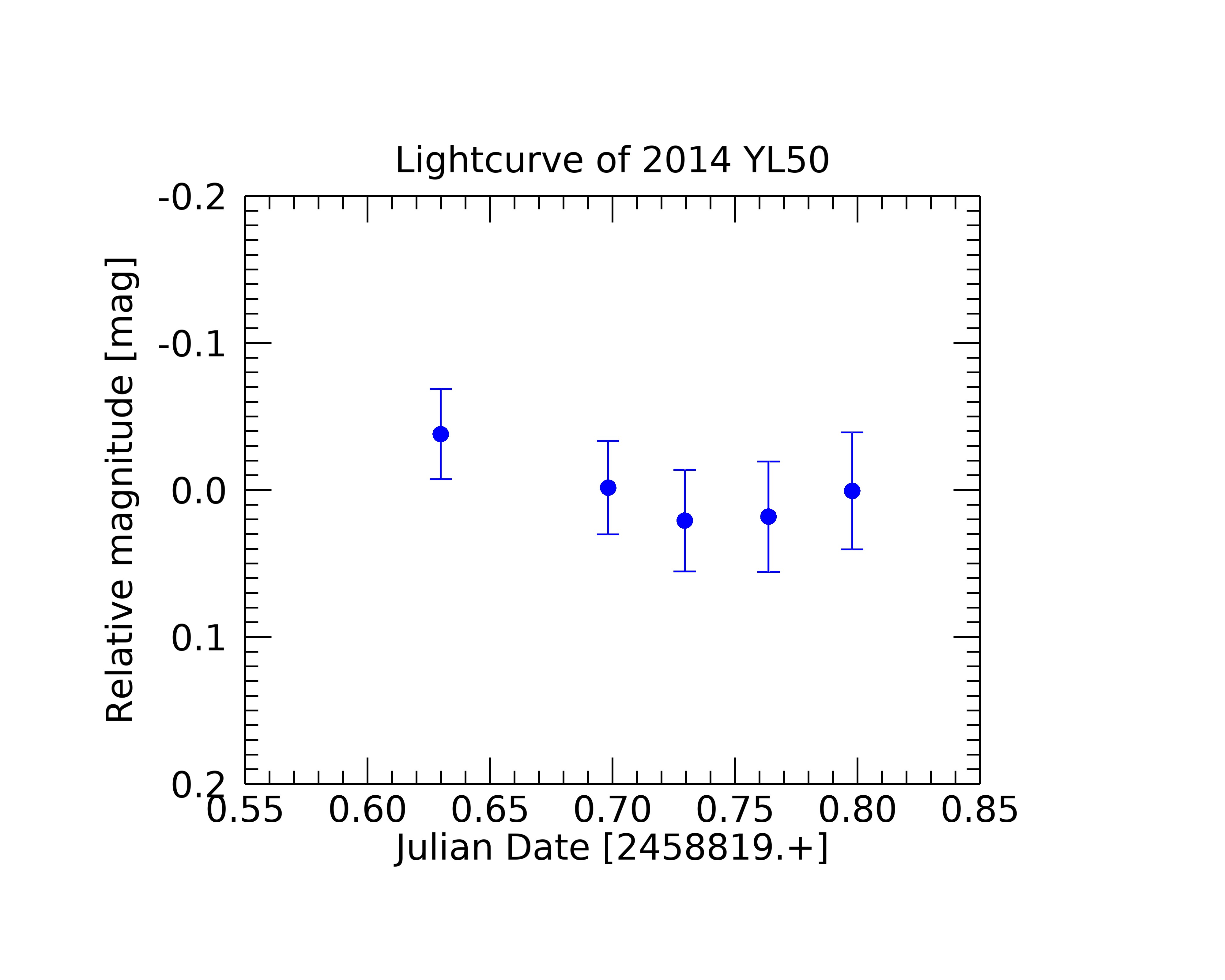} 
 \includegraphics[width=10cm,angle=0]{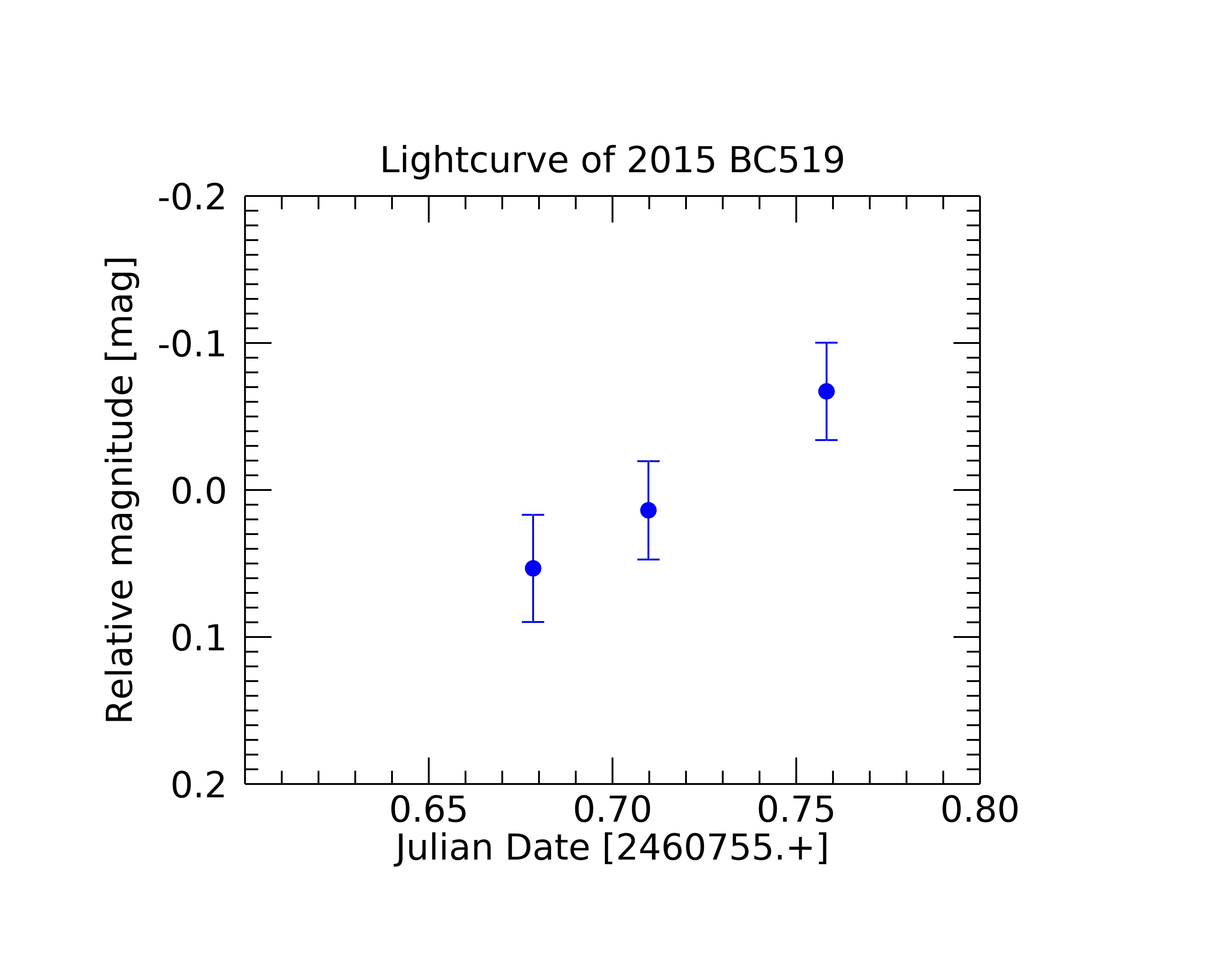}
         \includegraphics[width=10cm,angle=0]{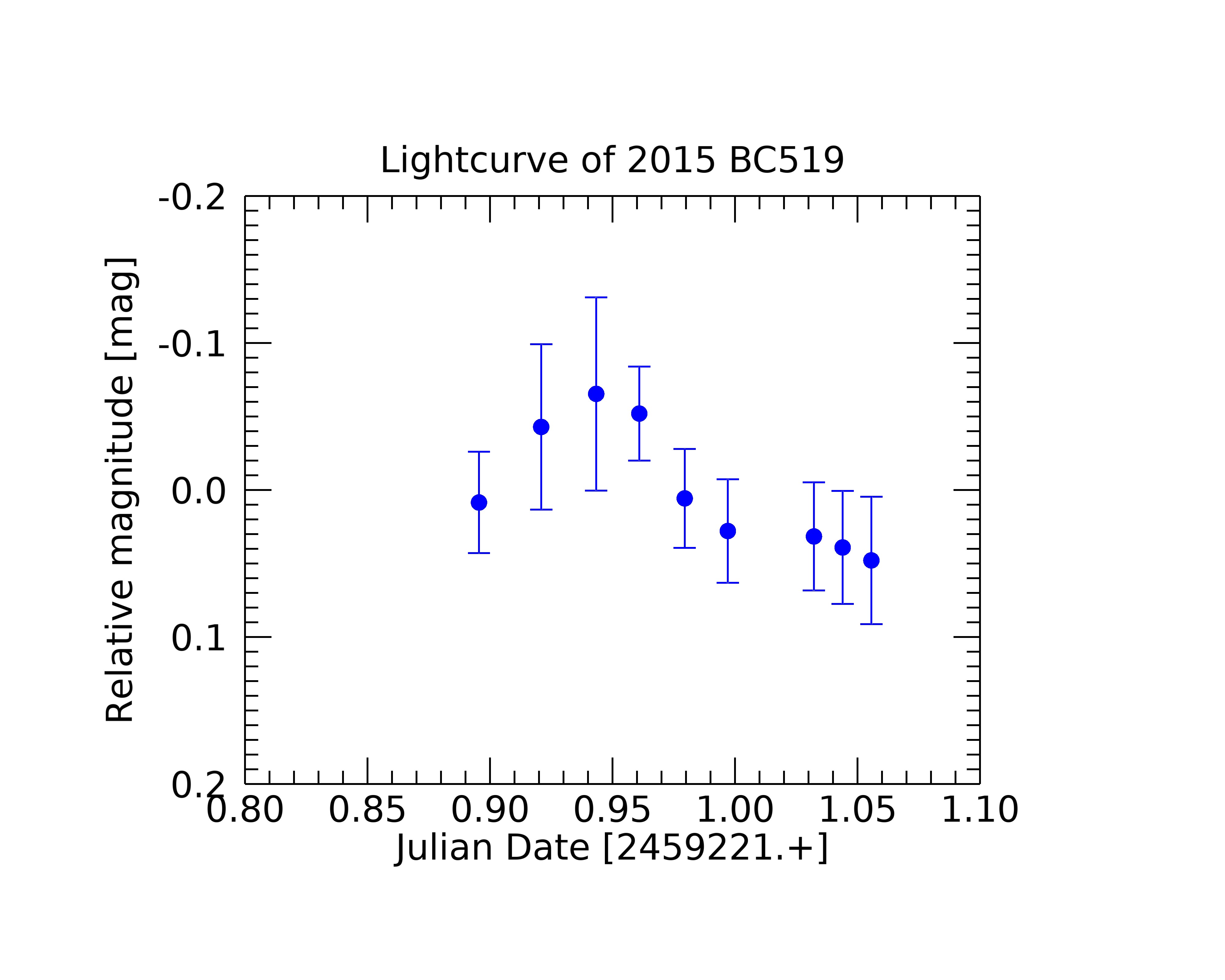}  
      \caption{ Continued}
\label{fig:LC523}
\end{figure}

      \begin{figure}
          \includegraphics[width=10cm,angle=0]{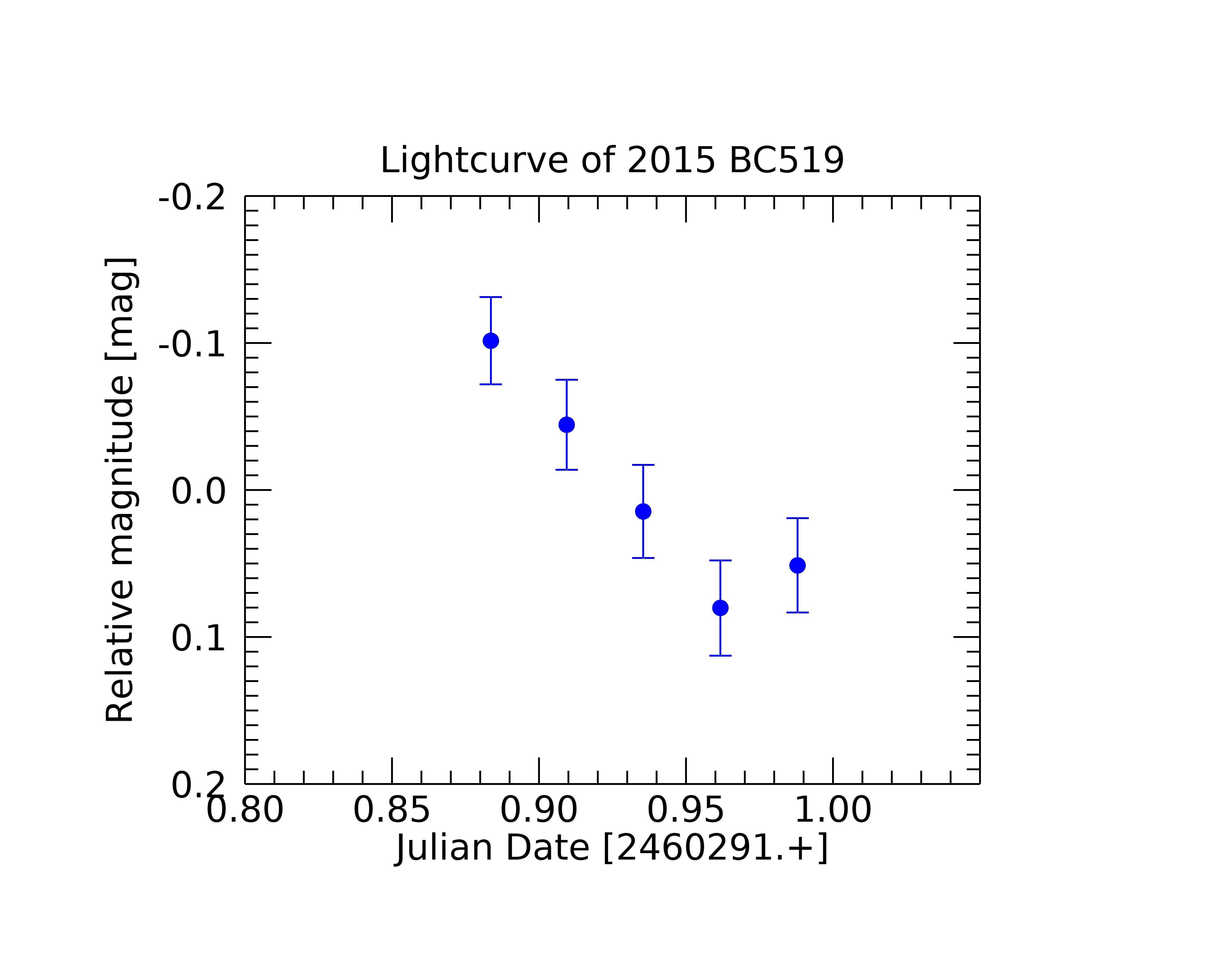} 
   \includegraphics[width=10cm,angle=0]{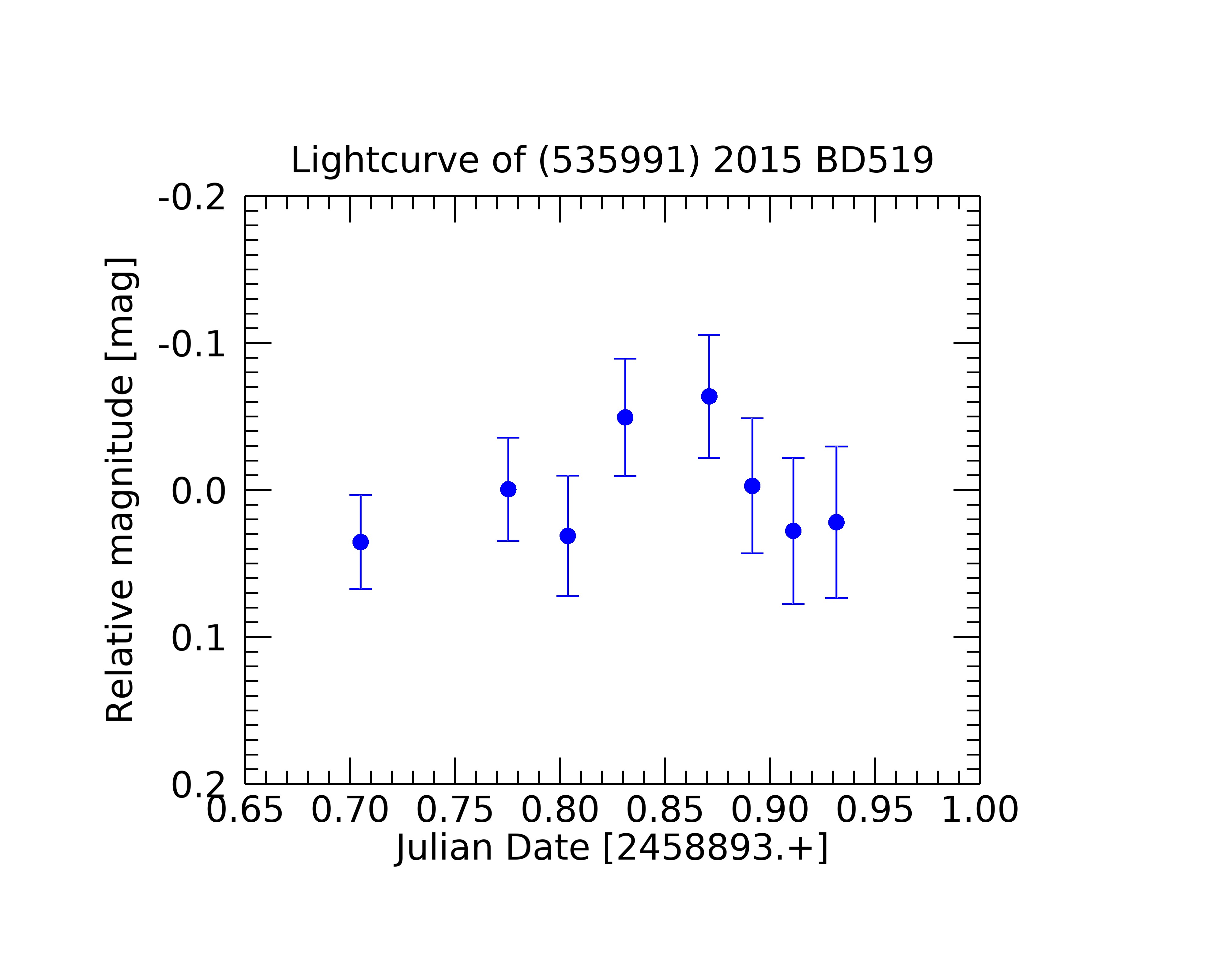} 
    \includegraphics[width=10cm,angle=0]{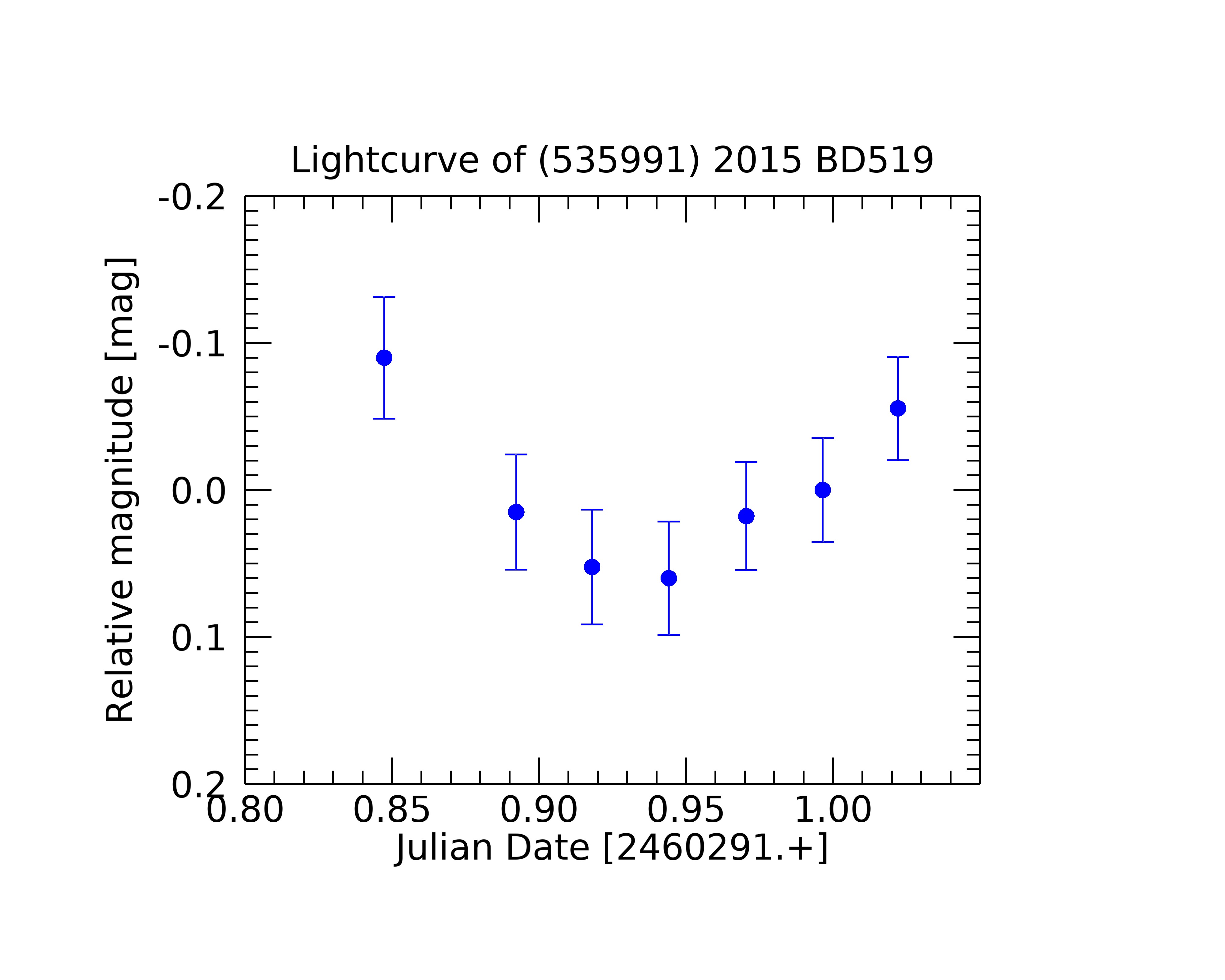}
 \includegraphics[width=10cm,angle=0]{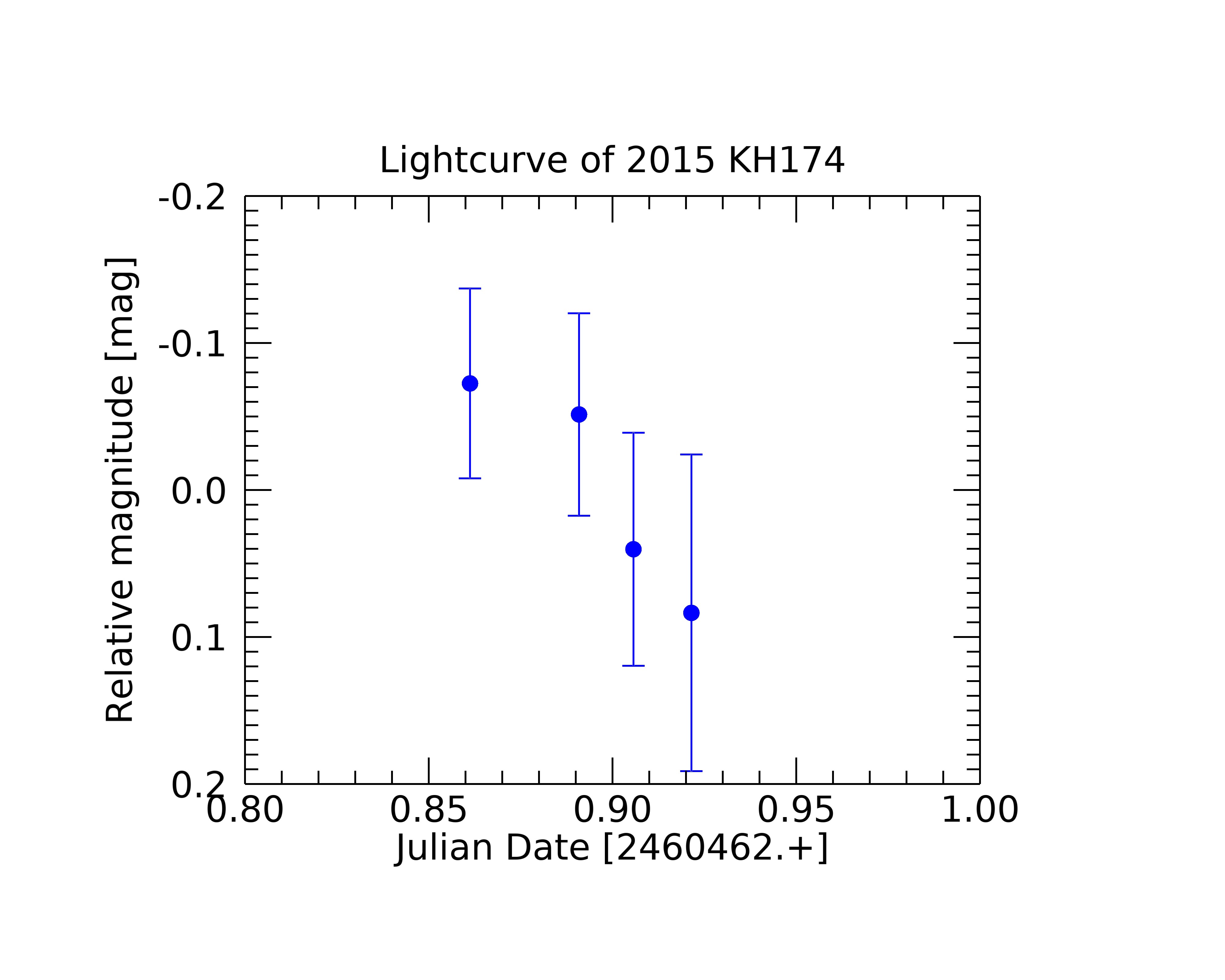} 
  \includegraphics[width=10cm,angle=0]{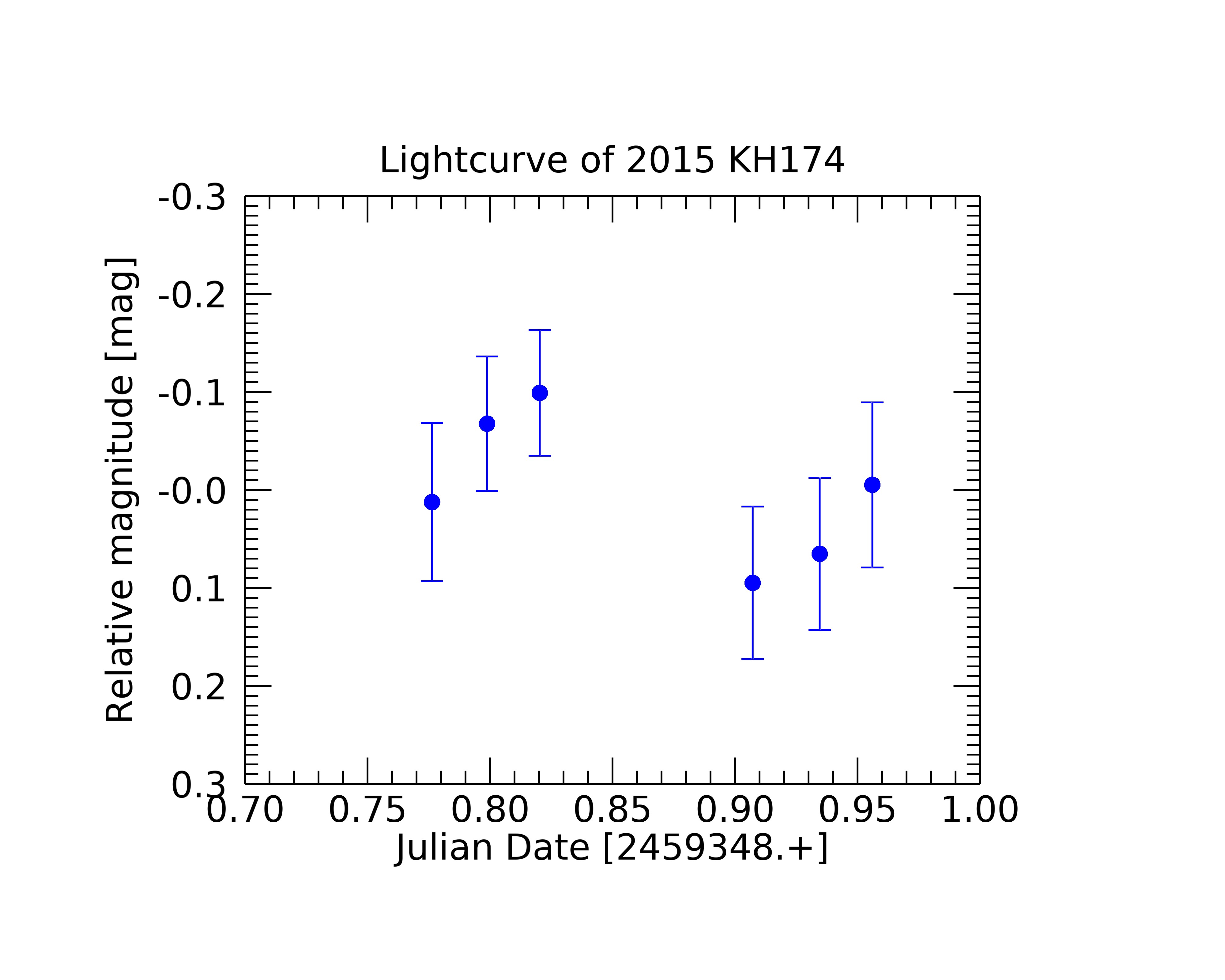} 
      \caption{ Continued}
\label{fig:LC524}
\end{figure}

\newpage
\clearpage
 \section{Appendix information}
\label{sec:appB}

 The photometry of all targets observed in this paper is available below. No light-time correction applied.  
 \startlongtable
\begin{deluxetable}{lccc}
 \tablecaption{\label{Tab:Summary_photo2}   }
\tablewidth{0pt}
\tablehead{
Object  & Julian Date & Relative Magnitude &  Error  \\
        &          &   [mag]           &  [mag]\\
}
\startdata 
(613490) 2006~RJ$_{103}$   &   &   &   \\
&	2460609.71423	&	0.06	&	0.03	\\
&	2460609.74464	&	-0.02	&	0.03	\\
&	2460609.78228	&	-0.07	&	0.03	\\
&	2460609.82085	&	-0.01	&	0.03	\\
&	2460609.84401	&	0.00	&	0.03	\\
&	2460609.88687	&	-0.05	&	0.03	\\
&	2460609.91764	&	-0.01	&	0.03	\\
&	2460609.95370	&	0.07	&	0.04	\\
&	2460609.97628	&	0.07	&	0.03	\\
&	2460610.00618	&	-0.02	&	0.03	\\
\hline 
(527604) 2007~VL$_{305}$ &   &   &   \\
&	2458817.78513	&	-0.07	&	0.03	\\
&	2458817.81067	&	0.00	&	0.03	\\
&	2458817.84061	&	0.06	&	0.03	\\
&	2458817.85079	&	0.02	&	0.04	\\
&	2458818.54024	&	0.00	&	0.04	\\
&	2458818.62529	&	-0.02	&	0.04	\\
&	2458818.67540	&	0.03	&	0.03	\\
&	2458818.74266	&	-0.03	&	0.03	\\
&	2458818.74691	&	-0.04	&	0.04	\\
&	2458818.80107	&	0.02	&	0.04	\\
&	2458818.82500	&	0.04	&	0.04	\\
&	2458819.55737	&	0.03	&	0.05	\\
&	2458819.72464	&	-0.03	&	0.04	\\
&	2458819.75892	&	-0.06	&	0.04	\\
&	2458819.80302	&	0.06	&	0.05	\\
\hline 
(666739) 2010~TS$_{191}$ &   &   &   \\
&	2459221.60422	&	0.07	&	0.04	\\
&	2459221.63101	&	0.03	&	0.06	\\
&	2459221.66707	&	-0.02	&	0.04	\\
&	2459221.80165	&	-0.07	&	0.05	\\
&	2459221.83554	&	-0.02	&	0.05	\\
\hline 
2010~TT$_{191}$ &   &   &   \\
&	2459221.64601	&	0.00	&	0.05	\\
&	2459221.68036	&	0.00	&	0.05	\\
&	2459221.74379	&	-0.03	&	0.06	\\
&	2459221.79437	&	0.03	&	0.06	\\
&	2459221.86220	&	-0.02	&	0.05	\\
&	2459221.88902	&	0.02	&	0.08	\\
&	2460353.66245	&	-0.06	&	0.06	\\
&	2460353.68291	&	0.01	&	0.06	\\
&	2460353.72179	&	0.03	&	0.06	\\
&	2460353.75942	&	0.03	&	0.07	\\
&	2460353.79742	&	0.00	&	0.09	\\
\hline 
(530664) 2011~SO$_{277}$ &   &   &   \\
&	2460586.68988	&	0.00	&	0.04	\\
&	2460586.69433	&	0.08	&	0.04	\\
&	2460586.73203	&	-0.02	&	0.04	\\
&	2460586.73649	&	0.02	&	0.05	\\
&	2460586.75698	&	0.05	&	0.04	\\
&	2460586.77712	&	0.00	&	0.03	\\
&	2460586.79274	&	0.03	&	0.03	\\
&	2460586.79719	&	-0.06	&	0.03	\\
&	2460586.81267	&	0.03	&	0.04	\\
&	2460586.81714	&	-0.02	&	0.03	\\
&	2460586.83275	&	-0.06	&	0.03	\\
&	2460586.83721	&	-0.07	&	0.03	\\
&	2460586.85275	&	0.00	&	0.03	\\
&	2460586.85722	&	0.03	&	0.04	\\
&	2460586.88078	&	-0.01	&	0.04	\\
\hline 
2012~UD$_{185}$ &   &   &   \\
&	2460609.70682	&	-0.01	&	0.04	\\
&	2460609.73724	&	0.00	&	0.04	\\
&	2460609.77473	&	0.02	&	0.04	\\
&	2460609.80533	&	-0.08	&	0.04	\\
&	2460609.83650	&	-0.01	&	0.04	\\
&	2460609.86695	&	0.05	&	0.04	\\
&	2460609.91032	&	0.00	&	0.04	\\
&	2460609.96119	&	0.04	&	0.04	\\
&	2460609.98380	&	0.00	&	0.04	\\
&	2460609.99879	&	0.01	&	0.04	\\
&	2460970.71956	&	-0.02	&	0.04	\\
&	2460970.75215	&	0.04	&	0.04	\\
&	2460970.78513	&	-0.02	&	0.04	\\
&	2460970.81761	&	-0.04	&	0.03	\\
&	2460970.84996	&	0.05	&	0.04	\\
&	2460970.88266	&	0.01	&	0.03	\\
&	2460970.96804	&	0.01	&	0.03	\\
&	2460970.98393	&	-0.03	&	0.03	\\
&	2460970.99975	&	-0.01	&	0.03	\\
&	2460971.00687	&	0.02	&	0.03	\\
\hline 
(427581) 2003~QB$_{92}$ &   &   &   \\
&	2460556.72194	&	-0.08	&	0.13	\\
&	2460556.74677	&	-0.06	&	0.13	\\
&	2460556.77262	&	0.01	&	0.14	\\
&	2460556.80802	&	-0.01	&	0.11	\\
&	2460556.87857	&	0.00	&	0.12	\\
&	2460556.91282	&	0.14	&	0.12	\\
\hline 
(308460) 2005~SC$_{278}$ &   &   &   \\
&	2460236.65163	&	-0.16	&	0.05	\\
&	2460236.65818	&	-0.06	&	0.04	\\
&	2460236.68216	&	0.00	&	0.04	\\
&	2460236.69474	&	0.00	&	0.04	\\
&	2460236.70724	&	-0.03	&	0.04	\\
&	2460236.72599	&	0.07	&	0.04	\\
&	2460236.75566	&	0.02	&	0.04	\\
&	2460236.79169	&	0.03	&	0.04	\\
&	2460236.82545	&	0.07	&	0.04	\\
&	2460236.84823	&	0.05	&	0.04	\\
&	2460236.88088	&	0.09	&	0.04	\\
&	2460236.90355	&	0.07	&	0.04	\\
&	2460236.92616	&	0.01	&	0.04	\\
&	2460236.94846	&	0.04	&	0.04	\\
&	2460236.96413	&	-0.12	&	0.04	\\
&	2460236.97972	&	-0.08	&	0.04	\\
&	2460588.70117	&	-0.04	&	0.05	\\
&	2460588.72799	&	0.01	&	0.05	\\
&	2460588.77104	&	0.00	&	0.04	\\
&	2460588.85758	&	0.03	&	0.04	\\
&	2460609.66099	&	0.04	&	0.04	\\
&	2460609.67654	&	0.04	&	0.04	\\
&	2460609.69173	&	0.09	&	0.04	\\
&	2460609.81316	&	0.01	&	0.04	\\
&	2460609.85170	&	-0.04	&	0.04	\\
&	2460609.89472	&	-0.10	&	0.04	\\
&	2460609.92516	&	-0.05	&	0.05	\\
&	2460609.96866	&	0.01	&	0.05	\\
&	2460637.60288	&	-0.01	&	0.04	\\
&	2460637.63215	&	-0.02	&	0.04	\\
&	2460637.66959	&	-0.02	&	0.03	\\
&	2460637.70720	&	-0.06	&	0.03	\\
&	2460637.74331	&	-0.03	&	0.03	\\
&	2460637.78585	&	0.03	&	0.03	\\
&	2460637.82194	&	0.06	&	0.04	\\
&	2460637.85983	&	0.03	&	0.04	\\
&	2460637.89437	&	0.02	&	0.05	\\
\hline 
2014~UC$_{225}$ &   &   &   \\
&	2460970.71158	&	0.04	&	0.05	\\
&	2460970.74391	&	-0.04	&	0.07	\\
&	2460970.77619	&	0.02	&	0.06	\\
&	2460970.80902	&	0.01	&	0.05	\\
&	2460970.84178	&	0.00	&	0.07	\\
&	2460970.87469	&	-0.03	&	0.04	\\
&	2460970.90672	&	0.04	&	0.06	\\
&	2460970.96029	&	0.05	&	0.05	\\
&	2460970.97594	&	-0.06	&	0.05	\\
&	2460970.99173	&	-0.04	&	0.05	\\
\hline 
(523955) 1998~UU$_{43}$ &   &   &   \\
&	2461080.64553	&	-0.06	&	0.05	\\
&	2461080.67029	&	-0.03	&	0.05	\\
&	2461080.70144	&	-0.02	&	0.05	\\
&	2461080.75045	&	0.03	&	0.05	\\
&	2461080.82495	&	0.08	&	0.07	\\
\hline 
(143685) 2003~SS$_{317}$ &   &   &   \\
&	2460291.70641	&	-0.02	&	0.04	\\
&	2460291.73585	&	0.01	&	0.04	\\
&	2460291.76514	&	-0.03	&	0.04	\\
&	2460291.82064	&	0.00	&	0.04	\\
&	2460291.86539	&	0.04	&	0.04	\\
&	2460637.62491	&	-0.02	&	0.04	\\
&	2460637.65439	&	0.02	&	0.04	\\
&	2460637.69138	&	-0.02	&	0.03	\\
&	2460637.72796	&	-0.01	&	0.03	\\
&	2460637.76431	&	0.01	&	0.04	\\
&	2460637.80040	&	0.00	&	0.04	\\
&	2460637.83860	&	0.03	&	0.04	\\
&	2460637.87397	&	0.00	&	0.04	\\
&	2460637.90957	&	0.00	&	0.05	\\
\hline 
2013~RH$_{109}$ &   &   &   \\
&	2460934.70087	&	-0.13	&	0.06	\\
&	2460934.72503	&	0.02	&	0.06	\\
&	2460934.83941	&	0.04	&	0.06	\\
&	2460934.92226	&	0.07	&	0.09	\\
&	2460939.66299	&	-0.02	&	0.07	\\
&	2460939.68856	&	-0.01	&	0.07	\\
&	2460939.71327	&	0.03	&	0.07	\\
&	2460939.75686	&	-0.06	&	0.06	\\
&	2460939.82518	&	0.02	&	0.06	\\
&	2460939.87497	&	0.05	&	0.07	\\
\hline 
2013~RQ$_{109}$ &   &   &   \\
&	2460588.76302	&	0.05	&	0.07	\\
&	2460588.80544	&	-0.03	&	0.06	\\
&	2460588.84941	&	0.03	&	0.06	\\
&	2460588.89223	&	-0.06	&	0.05	\\
&	2460588.92573	&	0.01	&	0.06	\\
\hline 
2013~RQ$_{157}$ &   &   &   \\
&	2460556.66653	&	0.10	&	0.09	\\
&	2460556.68312	&	0.00	&	0.08	\\
&	2460556.70069	&	0.03	&	0.08	\\
&	2460556.73056	&	-0.04	&	0.07	\\
&	2460556.75524	&	-0.04	&	0.07	\\
&	2460556.79165	&	-0.06	&	0.08	\\
&	2460556.82631	&	-0.07	&	0.07	\\
&	2460556.86147	&	0.10	&	0.09	\\
\hline 
2013~RW$_{124}$ &   &   &   \\
&	2460556.81660	&	-0.04	&	0.08	\\
&	2460556.85125	&	-0.01	&	0.08	\\
&	2460556.88704	&	0.11	&	0.08	\\
&	2460556.92144	&	0.04	&	0.08	\\
&	2460556.95749	&	-0.02	&	0.07	\\
&	2460556.98158	&	-0.09	&	0.07	\\
&	2460588.71962	&	0.10	&	0.09	\\
&	2460588.75387	&	-0.02	&	0.07	\\
&	2460588.79668	&	-0.03	&	0.07	\\
&	2460588.91679	&	-0.05	&	0.06	\\
\hline 
(533211) 2014~DU$_{143}$ &   &   &   \\
&	2460785.78309	&	0.07	&	0.05	\\
&	2460785.82425	&	0.02	&	0.05	\\
&	2460785.86414	&	0.03	&	0.05	\\
&	2460785.90417	&	0.02	&	0.06	\\
&	2460785.93371	&	-0.08	&	0.06	\\
&	2460785.95621	&	-0.05	&	0.07	\\
\hline 
(535019) 2014~WE$_{509}$ &   &   &   \\
&	2460438.72179	&	0.00	&	0.04	\\
&	2460438.75521	&	-0.05	&	0.04	\\
&	2460438.78490	&	-0.04	&	0.04	\\
&	2460438.81491	&	0.03	&	0.05	\\
&	2460438.84760	&	0.02	&	0.05	\\
&	2460438.87185	&	0.05	&	0.07	\\
&	2460785.75133	&	-0.05	&	0.04	\\
&	2460785.79015	&	0.00	&	0.04	\\
&	2460785.83143	&	0.01	&	0.04	\\
&	2460785.87130	&	0.04	&	0.06	\\
\hline 
(559178) 2015~BM$_{518}$ &   &   &   \\
&	2461124.83601	&	-0.02	&	0.03	\\
&	2461124.86283	&	0.00	&	0.03	\\
&	2461124.90094	&	0.00	&	0.03	\\
&	2461124.93840	&	0.02	&	0.04	\\
\hline 
2016~SJ$_{57}$ &   &   &   \\
&	2460291.71503	&	0.00	&	0.04	\\
&	2460291.74460	&	0.03	&	0.04	\\
&	2460291.78376	&	-0.02	&	0.04	\\
&	2460291.81162	&	-0.01	&	0.04	\\
\hline 
2013~RM$_{109}$ &   &   &   \\
&	2460556.67491	&	-0.03	&	0.07	\\
&	2460556.69122	&	0.07	&	0.07	\\
&	2460556.71386	&	0.02	&	0.07	\\
&	2460556.73859	&	0.03	&	0.07	\\
&	2460556.76356	&	0.00	&	0.06	\\
&	2460556.79950	&	-0.04	&	0.06	\\
&	2460556.83443	&	0.02	&	0.06	\\
&	2460556.90419	&	0.01	&	0.07	\\
&	2460556.93980	&	-0.08	&	0.07	\\
\hline 
(131696) 2001~XT$_{254}$ &   &   &   \\
&	2461080.76718	&	-0.07	&	0.05	\\
&	2461080.80849	&	-0.08	&	0.04	\\
&	2461080.84942	&	0.12	&	0.06	\\
&	2461080.88893	&	0.00	&	0.06	\\
&	2461080.92667	&	0.05	&	0.06	\\
&	2461080.96443	&	0.07	&	0.08	\\
&	2461081.00257	&	-0.08	&	0.07	\\
\hline 
(183964) 2004~DJ$_{71}$ &   &   &   \\
&	2461124.67937	&	0.01	&	0.05	\\
&	2461124.75721	&	0.00	&	0.04	\\
&	2461124.80178	&	-0.01	&	0.04	\\
&	2461124.83028	&	0.02	&	0.04	\\
&	2461124.86911	&	-0.03	&	0.03	\\
&	2461124.90716	&	0.02	&	0.04	\\
\hline 
(523624) 2008~CT$_{190}$ &   &   &   \\
&	2461124.65075	&	0.00	&	0.02	\\
&	2461124.67322	&	0.03	&	0.02	\\
&	2461124.69700	&	-0.01	&	0.02	\\
&	2461124.72484	&	-0.03	&	0.02	\\
&	2461124.75159	&	0.01	&	0.02	\\
&	2461124.78741	&	0.00	&	0.02	\\
&	2461124.88358	&	0.02	&	0.02	\\
&	2461124.92143	&	0.00	&	0.02	\\
&	2461124.95913	&	-0.01	&	0.02	\\
\hline 
(495297) 2013~TJ$_{159}$ &   &   &   \\
&	2460977.70748	&	0.05	&	0.10	\\
&	2460977.71263	&	-0.03	&	0.10	\\
&	2460977.74714	&	-0.01	&	0.11	\\
&	2460994.51709	&	0.14	&	0.04	\\
&	2460994.52183	&	0.18	&	0.04	\\
&	2460994.52686	&	0.25	&	0.04	\\
&	2460994.53190	&	0.14	&	0.04	\\
&	2460994.53692	&	0.12	&	0.03	\\
&	2460994.54194	&	0.07	&	0.03	\\
&	2460994.54698	&	0.02	&	0.03	\\
&	2460994.55200	&	-0.06	&	0.03	\\
&	2460994.55703	&	-0.07	&	0.03	\\
&	2460994.56206	&	-0.13	&	0.03	\\
&	2460994.56709	&	-0.19	&	0.03	\\
&	2460994.57212	&	-0.21	&	0.03	\\
&	2460994.59751	&	-0.27	&	0.02	\\
&	2461031.54431	&	0.15	&	0.12	\\
&	2461031.55436	&	0.10	&	0.12	\\
&	2461031.55939	&	0.11	&	0.10	\\
&	2461031.56441	&	0.13	&	0.12	\\
&	2461031.56943	&	0.14	&	0.12	\\
&	2461031.57948	&	0.13	&	0.12	\\
&	2461031.58453	&	0.12	&	0.11	\\
&	2461031.58954	&	0.03	&	0.10	\\
&	2461031.59456	&	-0.02	&	0.11	\\
&	2461031.59959	&	-0.09	&	0.10	\\
&	2461031.60462	&	-0.18	&	0.10	\\
&	2461031.60963	&	-0.15	&	0.10	\\
&	2461031.61466	&	-0.26	&	0.10	\\
&	2461031.61969	&	-0.21	&	0.10	\\
&	2461032.53948	&	-0.08	&	0.05	\\
&	2461032.54449	&	-0.04	&	0.05	\\
&	2461032.54951	&	0.03	&	0.05	\\
&	2461032.55455	&	0.11	&	0.05	\\
&	2461032.57396	&	0.17	&	0.05	\\
&	2461032.59634	&	0.02	&	0.05	\\
&	2461032.62215	&	-0.21	&	0.05	\\
&	2461060.53939	&	-0.07	&	0.07	\\
&	2461060.54412	&	0.02	&	0.06	\\
&	2461060.54913	&	0.06	&	0.06	\\
\hline 
(79978) 1999~CC$_{158}$ &   &   &   \\
&	2461080.81671	&	-0.01	&	0.03	\\
&	2461080.85749	&	0.07	&	0.04	\\
&	2461080.89704	&	0.05	&	0.04	\\
&	2461080.93487	&	-0.04	&	0.04	\\
&	2461080.97254	&	-0.03	&	0.05	\\
&	2461081.01074	&	-0.04	&	0.05	\\
&	2461081.03253	&	0.00	&	0.06	\\
\hline 
2015~AR$_{293}$ &   &   &   \\
&	2460796.69685	&	-0.04	&	0.08	\\
&	2460796.70406	&	-0.16	&	0.09	\\
&	2460796.71965	&	-0.10	&	0.10	\\
&	2460796.72709	&	0.03	&	0.05	\\
&	2460796.73422	&	0.06	&	0.04	\\
&	2460796.80952	&	0.02	&	0.10	\\
&	2460796.81945	&	-0.09	&	0.04	\\
&	2460796.84174	&	-0.15	&	0.04	\\
&	2460796.86327	&	-0.10	&	0.03	\\
&	2460796.87281	&	-0.05	&	0.04	\\
&	2460815.71197	&	-0.08	&	0.04	\\
&	2460815.74156	&	-0.03	&	0.04	\\
&	2460815.77214	&	0.11	&	0.05	\\
&	2460815.80050	&	0.11	&	0.05	\\
&	2460815.82813	&	-0.11	&	0.04	\\
&	2461067.87311	&	0.06	&	0.05	\\
&	2461067.90751	&	0.04	&	0.04	\\
&	2461067.93368	&	-0.10	&	0.03	\\
\hline 
(38084) 1999~HB$_{12}$ &   &   &   \\
&	2458642.76091	&	0.03	&	0.03	\\
&	2458642.78314	&	-0.02	&	0.03	\\
&	2458642.80500	&	-0.05	&	0.03	\\
&	2458642.82679	&	-0.06	&	0.03	\\
&	2458642.84895	&	0.05	&	0.03	\\
&	2458642.87064	&	0.06	&	0.03	\\
\hline 
(524365) 2001~XQ$_{254}$ &   &   &   \\
&	2458893.74416	&	-0.06	&	0.04	\\
&	2458893.78218	&	-0.03	&	0.04	\\
&	2458893.80999	&	0.10	&	0.06	\\
&	2460353.77279	&	-0.06	&	0.05	\\
&	2460353.81090	&	0.16	&	0.07	\\
&	2460353.85042	&	0.10	&	0.06	\\
&	2460353.88265	&	-0.03	&	0.06	\\
&	2460353.91495	&	-0.15	&	0.05	\\
&	2460353.94627	&	-0.10	&	0.06	\\
&	2460353.97787	&	-0.01	&	0.07	\\
&	2460353.99819	&	0.00	&	0.07	\\
&	2460354.01836	&	0.08	&	0.09	\\ 
&	2460706.74666	&	-0.10	&	0.07	\\
&	2460706.78496	&	0.01	&	0.07	\\
&	2460706.82453	&	0.19	&	0.08	\\
&	2460706.85815	&	0.09	&	0.06	\\
&	2460706.89330	&	-0.18	&	0.05	\\
&	2460706.92797	&	-0.11	&	0.05	\\
&	2460707.02333	&	0.11	&	0.08	\\
&	2460767.66577	&	0.04	&	0.10	\\
&	2460767.70379	&	-0.16	&	0.08	\\
&	2460767.73393	&	-0.06	&	0.09	\\
&	2460767.76459	&	-0.12	&	0.09	\\
&	2460767.79520	&	0.11	&	0.10	\\
&	2460767.85785	&	-0.08	&	0.09	\\ 
\hline 
2005~XN$_{113}$ &   &   &   \\
&	2458759.81612	&	0.08	&	0.04	\\
&	2458759.84389	&	0.09	&	0.04	\\
&	2458759.87045	&	0.02	&	0.04	\\
&	2458759.89414	&	0.03	&	0.04	\\
&	2458759.91875	&	-0.03	&	0.03	\\
&	2458759.94317	&	-0.05	&	0.03	\\
&	2458759.96718	&	-0.09	&	0.03	\\
&	2458759.99112	&	-0.05	&	0.03	\\
&	2458760.00918	&	-0.01	&	0.04	\\
&	2458762.77596	&	0.00	&	0.04	\\
&	2458762.80912	&	0.04	&	0.03	\\
&	2458762.84769	&	0.05	&	0.03	\\
&	2458762.88629	&	0.04	&	0.03	\\
&	2458762.92426	&	-0.03	&	0.03	\\
&	2458762.95687	&	-0.05	&	0.04	\\
&	2458762.98761	&	-0.04	&	0.03	\\
&	2460706.63421	&	0.00	&	0.03	\\
&	2460706.67031	&	0.08	&	0.04	\\
&	2460706.70955	&	0.04	&	0.03	\\
&	2460706.75605	&	-0.01	&	0.03	\\
&	2460706.80492	&	-0.11	&	0.04	\\
\hline 
2009~YG$_{19}$ &   &   &   \\
&	2460233.88001	&	0.07	&	0.02	\\
&	2460233.91367	&	-0.07	&	0.02	\\
&	2460233.93754	&	-0.07	&	0.02	\\
&	2460233.96213	&	0.08	&	0.02	\\
\hline 
(531017) 2012~BA$_{155}$ &   &   &   \\
&	2458542.61600	&	0.01	&	0.03	\\
&	2458542.63476	&	0.00	&	0.03	\\
&	2458542.67398	&	-0.03	&	0.03	\\
&	2458542.74650	&	0.05	&	0.03	\\
&	2458542.81904	&	-0.01	&	0.03	\\
&	2458543.60783	&	-0.02	&	0.02	\\
&	2458543.66846	&	-0.03	&	0.02	\\
&	2458543.80556	&	-0.02	&	0.02	\\
&	2458544.68384	&	0.04	&	0.02	\\
\hline 
2013~LZ$_{36}$ &   &   &   \\
&	2458988.80288	&	-0.08	&	0.05	\\
&	2458988.83520	&	0.04	&	0.05	\\
&	2458988.90553	&	0.07	&	0.04	\\
&	2458988.93303	&	0.09	&	0.05	\\
&	2458988.94663	&	0.01	&	0.05	\\
&	2458988.95258	&	-0.05	&	0.05	\\
&	2458988.95852	&	-0.06	&	0.05	\\
&	2458988.96450	&	-0.03	&	0.07	\\
\hline 
2013~GY$_{136}$ &   &   &   \\
&	2459348.74814	&	0.04	&	0.05	\\
&	2459348.76905	&	0.03	&	0.05	\\
&	2459348.82774	&	0.02	&	0.04	\\
&	2459348.85806	&	0.00	&	0.04	\\
&	2459348.88678	&	-0.03	&	0.04	\\
&	2459348.91448	&	-0.06	&	0.05	\\
\hline 
2013~RZ$_{108}$ &   &   &   \\
&	2460637.61755	&	-0.11	&	0.07	\\
&	2460637.64678	&	-0.11	&	0.07	\\
&	2460637.68393	&	0.05	&	0.07	\\
&	2460637.72070	&	0.20	&	0.08	\\
&	2460637.75716	&	0.25	&	0.07	\\
&	2460637.79338	&	-0.19	&	0.06	\\
&	2460637.82894	&	-0.07	&	0.06	\\
&	2460934.82973	&	0.11	&	0.07	\\
&	2460934.87209	&	0.04	&	0.07	\\
&	2460934.96038	&	-0.04	&	0.07	\\
&	2460934.98500	&	-0.07	&	0.06	\\
&	2460939.80835	&	-0.07	&	0.06	\\
&	2460939.86519	&	-0.02	&	0.06	\\
&	2460939.91453	&	0.04	&	0.07	\\
&	2460939.94846	&	0.20	&	0.07	\\
&	2460980.69089	&	0.06	&	0.06	\\
&	2460980.72588	&	-0.14	&	0.05	\\
&	2460980.76130	&	-0.11	&	0.05	\\
&	2460980.79726	&	-0.14	&	0.06	\\
&	2460980.83323	&	-0.01	&	0.06	\\
&	2460980.86888	&	0.09	&	0.05	\\
&	2460981.00338	&	0.06	&	0.05	\\
&	2460981.01167	&	-0.01	&	0.09	\\
\hline 
(523713) 2014~JX$_{80}$ &   &   &   \\
&	2458642.85576	&	0.12	&	0.02	\\
&	2458642.87777	&	0.06	&	0.02	\\
&	2458642.89211	&	0.04	&	0.02	\\
&	2458642.90607	&	-0.01	&	0.02	\\
&	2458642.92003	&	-0.04	&	0.02	\\
&	2458642.93441	&	-0.06	&	0.02	\\
&	2458642.94837	&	-0.11	&	0.02	\\
&	2459077.70402	&	0.02	&	0.02	\\
&	2459077.71783	&	0.01	&	0.02	\\
&	2459077.73044	&	-0.01	&	0.02	\\
&	2459077.77078	&	-0.02	&	0.02	\\
&	2460123.81682	&	0.07	&	0.04	\\
&	2460123.85156	&	0.04	&	0.03	\\
&	2460123.88337	&	0.00	&	0.03	\\
&	2460123.90734	&	-0.03	&	0.03	\\
&	2460123.92976	&	-0.08	&	0.03	\\
\hline 
2014~SW$_{373}$ &   &   &   \\
&	2460939.81652	&	0.06	&	0.05	\\
&	2460939.85689	&	0.02	&	0.04	\\
&	2460939.90629	&	-0.05	&	0.04	\\
&	2460939.94006	&	-0.04	&	0.04	\\
&	2460939.95662	&	-0.01	&	0.04	\\
&	2460939.98386	&	0.02	&	0.04	\\
\hline 
(544430) 2014~UW$_{224}$ &   &   &   \\
&	2460353.69008	&	-0.07	&	0.03	\\
&	2460353.72879	&	-0.06	&	0.03	\\
&	2460353.84317	&	-0.04	&	0.03	\\
&	2460353.87595	&	0.03	&	0.03	\\
&	2460353.90817	&	-0.02	&	0.03	\\
&	2460353.93999	&	0.04	&	0.04	\\
&	2460353.97162	&	0.12	&	0.05	\\
&	2461080.65387	&	-0.04	&	0.05	\\
&	2461080.67854	&	-0.03	&	0.05	\\
&	2461080.70988	&	0.00	&	0.04	\\
&	2461080.75884	&	0.01	&	0.03	\\
&	2461080.80020	&	0.05	&	0.03	\\
&	2461080.84121	&	0.09	&	0.04	\\
&	2461080.88081	&	0.02	&	0.04	\\
&	2461080.91868	&	0.02	&	0.04	\\
&	2461080.95644	&	-0.06	&	0.05	\\
&	2461080.99442	&	-0.08	&	0.06	\\
\hline 
2014~YL$_{50}$ &   &   &   \\
&	2458819.62942	&	-0.04	&	0.03	\\
&	2458819.69811	&	0.00	&	0.03	\\
&	2458819.72920	&	0.02	&	0.03	\\
&	2458819.76362	&	0.02	&	0.04	\\
&	2458819.79779	&	0.00	&	0.04	\\
\hline 
2015~BC$_{529}$ &   &   &   \\
&	2459221.89589	&	0.01	&	0.03	\\
&	2459221.92042	&	-0.04	&	0.06	\\
&	2459221.94376	&	-0.07	&	0.07	\\
&	2459221.96095	&	-0.05	&	0.03	\\
&	2459221.97918	&	0.01	&	0.03	\\
&	2459221.99689	&	0.03	&	0.04	\\
&	2459222.03199	&	0.03	&	0.04	\\
&	2459222.04401	&	0.04	&	0.04	\\
&	2459222.05597	&	0.05	&	0.04	\\
&	2460291.88364	&	-0.10	&	0.03	\\
&	2460291.90942	&	-0.04	&	0.03	\\
&	2460291.93547	&	0.01	&	0.03	\\
&	2460291.96169	&	0.08	&	0.03	\\
&	2460291.98796	&	0.05	&	0.03	\\
&	2460755.67841	&	0.05	&	0.04	\\
&	2460755.70980	&	0.01	&	0.03	\\
&	2460755.75827	&	-0.07	&	0.03	\\
\hline 
(535991) 2015~BD$_{529}$ &   &   &   \\
&	2458893.70503	&	0.04	&	0.03	\\
&	2458893.77583	&	0.00	&	0.04	\\
&	2458893.80328	&	0.03	&	0.04	\\
&	2458893.83081	&	-0.05	&	0.04	\\
&	2458893.87091	&	-0.06	&	0.04	\\
&	2458893.89160	&	0.00	&	0.05	\\
&	2458893.91144	&	0.03	&	0.05	\\
&	2458893.93133	&	0.02	&	0.05	\\
&	2460291.84734	&	-0.09	&	0.04	\\
&	2460291.89227	&	0.02	&	0.04	\\
&	2460291.91809	&	0.05	&	0.04	\\
&	2460291.94416	&	0.06	&	0.04	\\
&	2460291.97052	&	0.02	&	0.04	\\
&	2460291.99649	&	0.00	&	0.04	\\
&	2460292.02213	&	-0.06	&	0.04	\\
\hline 
2015~KH$_{174}$ &   &   &   \\
&	2459348.77652	&	0.01	&	0.08	\\
&	2459348.79844	&	-0.07	&	0.07	\\
&	2459348.82060	&	-0.10	&	0.06	\\
&	2459348.90729	&	0.09	&	0.08	\\
&	2459348.93485	&	0.07	&	0.08	\\
&	2459348.95565	&	-0.01	&	0.08	\\
&	2460462.86123	&	-0.07	&	0.06	\\
&	2460462.89091	&	-0.05	&	0.07	\\
&	2460462.90570	&	0.04	&	0.08	\\
&	2460462.92149	&	0.08	&	0.11	\\
\enddata
\end{deluxetable}


\bibliography{biblio}{}
\bibliographystyle{aasjournalv7}



\end{document}